\documentclass[12pt]{article}
\usepackage{amsmath,amsfonts,amssymb}
\usepackage{bbm}
\usepackage[letterpaper, hmargin=1.0in, vmargin=0.75in]{geometry}

\numberwithin{equation}{section}

\newcommand{\be}{\begin{eqnarray}}
\newcommand{\ee}{\end{eqnarray}}
\newcommand{\non}{\nonumber}
\newcommand{\id}{\mathbb{I}}
\newcommand{\tr}{\mathop{\rm tr}\nolimits}
\newcommand{\str}{\mathop{\rm str}\nolimits}
\newcommand{\diag}{\mathop{\rm diag}\nolimits}

\newcommand{\SSS}{\ensuremath{\mathbb{S}}}
\newcommand{\TTT}{\ensuremath{\mathbb{T}}}
\newcommand{\FFF}{\ensuremath{\mathbb{F}}}
\newcommand{\ttt}{\ensuremath{\mathbbm{t}}}

\newcommand{\EQ}{\begin{equation}}
\newcommand{\EN}{\end{equation}}

\newcommand{\bear}{\begin{eqnarray}}
\newcommand{\ear}{\end{eqnarray}}
\newcommand{\bt} { \begin{tabular} }
\newcommand{\et}{ \end{tabular} }
\newcommand{\bc} { \begin{center} }
\newcommand{\ec}{ \end{center} }

\newcommand{\btb} { \begin{table} }
\newcommand{\etb}{ \end{table} }

\newcommand{\ihalf}{\sfrac{i}{2}}
\newcommand{\sfrac}[2]{{\textstyle\frac{#1}{#2}}}
\newcommand{\indup}[1]{_{\mathrm{#1}}}
\newcommand{\lrbrk}[1]{\left(#1\right)}

\newcommand{\vect}[1]{\mathbf{#1}}
\newcommand{\mat}[1]{\mathbf{#1}}

\newcommand{\trans}{{\scriptscriptstyle\mathrm{T}}}
\newcommand{\earel}[1]{\mathrel{}&\hspace{-2\arraycolsep}#1\hspace{-2\arraycolsep}&\mathrel{}}
\newcommand{\eq}{\earel{=}}

\begin{document}

\begin{titlepage}
\strut\hfill UMTG--268
\vspace{.5in}
\begin{center}

{\LARGE Twisted Bethe equations from a twisted $S$-matrix}

\vspace{1in}
Changrim Ahn \footnote{
Department of Physics and Institute for the
Early Universe, Ewha Womans University, DaeHyun
11-1, Seoul 120-750, S. Korea; ahn@ewha.ac.kr},
Zoltan Bajnok \footnote{
Theoretical Physics Research Group,
Hungarian Academy of Sciences, 1117 Budapest, P\'azm\'any s. 1/A
Hungary; bajnok@elte.hu}, 
Diego Bombardelli ${}^{1,}$\footnote{Physics Department and Theoretical Physics Center, 
University of Porto, Rua do Campo Alegre 687, 4169-007 Porto, 
Portugal; diego.bombardelli@fc.up.pt} and 
Rafael I. Nepomechie \footnote{
Physics Department, P.O. Box 248046,
University of Miami, Coral Gables, FL 33124, USA; nepomechie@physics.miami.edu}

\end{center}

\vspace{.5in}

\begin{abstract}
All-loop asymptotic Bethe equations for a 3-parameter deformation of
$AdS_{5}/CFT_{4}$ have been proposed by Beisert and Roiban.  We
propose a Drinfeld-Reshetikhin twist of the $AdS_{5}/CFT_{4}$
$S$-matrix, together with $c$-number diagonal twists of the boundary
conditions, from which we derive these Bethe equations.  Although the
undeformed $S$-matrix factorizes into a product of two $su(2|2)$
factors, the deformed $S$-matrix cannot be so factored.
Diagonalization of the corresponding transfer matrix requires a
generalization of the conventional algebraic Bethe ansatz approach,
which we first illustrate for the simpler case of the twisted $su(2)$
principal chiral model.  We also demonstrate that the transfer matrix
is spectrally equivalent to a transfer matrix which is constructed
using instead untwisted $S$-matrices and boundary conditions with
operatorial twists.
\end{abstract}

\end{titlepage}

\setcounter{footnote}{0}

\section{Introduction}

One of the major triumphs of theoretical physics in this century has
been the discovery and exploitation of integrability in AdS/CFT. (For
recent reviews, see for example \cite{reviews}.)  This integrability
is remarkably robust.  In particular, it seems to persist for the
so-called $\beta$-deformed $AdS_{5}/CFT_{4}$ theory 
\cite{LeiStr}-\cite{AdLvT}, and even for a 3-parameter deformation 
\cite{LunMal}-\cite{BR}.
However, much remains to be understood about the integrability of the
deformed theory, and we have sought to make progress
toward that end.

We focus here on the problem of deriving the Bethe equations.  For the
undeformed theory, the all-loop asymptotic Bethe equations
\cite{BeiSta} have been derived \cite{Beisert, MM} from the
$AdS_{5}/CFT_{4}$ $S$-matrix \cite{Staudacher}-\cite{Arutyunov:2009ga}.  For the
3-parameter deformed theory, corresponding all-loop asymptotic Bethe
equations were proposed by Beisert and Roiban in \cite{BR}.  A long
outstanding question has been whether it is possible to derive these
deformed Bethe equations from some deformed $S$-matrix.
\footnote{Recently these equations were obtained from a twisted
transfer matrix solution of the Y-system \cite{GLM} corresponding to
operatorial twisted boundary conditions \cite{AdLvT}.} Here we
answer this question in the affirmative: the deformed Bethe
equations {\em can} be derived from a certain Drinfeld-Reshetikhin twist
\cite{drinfeld1, drinfeld2} of the $AdS_{5}/CFT_{4}$ $S$-matrix, together with
$c$-number diagonal twists of the boundary conditions.

A key point is that, although the undeformed $S$-matrix factorizes
into a product of two $su(2|2)$ factors, the deformed $S$-matrix
cannot be so factored.  Indeed, the twist matrix connects the two
$su(2|2)$ factors, and cannot be factorized into a product of separate
twist matrices for the two $su(2|2)$ factors.  To our knowledge, such
``non-factoring'' twists have not been considered previously in the
literature, and it is not obvious how to diagonalize the corresponding
transfer matrix.  Indeed, since the transfer matrix no longer splits
into a product of commuting left and right pieces, one would naively expect that
such a twist leads to very complicated Bethe equations.  Hence, before
addressing the problem of actual interest, we first consider the
simpler case of the $su(2)$ principal chiral model with a
non-factoring twist.  We develop techniques for this model which we
subsequently use to solve the twisted $AdS_{5}/CFT_{4}$ problem.

The outline of this paper is as follows.  In Sec.  \ref{sec:heuristic}
we present a heuristic argument to infer the specific non-factoring
Drinfeld-Reshetikhin twist of the $su(2|2)^{2}$ $S$-matrix and twisted boundary
conditions which should lead to the twisted Bethe equations in
\cite{BR}.  In Sec.  \ref{sec:PCM} we consider the $su(2)$ principal
chiral model with a similar non-factoring twist.  We develop an
algebraic Bethe ansatz method for diagonalizing the transfer matrix
and deriving the Bethe equations.  In Sec.  \ref{sec:Hubbard} we
consider two copies of the Hubbard model with a non-factoring twist,
which is technically very similar to the twisted $AdS_{5}/CFT_{4}$
problem.  Using the method of the previous section together with the
algebraic Bethe ansatz for the Hubbard model developed by Martins and
Ramos \cite{MR}, we diagonalize the transfer matrix and derive the
corresponding Bethe equations.  Finally, in Sec.  \ref{sec:AdS/CFT},
we consider the problem of actual interest; namely, the
$AdS_{5}/CFT_{4}$ $S$-matrix with a non-factoring twist.  We obtain
the eigenvalues of the transfer matrix from the preceding results, and
write down the corresponding Bethe equations.  In Sec.  \ref{sec:BR}
we show that these Bethe equations agree with those proposed in
\cite{BR}.  We close in Sec.  \ref{sec:disc} with a brief discussion
of our results.  Some technical details are treated in Appendices
\ref{sec:exchange} and \ref{sec:crossing}.  We show in Appendix
\ref{sec:Luscher} that the proposed twisted $S$-matrix and twisted
boundary conditions reproduce the wrapping correction not only for the
Konishi operator \cite{ABBN} but also for generic multiparticle states
both in the $su(2)$ and $sl(2)$ sectors analyzed in \cite{GLM, AdLvT}.
We demonstrate in Appendix \ref{sec:optwist} that the transfer matrix
is spectrally equivalent to a transfer matrix which is constructed
using instead {\em untwisted} $S$-matrices and boundary conditions with
{\em operatorial} twists. It is the latter type of transfer
matrix which is considered in \cite{AdLvT}.
Finally, in Appendix \ref{sec:sl2} we transform our twisted
Bethe ansatz results from the ``$su(2)$'' grading to the ``$sl(2)$''
grading.

\section{Deformation from a $psu(2,2\vert 4)$ perspective}\label{sec:heuristic}

The first indication of the equivalence between type-$IIB$ string
theory on an $AdS_{5}\times S^{5}$ background and $\mathcal{N}=4$
supersymmetric four-dimensional Yang-Mills theory was their common
global symmetry: namely, $psu(2,2\vert 4)$.  On the string-theory
side it is the supersymmetric extension of the isometries of the
background geometry, while on the gauge-theory side it is
the model's superconformal invariance.  This common symmetry algebra
enables one to compare observables: both the energy levels of the string
states and the anomalous dimensions of gauge-invariant operators are
organized in the same $psu(2,2\vert 4)$ multiplet.  Recent studies
suggest that the complete solution of the spectral problem can be
formulated in terms of the $Y$-system of $psu(2,2\vert 4)$, too
\cite{Gromov:2010vb}.  Unfortunately, however, there has not been much
progress yet in solving directly the $Y$-system beyond the leading
weak-coupling order \cite{Gromov:2009tv}, or outside of the string
semiclassical domain \cite{Gromov:2009tq}.

Alternatively, quantization based on the light-cone gauge has proved
to be successful so far: In solving the string $\sigma$-model one
chooses a generalized light-cone gauge, which turns the model into a
massive integrable quantum field theory in a finite volume (prescribed
by the light-cone momentum), where excitations satisfy the
level-matching condition.  In the infinite volume limit these
excitations scatter via a factorizing scattering matrix, which can be
uniquely determined from the remaining $psu(2,2\vert 4)\to su(2\vert
2)\otimes su(2\vert 2)$ global symmetry together with crossing
symmetry \cite{Staudacher}-\cite{Arutyunov:2009ga}.  The resulting $S$-matrix can be
used for any value of the coupling, and defines the theory completely: The full
particle spectrum can be read off from its singularity structure, it
governs the finite-size corrections to the energies, and via the
Thermodynamic Bethe Ansatz it describes the complete spectrum for any
finite volume.  Unfortunately, the $psu(2,2\vert 4)$ symmetry is
broken in this description by the light-cone gauge and is realized
only implicitly in the spectrum.  The analogue of this phenomenon can
be found on the gauge-theory side: In calculating the anomalous
dimension of an operator, a BPS ``vacuum'' state is chosen $\tr
(Z^{L})$ which breaks the superconformal $psu(2,2\vert4)$ symmetry 
down to $su(2\vert 2)\otimes su(2\vert 2)$.  The broken symmetry then controls
the scattering of the ``excitations''
$\tr(Z^{L-k-2}\chi_{1}Z^{k}\chi_{2})$ over the background, and
determines their scattering matrix.  The boundary condition is
provided again by the physical meaning of the trace: namely, the
total momentum has to vanish.  Similarly to the string case, the
$psu(2,2\vert 4)$ symmetry is not manifested in the scattering matrix,
but rather in the structure of the one-loop Bethe ansatz and
implicitly in the anomalous dimensions of the fields.

Integrable deformations of the $psu(2,2\vert 4)$ structures appear
both on the string-theory and on the gauge-theory sides.  In both
cases, the $su(4)$ part of the symmetry is Drinfeld-Reshetikhin twisted
\cite{drinfeld1, drinfeld2} by three parameters/charges corresponding to the
Cartan generators.  On the gauge-theory side, the authors of \cite{BR}
present the deformed $psu(2,2\vert 4)$ one-loop Bethe ansatz equations
and conjecture the all-loop generalizations.  These Bethe ans\"atze
can be equivalently described by the asymptotical solution of a
twisted $Y$-system, which presumably originates from the twisted
$psu(2,2\vert 4)$ symmetry.

As the $S$-matrix approach has turned out to be very powerful in the
undeformed case, we pursue it for the deformed case too.  For this we
must understand how a twist of the broken $psu(2,2\vert 4)$ symmetry
shows up at the unbroken $su(2\vert 2)\otimes su(2\vert 2)$ level.  To
this end, we first analyze a toy model: an $su(4)$-invariant spin
chain.  There we identify two effects: a twist of the scattering
matrix of the excitations, and a twist of the boundary conditions.
Implementing the analogous twist at the $psu(2,2\vert 4)$ level, we
can infer the form of the twisted AdS/CFT scattering matrix together
with the twisted boundary conditions.  In Secs.
\ref{sec:AdS/CFT}-\ref{sec:disc} we test our proposal against the
Bethe equations of \cite{BR} and the L\"uscher corrections.

\subsection{Twisted $su(4)$ spin chain}

We first recall how to break the $su(4)$ symmetry down to $su(3)$
by choosing a pseudovacuum, and how the $su(3)$-invariant scattering
matrix appears in this context. We then turn to the twisted problem. 

Suppose we would like to solve the spectral problem for an $su(4)$
spin chain with $N$ sites. It is defined in terms of the $su(4)$-invariant $S$-matrix
\be 
\SSS(u)= u\mathbb{I}\otimes\mathbb{I}+i\mathcal{P} \,,
\ee 
where $\mathbb{I}$ is the 4-dimensional identity matrix, and $\mathcal{P}$
is the $16\times 16$ permutation matrix. 
We are interested
in the eigenvalues of the transfer matrix 
\be
\ttt(u)=\tr_{a}(\TTT_{a}(u))=\tr_{a}(\prod_{j=1}^{N}\SSS_{aj}(u))\,.
\ee 
These eigenvalues can be calculated by the nested algebraic 
Bethe ansatz method. Here we are
interested only in the first step of the nesting. This means to choose
a pseudovacuum state 
\be 
\vert 0\rangle=\vert 1,\dots,1\rangle\equiv\vert 1^{N}\rangle \,,
\ee 
and to analyze the excitations ($2,3,4)$ over this background, which are
invariant only under the unbroken $su(3)$ subgroup.  In keeping with this 
residual symmetry, we decompose the monodromy matrix as 
\be
\TTT(u)=\left(\begin{array}{cccc}
A(u) & B_{2}(u) & B_{3}(u) & B_{4}(u)\\
C_{2}(u) & D_{22}(u) & D_{23}(u) & D_{24}(u)\\
C_{3}(u) & D_{32}(u) & D_{33}(u) & D_{34}(u)\\
C_{4}(u) & D_{42}(u) & D_{43}(u) & D_{44}(u)\end{array}\right)\,,
\ee 
where the $C_{i}$'s together with 
$D_{i\neq j}$ annihilate the pseudovacuum; i.e.,
$C_{i}(u)\vert0\rangle=D_{i\neq j}(u)\vert0\rangle=0$. The diagonal elements
of the monodromy matrix (which contribute to the trace) act 
diagonally \footnote{We use the convention 
$\SSS(u)_{kl}^{mn}= u 
\delta_{k}^{m}\delta_{l}^{n}+i\delta_{k}^{n}\delta_{l}^{m}$.}
\be
A(u)\vert 0\rangle= a(u) \vert 0\rangle\,, \qquad
D_{kk}(u)\vert 0\rangle=
\SSS_{k1}^{k1}(u)^{N}\vert 0\rangle= d_{k}(u)\vert 0\rangle \,,
\ee
and the $B_{i}$'s create the three $su(3)$ excitations. A general multiparticle
$B$-state has the form 
\be 
B_{i_{1}}(v_{1})\dots B_{i_{K}}(v_{K})\vert 0\rangle \,,
\ee 
and we would like to diagonalize the action of $A$ and $D_{kk}$ on
these states, as this is needed to obtain the transfer matrix
eigenvalues.  To do so, we need the commutation relations of the
various operators, which can be obtained from the $\SSS \TTT 
\TTT=\TTT \TTT \SSS$ relations.
The $B$-particles are exchanged as
\be 
B_{i}(u)B_{j}(v)= (u-v)B_{j}(v)B_{i}(u)
+iB_{i}(v)B_{j}(u) ={\cal S}_{ij}^{kl}(u-v)B_{l}(v)B_{k}(u) \,,
\ee 
showing that they scatter on each other with the $su(3)$-invariant
$S$-matrix, which we denote by ${\cal S}$. The action of $A$ on the multiparticle
state can be computed from 
\be 
A(u)B_{j}(v)=\frac{v-u+i}{v-u}B_{j}(v)A(u)-\frac{i}{v-u}B_{j}(u)A(v)\,.
\ee 
In computing the eigenvalue, we focus on the first ``wanted'' term and
neglect the second ``unwanted'' one, as its contribution will vanish
when $v_{i}$ satisfy the Bethe equations. Similarly, the wanted terms
resulting from acting with $D_{ki}$ are 
\be
D_{ki}(u)B_{j}(v) & = & \frac{1}{(u-v)}\left[(u-v)B_{j}(v)D_{ki}(u)
+iB_{i}(v)D_{kj}(u)\right]+\dots \non \\
 & \propto & {\cal S}_{ij}^{lm}(u-v)B_{m}(v)D_{kl}(u)+\dots
\ee  
Clearly, the action of $D_{ki}$ mixes up the $su(3)$ indices of a
given state, and we have to diagonalize the following expression 
\be 
\lefteqn{D_{kk}(u)B_{i_{1}}(v_{1})\dots B_{i_{K}}(v_{K})\vert 
0\rangle}\non \\
&&\propto {\cal S}_{k, i_{1}}^{k_{1} j_{1}}(u-v_{1})\dots {\cal S}_{k_{K-1} 
i_{K}}^{k, j_{K}}(u-v_{K})
B_{j_{1}}(v_{1})\dots B_{j_{K}}(v_{K})D_{kk}(u)\vert 0\rangle+\dots 
\ee 
As $D_{k\neq i}\vert 0\rangle=0$, the nonvanishing elements form a trace
of the reduced $su(3)$ transfer matrix
\be 
t(u)=\sum_{k=2}^{4}{\cal S}_{k, i_{1}}^{k_{1}j_{1}}(u-v_{1})\dots 
{\cal S}_{k_{K-1}i_{K}}^{k, j_{K}}(u-v_{K}) d_{k}(u)\,,
\ee 
which must be diagonalized in order to finally solve the $su(4)$ eigenvalue
problem. However, we shall not pursue this problem further here.

We now would like to instead twist the $su(4)$ scattering matrix by a 
Drinfeld-Reshetikhin twist \footnote{The general notion of twisting 
for (quasi-triangular) quasi-Hopf algebras was introduced by Drinfeld 
\cite{drinfeld1}. Reshetikhin considered \cite{drinfeld2} specific 
twists with elements $F$ constructed from the Cartan 
generators. It is the latter type of twist which we use here. For other 
applications of such twists, see for example \cite{drinfeld3}; and 
for work related to quantized braided algebras, see \cite{referee}.}
\be 
\SSS\to\tilde{\SSS}=\FFF\, \SSS\, \FFF\,, \qquad 
\FFF=e^{\frac{i}{2}\sum_{i,j=1}^{3}\gamma_{ij}(H_{i}\otimes H_{j}-H_{j}\otimes 
H_{i})} \,,
\ee 
where $H_{i}$ are the Cartan elements of $su(4)$: 
$(H_{i})_{kj}=\frac{1}{2}(\delta_{i,k}\delta_{i,j}-\delta_{i+1,j}\delta_{i+1,k})$.
Due to the special form of the scattering matrix, only the diagonal
elements are twisted 
\be 
\tilde{\SSS}(u)_{kl}^{mn}= u\Gamma_{kl}\delta_{k}^{m}\delta_{l}^{n}
+i\delta_{k}^{n}\delta_{l}^{m}\,,
\ee 
which can be encoded in the matrix $\Gamma$. We are interested in
the eigenvalues of the twisted transfer matrix
\be 
\tilde{\ttt}(u)=\tr_{a}(\tilde{\TTT}_{a}(u))=\tr_{a}(\prod_{j=1}^{N}\tilde{\SSS}_{aj}(u))\,.
\ee 
The pseudovacuum state $\vert 0\rangle=\vert 1,\dots,1\rangle=\vert 1^{N}\rangle$
is annihilated by the generators 
$\tilde{C}_{i}$ and $\tilde{D}_{i\neq j}$,
and it is an eigenstate of the diagonal elements
\be 
\tilde{A}(u)\vert 0\rangle=a(u)\vert 0\rangle\,, \qquad
\tilde{D}_{kk}(u)\vert 
0\rangle= \tilde{\SSS}_{k1}^{k1}(u)^{N}\vert 
0\rangle=\tilde{d}_{k}(u)\vert 0\rangle\,.
\ee 
The eigenvalues are now different for each $k$, and depend on
the twist as $\tilde{d}_{k}(u)=(\Gamma_{k1})^{N}d_{k}(u)$. This indicates
that the reduced symmetry is not even $su(3)$, which can be seen also
from the way that the twisted creation operators $\tilde{B}_{i}$ are exchanged: 
\be 
\tilde{B}_{i}(u)\tilde{B}_{j}(v)= 
 (u-v)\Gamma_{ij}\tilde{B}_{j}(v)\tilde{B}_{i}(u)+i\tilde{B}_{i}(v)\tilde{B}_{j}(u) 
=\tilde{\cal S}_{ij}^{kl}(u-v)\tilde{B}_{l}(v)\tilde{B}_{k}(u)\,,
\ee 
exactly with the reduced twisted scattering matrix elements. From 
the point of view  of
the reduced transfer matrix, the relevant commutation
relation is twisted as follows
\be 
\tilde{D}_{ki}(u)\tilde{B}_{j}(v)  \propto  
(\Gamma_{k 1})^{-1}\tilde{\cal S}_{ij}^{lm}(u-v)\tilde{B}_{m}(v)\tilde{D}_{kl}(u)+\dots
\ee 
It will result in the twisted reduced transfer matrix 
\be 
\tilde{t}(u)  &=&  \sum_{k=2}^{4}\tilde{\cal S}_{k, i_{1}}^{k_{1} 
j_{1}}(u-v_{1})
\dots\tilde{\cal S}_{k_{K-1} i_{K}}^{k, j_{K}}(u-v_{K}) 
(\Gamma_{k 1})^{-K}\tilde{d}_{k}(u) \non \\
&=& \sum_{k=2}^{4}\tilde{\cal S}_{k, i_{1}}^{k_{1}j_{1}}(u-v_{1}) \dots
\tilde{\cal S}_{k_{K-1}i_{K}}^{k, j_{K}}(u-v_{K}) (\Gamma_{k 
1})^{N-K} d_{k}(u)\,,
\ee 
which must still be diagonalized at the $su(3)$ level. 

Focusing on the effect of the twist, we can see the emergence of two
main features: (i) the reduced twisted scattering matrix $\tilde{\cal S}$
is a reduction of the twisted scattering matrix $\tilde{\SSS}$; and (ii)
there is a twisted boundary condition which depends on the number of
sites $N$ and the number of particles $K$.  The twisted boundary
condition contains the twist factors $\Gamma_{k1}$, which are
unphysical from the reduced-space point of view, as it is spanned
only by ($2,3,4$).  Let us see how we can implement a similar twist in
the AdS/CFT realm, where the full $psu(2,2\vert 4)$ scattering matrix
is not yet known.

\subsection{Proposed twists for AdS/CFT}

We would like to develop the AdS/CFT case in parallel to our $su(4)$ 
example. The
full symmetry of the model which we wish to twist is the implicit
$psu(2,2\vert 4)$ symmetry. According to \cite{BR}, one should twist 
the $su(4)$ $R$-symmetry subgroup (which corresponds to the isometries of the $S^{5}$ part
of $AdS_{5}\times S^{5}$) by its charges $R_{1},R_{2},R_{3}$. 
Let us suppose that the full $psu(2,2\vert 4)$
scattering matrix $\SSS$ were known. Let us twist it with the charges as follows
\be 
\tilde{\SSS}=\FFF\, \SSS\, \FFF\,, \qquad 
\FFF=e^{\frac{i}{2}\sum_{i,j=1}^{3}\gamma_{ij}(R_{i}\otimes R_{j}-R_{j}\otimes 
R_{i})} \,.
\ee 
Here $\SSS$ describes how in the gauge-theory side the excitations
$X_{1}=X,X_{2}=Y,X_{3}=Z,\Psi,D,\dots$ scatter on each other; and
the action of the charges on the scalars are given by 
\be 
R_{i}\vert X_{j}\rangle=\delta_{ij}\vert X_{j}\rangle\,.
\ee 
In describing the AdS/CFT integrable model, the pseudovacuum is usually
chosen to be $\vert 0\rangle=\vert Z^{J}\rangle$. We are interested
in the remaining degrees of freedom, which we  consider to be 
excitations over this background. Clearly, in the undeformed case this
choice of the vacuum breaks the $psu(2,2\vert 4)$ symmetry down to 
$su(2\vert2)\otimes su(2\vert2)$,
and this reduced symmetry is what labels the excitations:
\be 
(1,2,3,4)\otimes(\dot{1},\dot{2},\dot{3},\dot{4})\,,
\ee 
where as usual the $R$-symmetry acts in the first two components.
Choosing $X=1\dot{1}$ and $Y=2\dot{1}$, one can see that
\be
R_{1}=1\otimes h + h\otimes1\,, \quad 
R_{2}=1\otimes h - h\otimes1\,, \quad R_{3}=0\,,
\ee 
where $h=\diag(\frac{1}{2},-\frac{1}{2},0,0)$. 

The two effects of the twist implemented on $\SSS$ by $\FFF$
read on the reduced level as follows. First, consider how the reduced
scattering matrix is twisted. Since $R_{3}=0$, the twist factors given
by $\gamma_{13}$ and $\gamma_{23}$ have no effect, while the $\gamma_{12}$
twist propagates directly through. This means that the reduction of
the unknown twisted scattering matrix $\tilde \SSS$ must be simply 
the Drinfeld-Reshetikhin
twist of the reduced scattering matrix,
\be 
\tilde{\cal S}=F {\cal S} F\, \qquad F=e^{i\gamma_{12}(R_{1}\otimes R_{2}-R_{2}\otimes R_{1})}
=e^{2i\gamma_{12}(h\otimes1\otimes1\otimes h - 1\otimes h\otimes h\otimes1)}\,,
\label{nonfactoring}
\ee 
where ${\cal S}$ is the $su(2\vert2)\otimes su(2\vert2)$-invariant $AdS/CFT$
scattering matrix. 
Second, consider the twists of the boundary conditions for the
particles, which come from the charge of the background and the
commutation relations of the operators.  Both have the form
\be
e^{2i(\gamma_{13}R_{1}+\gamma_{23}R_{2})J}=
e^{2i(\gamma_{13}-\gamma_{23})(h\otimes1)J+2i(\gamma_{13}+\gamma_{23})(1\otimes h)J}
= e^{2i(\gamma_{13}-\gamma_{23}) J h} \otimes e^{2i(\gamma_{13}+\gamma_{23}) J h}\,,
\label{TBC}
\ee 
where $J$, being the $R_{3}$ charge of the background, is related
to the volume of our theory. 

As previewed in the Introduction, the twist matrix $F$ in
(\ref{nonfactoring}) connects the two $su(2|2)$ factors in ${\cal S}$, and
cannot be factorized into a product of separate twist matrices for the
two $su(2|2)$ factors.  In order to learn how to handle such twists,
we consider in the following section the analogous problem for a
simpler model.

\section{Twisting the $su(2)$ principal chiral model}\label{sec:PCM}

We define the $su(2)$-invariant $S$-matrix $S(u)$ by 
\be
S(u) = S_{0}(u) \left[ u\, \id\otimes \id  + i {\cal P} \right]
\ee
where $\id$ is the $2 \times 2$ unit matrix, ${\cal P}$ is the $4
\times 4$ permutation matrix, and $S_{0}(u)$ is some scalar factor 
whose explicit value will not concern us here. 
This $S$-matrix acts on $V \otimes V$, where $V$ is a 
2-dimensional vector space.  
The $S$-matrix ${\cal S}(u)$ of the $su(2)$ principal chiral model is
given by a tensor product of two copies of $S(u)$ \cite{ZZ1, ZZ2}.
That is,
\be
{\cal S}_{a\, \dot a\, b\, \dot b }(u) = S_{ab}(u)\, S_{\dot a  \dot b}(u) \,.
\ee 
Our convention is to arrange the four vector spaces on which ${\cal S}$ acts in the order $V_{a} \otimes 
V_{\dot a} \otimes  V_{b} \otimes V_{\dot b}$. Hence,
\be
{\cal S}_{1234}(u) = S_{13}(u)\,  S_{24}(u) \,.
\ee
Starting from $S_{12}=S \otimes \id \otimes \id$, one can sequentially construct
\be
S_{13} &=& {\cal P}_{23} S_{12} {\cal P}_{23} \,, \non \\
S_{23} &=& {\cal P}_{12} S_{13} {\cal P}_{12} \,, \non \\
S_{24} &=& {\cal P}_{34} S_{23} {\cal P}_{34} \,,
\ee
where ${\cal P}_{12} = {\cal P}\otimes \id \otimes \id$, etc.

In view of our proposal (\ref{nonfactoring}), we consider the Drinfeld-Reshetikhin twist 
\be 
\tilde {\cal S}(u) = F\, {\cal S}(u)\, F \,, 
\ee 
where the twist matrix $F$ is given by
\be 
F = e^{i\gamma_{1} (h \otimes \id \otimes \id \otimes h - 
\id \otimes h \otimes h \otimes \id )} \,, 
\label{twist1}
\ee 
where $\gamma_{1}$ is a twist parameter, and 
$h$ is the diagonal matrix
\be
h = \diag(\frac{1}{2},-\frac{1}{2}) \,.
\label{hPCM}
\ee
As already mentioned, $\id$ is the $2 \times 2$ unit matrix.  Note that $F$ cannot be factored
between the two $su(2)$ factors.

We consider the transfer 
matrix \footnote{Ultimately, we shall need the eigenvalues of an inhomogeneous transfer matrix, 
with inhomogeneities $\theta_{j}$ at each site $j$. But once one 
understands how to solve the homogeneous problem, it is trivial to 
generalize to the inhomogeneous case.}
\be
\tilde t(u) = \tr_{a \dot a} M_{a \dot a} \tilde  {\cal T}_{a \dot a}(u) \,,
\label{twistedtransferPCM}
\ee
where the monodromy matrix is constructed from the twisted $S$-matrix 
as follows
\be
\tilde {\cal T}_{a \dot a}(u)
= \tilde {\cal S}_{a\, \dot a\, 1\, \dot 1 
}(u) \cdots \tilde {\cal S}_{a\, \dot a\, L\, \dot L 
}(u) \,.
\label{twistedmonodromyPCM}
\ee
The matrix $M_{a \dot a}$, which acts only in the auxiliary space and serves to 
twist the boundary conditions, is given by (see (\ref{TBC}))
\be
M = e^{i \gamma_{2} h} \otimes e^{i \gamma_{3} h} = 
\diag\left( e^{i (\gamma_{2}+\gamma_{3})/2}, e^{i (\gamma_{2}-\gamma_{3})/2}, 
e^{i(\gamma_{3}-\gamma_{2})/2}, e^{-i(\gamma_{2}+\gamma_{3})/2}\right) \,,
\label{MPCM}
\ee 
where $\gamma_{2}, \gamma_{3}$ are additional twist parameters. 
The twisted $S$-matrix $\tilde {\cal S}(u)$ by construction 
\cite{drinfeld2}
obeys the Yang-Baxter equation, and therefore, the twisted monodromy 
matrix obeys the usual intertwining relation \footnote{Our case does
not seem to be related to quantized braided algebras where the
Yang-Baxter equation is also braided.}
\be
\tilde {\cal S}_{a\, \dot a\, b\, \dot b }(u-v) \, 
\tilde {\cal T}_{a \dot a}(u)\,
\tilde {\cal T}_{b \dot b}(v) =
\tilde {\cal T}_{b \dot b}(v)\,
\tilde {\cal T}_{a \dot a}(u)\,
\tilde {\cal S}_{a\, \dot a\, b\, \dot b }(u-v) \,.
\ee
Also $\left[\tilde {\cal S}(u)
\,, M \otimes M \right]=0$; and therefore \cite{Sk}, the transfer matrix
(\ref{twistedtransferPCM}) has the commutativity property
$\left[\tilde t (u) \,, \tilde t (v) \right]=0$.
The main problem is to determine the eigenvalues of the transfer
matrix.  Since it is not evident how to solve this problem, it is
helpful to begin with the untwisted case.

\subsection{Untwisted case: conventional approach}\label{subsec:conventional}

We now consider the untwisted case; i.e., $\gamma_{i} = 0$, and 
therefore both $F$ and $M$ are 1.
In this case, the monodromy matrix is given by
\be
{\cal T}_{a \dot a}(u)
&=& {\cal S}_{a\, \dot a\, 1\, \dot 1 
}(u) \cdots  {\cal S}_{a\, \dot a\, L\, \dot L 
}(u) \non \\
&=&  S_{a 1}(u) S_{\dot a \dot 1}(u)  \ldots    S_{a L}(u) S_{\dot a 
\dot L}(u) \non \\
&=& S_{a 1}(u) \ldots  S_{a L}(u)\ S_{\dot a \dot 1}(u) \ldots  
S_{\dot a \dot L}(u) \non \\
&=& T_{a}(u)\ T_{\dot a}(u)  \,. \label{monoprod}
\ee
Hence,
\be
{\cal T} = \left( \begin{array}{cc}
A & B \\
C & D 
\end{array}\right) \otimes \left( \begin{array}{cc}
\dot A & \dot B \\
\dot C & \dot D 
\end{array}\right) 
= \left( \begin{array}{cccc}
A \dot A & A \dot B & B \dot A & B  \dot B \\
A \dot C & A \dot D & B \dot C & B  \dot D \\
C \dot A & C \dot B & D \dot A & D  \dot B \\
C \dot C & C \dot D & D \dot C & D  \dot D
\end{array}\right) 
\label{monodromy}
\ee 

The transfer matrix therefore factors into a product of two commuting pieces
\be
t(u) = \tr_{a \dot a}  {\cal T}_{a \dot a}(u) = 
\Big[ A(u) + D(u) \Big]
\Big[ \dot A(u) + \dot D(u) \Big] \,.
\ee 

We make the ansatz that the eigenstates of the transfer matrix are given by
\be
|\Lambda \rangle =
\prod_{j=1}^{m} B(u_{j}) |0\rangle \otimes 
\prod_{k=1}^{\dot m} \dot B(\dot u_{k})   |\dot 0\rangle \,,
\label{ansatz1}
\ee
where $|0\rangle$ and $|\dot 0\rangle$ are states with all spins up.

Evidently, the problem has factored into two copies of the XXX
spin-1/2 chain, whose solution is well known.  The eigenvalues
$\Lambda(u)$ of the transfer matrix can therefore be easily written
down,
\be
\Lambda(u) &=& S_{0}(u)^{2L}\left[ (u+i)^{L}\prod_{j=1}^{m} 
\left(\frac{u-u_{j}-i}{u-u_{j}}\right) + u^{L} \prod_{j=1}^{m} 
\left(\frac{u-u_{j}+i}{u-u_{j}}\right) \right] \non  \\
& & \times \left[ (u+i)^{L}\prod_{k=1}^{\dot m} 
\left(\frac{u-\dot u_{k}-i}{u-\dot u_{k}}\right) + u^{L} \prod_{k=1}^{\dot m} 
\left(\frac{u-\dot u_{k}+i}{u-\dot u_{k}}\right) \right] \,.
\label{eigenvals1}
\ee
In principle, one should derive the Bethe equations by carefully tracking the 
``unwanted'' terms, and demanding that they cancel; however, this is 
a tedious computation. In practice, it is much simpler to impose
the requirement that the poles of the eigenvalues should cancel.\footnote{Actually, 
since here the eigenvalues (\ref{eigenvals1}) factor into a product
\be
\Lambda(u) = \lambda(u) \dot \lambda(u) \,, \non 
\ee
we encounter the following interesting subtlety. One possibility 
(which we believe is the correct one) is 
to {\em separately} require the cancellation of poles in $\lambda(u)$ and  
$\dot \lambda(u)$, which leads to (\ref{BAEs1}). Alternatively, one could require only the 
cancellation of poles in $\Lambda(u)$, which is a weaker condition.
This leads to additional 
(spurious) Bethe ansatz-like equations which couple the left 
and right Bethe roots. We conclude that, although the trick of 
obtaining the Bethe equations by requiring cancellation of poles can save a lot of 
effort, it should be applied with care.}
In this way, we readily obtain the following Bethe equations:
\footnote{These equations may look strange. However, they can be recast in 
the more familiar (symmetric) form by shifting all the Bethe roots by 
$i/2$; i.e., $u_{j} \mapsto u_{j}-i/2\,,  \dot u_{j} \mapsto \dot u_{j}-i/2$.}
\be
\left( \frac{u_{j} +i}{u_{j}}\right)^{L} = \prod_{j' \ne j}^{m}
\frac{u_{j} - u_{j'}+i}{u_{j} - u_{j'}-i} \,, \qquad
\left( \frac{\dot u_{k} +i}{\dot u_{k}}\right)^{L} = \prod_{k' \ne k}^{\dot m}
\frac{\dot u_{k} - \dot u_{k'}+i}{\dot u_{k} - \dot u_{k'}-i} \,.
\label{BAEs1}
\ee 
For further details, see Appendix A in \cite{ZZ2} and Appendix C in \cite{GKV}.

\subsection{Untwisted case: new approach}

We have described above the ``obvious'' way to solve the untwisted problem. 
However, this approach cannot be used to solve the twisted problem, since 
then the transfer matrix does not factor into left and right pieces. 
So, now we want to solve the untwisted problem again but in a different way, {\em without} 
exploiting the factorizability of the transfer matrix. The basic idea is to 
develop an algebraic Bethe ansatz for the ``full'' monodromy matrix 
(\ref{monodromy}).

Using the well-known exchange relations between $A, B, C, D$, 
together with the result (\ref{monodromy}),
it is not difficult to show that  (see Appendix \ref{sec:exchange})
\be
{\cal T}_{11}(u)\, {\cal T}_{13}(v) &=& 
\frac{u-v-i}{u-v} {\cal T}_{13}(v) \, {\cal T}_{11}(u) +
\frac{i}{u-v} {\cal T}_{13}(u)\, {\cal T}_{11}(v) \,, \non \\
{\cal T}_{11}(u)\, {\cal T}_{12}(v) &=& 
\frac{u-v-i}{u-v} {\cal T}_{12}(v) \, {\cal T}_{11}(u) +
\frac{i}{u-v} {\cal T}_{12}(u)\, {\cal T}_{11}(v) \,, \non \\
{\cal T}_{22}(u)\, {\cal T}_{12}(v) &=& 
\frac{u-v+i}{u-v} {\cal T}_{12}(v) \, {\cal T}_{22}(u) -
\frac{i}{u-v} {\cal T}_{12}(u)\, {\cal T}_{22}(v) \,, \non \\
{\cal T}_{33}(u)\, {\cal T}_{13}(v) &=& 
\frac{u-v+i}{u-v} {\cal T}_{13}(v) \, {\cal T}_{33}(u) -
\frac{i}{u-v} {\cal T}_{13}(u)\, {\cal T}_{33}(v)  \,,
\label{exchange1}
\ee 
where now the subscripts refer to matrix elements of the monodromy 
matrix regarded as a $4 \times 4$ matrix of operators.  In each 
exchange relation, the first (``diagonal'') term gives the ``wanted'' contribution, 
and the second term gives ``unwanted'' contributions.
With more effort, one can also show that  (see again Appendix \ref{sec:exchange})
\be
{\cal T}_{22}(u)\, {\cal T}_{13}(v) &=& 
\frac{u-v-i}{u-v} {\cal T}_{13}(v) \, {\cal T}_{22}(u) +
\frac{i(u-v-i)}{(u-v)^{2}}{\cal T}_{14}(v) \, {\cal T}_{21}(u)  \non \\
&+&
\frac{i}{u-v} {\cal T}_{24}(u)\, {\cal T}_{11}(v)
- \frac{i}{u-v} {\cal T}_{12}(u)\, {\cal T}_{23}(v) -
\frac{1}{(u-v)^{2}}{\cal T}_{14}(u) \, {\cal T}_{21}(v) \,, \non \\
{\cal T}_{33}(u)\, {\cal T}_{12}(v) &=& 
\frac{u-v-i}{u-v} {\cal T}_{12}(v) \, {\cal T}_{33}(u) +
\frac{i(u-v-i)}{(u-v)^{2}}{\cal T}_{14}(v) \, {\cal T}_{31}(u) \non \\
&+&
\frac{i}{u-v} {\cal T}_{34}(u)\, {\cal T}_{11}(v) 
- \frac{i}{u-v} {\cal T}_{13}(u)\, {\cal T}_{32}(v) -
\frac{1}{(u-v)^{2}}{\cal T}_{14}(u) \, {\cal T}_{31}(v) \,, \non \\
{\cal T}_{44}(u)\, {\cal T}_{12}(v) &=& 
\frac{u-v+i}{u-v} {\cal T}_{12}(v) \, {\cal T}_{44}(u) +
\frac{i(u-v+i)}{(u-v)^{2}}{\cal T}_{14}(v) \, {\cal T}_{42}(u) \non \\
&-&
\frac{i}{u-v} {\cal T}_{34}(u)\, {\cal T}_{22}(v) 
- \frac{i}{u-v} {\cal T}_{24}(u)\, {\cal T}_{32}(v) +
\frac{1}{(u-v)^{2}}{\cal T}_{14}(u) \, {\cal T}_{42}(v) \,, \non \\
{\cal T}_{44}(u)\, {\cal T}_{13}(v) &=& 
\frac{u-v+i}{u-v} {\cal T}_{13}(v) \, {\cal T}_{44}(u) +
\frac{i(u-v+i)}{(u-v)^{2}}{\cal T}_{14}(v) \, {\cal T}_{43}(u) \label{exchange2} \\
&-&
\frac{i}{u-v} {\cal T}_{24}(u)\, {\cal T}_{33}(v)
- \frac{i}{u-v} {\cal T}_{34}(u)\, {\cal T}_{23}(v) +
\frac{1}{(u-v)^{2}}{\cal T}_{14}(u) \, {\cal T}_{43}(v) \,, 
\non
\ee 
where again only the first (diagonal) term gives the ``wanted'' contribution.

We make the ansatz that the eigenstates of the transfer matrix are
given by (cf., (\ref{ansatz1}))
\be
|\Lambda \rangle = \prod_{j=1}^{m} {\cal T}_{13}(u_{j})  
\prod_{k=1}^{\dot m}{\cal T}_{12}(\dot u_{k})   
\left( |0\rangle \otimes |\dot 0\rangle \right) \,,
\label{ansatz2}
\ee

We observe that the vacuum state is an eigenstate of the diagonal
elements of the monodromy matrix,
\be
{\cal T}_{11}(u) \left( |0\rangle \otimes |\dot 0\rangle \right) &=& 
S_{0}(u)^{2L}\, (u+i)^{2L} 
\left( |0\rangle \otimes |\dot 0\rangle \right) \,, \non \\
{\cal T}_{22}(u) \left( |0\rangle \otimes |\dot 0\rangle \right) &=& 
S_{0}(u)^{2L}\, (u+i)^{L} u^{L}
\left( |0\rangle \otimes |\dot 0\rangle \right) \,, \non \\
{\cal T}_{33}(u) \left( |0\rangle \otimes |\dot 0\rangle \right) &=& 
S_{0}(u)^{2L}\, (u+i)^{L} u^{L}
\left( |0\rangle \otimes |\dot 0\rangle \right) \,, \non \\
{\cal T}_{44}(u) \left( |0\rangle \otimes |\dot 0\rangle \right) &=&  
S_{0}(u)^{2L}\, u^{2L}
\left( |0\rangle \otimes |\dot 0\rangle \right) \,, 
\label{vaceigenval}
\ee

The transfer matrix is evidently given by the sum of the diagonal
elements of the monodromy matrix,
\be
t(u) = {\cal T}_{11}(u) +{\cal T}_{22}(u) +{\cal T}_{33}(u) +{\cal 
T}_{44}(u) \,.
\ee
Acting with this operator on the states (\ref{ansatz2}),
we use (in the standard way) the first term of the exchange 
relations (\ref{exchange1}), (\ref{exchange2}) together with 
(\ref{vaceigenval}) to obtain the eigenvalue,
\be
t(u) |\Lambda \rangle &=& S_{0}(u)^{2L} \Big\{
(u+i)^{2L}\prod_{j=1}^{m} \left(\frac{u-u_{j}-i}{u-u_{j}}\right)  
\prod_{k=1}^{\dot m} \left(\frac{u-\dot u_{k}-i}{u-\dot u_{k}}\right) 
\non \\
&&+
(u+i)^{L}u^{L} \prod_{j=1}^{m} \left(\frac{u-u_{j}-i}{u-u_{j}}\right)  
\prod_{k=1}^{\dot m} \left(\frac{u-\dot u_{k}+i}{u-\dot u_{k}}\right)
\non \\
&& +
(u+i)^{L}u^{L} \prod_{j=1}^{m} \left(\frac{u-u_{j}+i}{u-u_{j}}\right)  
\prod_{k=1}^{\dot m} \left(\frac{u-\dot u_{k}-i}{u-\dot u_{k}}\right)
\non \\
&&+
u^{2L} \prod_{j=1}^{m} \left(\frac{u-u_{j}+i}{u-u_{j}}\right)  
\prod_{k=1}^{\dot m} \left(\frac{u-\dot u_{k}+i}{u-\dot u_{k}}\right) 
\Big\} |\Lambda \rangle  + \mbox{ ``unwanted'' } \,.
\ee
We observe that the eigenvalue coincides with our previous result 
(\ref{eigenvals1}), and so we again obtain the Bethe equations 
(\ref{BAEs1}).

\subsection{Twisted case}

We are finally ready to tackle the twisted case. We have verified that only the 
diagonal elements of the monodromy matrix  $\tilde {\cal T}$ 
(\ref{twistedmonodromyPCM}) are 
affected by the twist (\ref{twist1}). (This result is a 
consequence of the special structure of the $S$-matrix.)
Hence, we can hope that the exchange 
relations  (\ref{exchange1}), (\ref{exchange2}) suffer only  
deformations of the coefficients; and this is exactly what we find.
Indeed, with the help of Mathematica, we find 
\be
\tilde {\cal T}_{11}(u)\, \tilde {\cal T}_{13}(v) &=& 
e^{i \gamma_{1}} \left( \frac{u-v-i}{u-v} \tilde {\cal T}_{13}(v) \, \tilde {\cal T}_{11}(u) +
\frac{i}{u-v} \tilde {\cal T}_{13}(u)\, \tilde {\cal T}_{11}(v) \right) \,, \non \\
\tilde {\cal T}_{11}(u)\, \tilde {\cal T}_{12}(v) &=& 
e^{-i \gamma_{1}} \left( \frac{u-v-i}{u-v} \tilde {\cal T}_{12}(v) \, \tilde {\cal T}_{11}(u) +
\frac{i}{u-v} \tilde {\cal T}_{12}(u)\, \tilde {\cal T}_{11}(v) \right) \,, \non \\
\tilde {\cal T}_{22}(u)\, \tilde {\cal T}_{12}(v) &=& 
e^{-i \gamma_{1}} \left(\frac{u-v+i}{u-v} \tilde {\cal T}_{12}(v) \, \tilde {\cal T}_{22}(u) -
\frac{i}{u-v} \tilde {\cal T}_{12}(u)\, \tilde {\cal T}_{22}(v) \right) \,, \non \\
\tilde {\cal T}_{33}(u)\, \tilde {\cal T}_{13}(v) &=& 
e^{i \gamma_{1}} \left(\frac{u-v+i}{u-v} \tilde {\cal T}_{13}(v) \, \tilde {\cal T}_{33}(u) -
\frac{i}{u-v} \tilde {\cal T}_{13}(u)\, \tilde {\cal T}_{33}(v)  \right) \,,
\label{exchange1twisted}
\ee 
and
\be
\tilde {\cal T}_{22}(u)\, \tilde {\cal T}_{13}(v) &=& 
e^{-i \gamma_{1}} \Big( \frac{u-v-i}{u-v} \tilde {\cal T}_{13}(v) \, \tilde {\cal T}_{22}(u) +
\frac{i(u-v-i)}{(u-v)^{2}}\tilde {\cal T}_{14}(v) \, \tilde {\cal T}_{21}(u) \non \\
&+&
e^{i \gamma_{1}}\frac{i}{u-v} \tilde {\cal T}_{24}(u)\, \tilde {\cal T}_{11}(v) 
- \frac{i}{u-v} \tilde {\cal T}_{12}(u)\, \tilde {\cal T}_{23}(v) -
\frac{1}{(u-v)^{2}}\tilde {\cal T}_{14}(u) \, \tilde {\cal T}_{21}(v) \Big) \,, \non \\
\tilde {\cal T}_{33}(u)\, \tilde {\cal T}_{12}(v) &=& 
e^{i \gamma_{1}} \Big( \frac{u-v-i}{u-v} \tilde {\cal T}_{12}(v) \, \tilde {\cal T}_{33}(u) +
\frac{i(u-v-i)}{(u-v)^{2}}\tilde {\cal T}_{14}(v) \, \tilde {\cal T}_{31}(u) \non \\
&+&
e^{-i \gamma_{1}}\frac{i}{u-v} \tilde {\cal T}_{34}(u)\, \tilde {\cal T}_{11}(v) 
- \frac{i}{u-v} \tilde {\cal T}_{13}(u)\, \tilde {\cal T}_{32}(v) -
\frac{1}{(u-v)^{2}}\tilde {\cal T}_{14}(u) \, \tilde {\cal T}_{31}(v) \Big) \,, \non \\
\tilde {\cal T}_{44}(u)\, \tilde {\cal T}_{12}(v) &=& 
e^{i \gamma_{1}} \frac{u-v+i}{u-v} \tilde {\cal T}_{12}(v) \, \tilde {\cal T}_{44}(u) +
\frac{i(u-v+i)}{(u-v)^{2}}\tilde {\cal T}_{14}(v) \, \tilde {\cal T}_{42}(u) \non \\
&-&
\frac{i}{u-v} \tilde {\cal T}_{34}(u)\, \tilde {\cal T}_{22}(v) 
- \frac{i}{u-v} \tilde {\cal T}_{24}(u)\, \tilde {\cal T}_{32}(v) +
\frac{1}{(u-v)^{2}}\tilde {\cal T}_{14}(u) \, \tilde {\cal T}_{42}(v)  \,, \non \\
\tilde {\cal T}_{44}(u)\, \tilde {\cal T}_{13}(v) &=& 
e^{-i \gamma_{1}}\frac{u-v+i}{u-v} \tilde {\cal T}_{13}(v) \, \tilde {\cal T}_{44}(u) +
\frac{i(u-v+i)}{(u-v)^{2}}\tilde {\cal T}_{14}(v) \, \tilde {\cal T}_{43}(u) \non \\
&-&
\frac{i}{u-v} \tilde {\cal T}_{24}(u)\, \tilde {\cal T}_{33}(v) 
- \frac{i}{u-v} \tilde {\cal T}_{34}(u)\, \tilde {\cal T}_{23}(v) +
\frac{1}{(u-v)^{2}}\tilde {\cal T}_{14}(u) \, \tilde {\cal T}_{43}(v) \,. 
\label{exchange2twisted}
\ee 

Moreover, the vacuum eigenvalues (\ref{vaceigenval}) become
\be
\tilde {\cal T}_{11}(u) \left( |0\rangle \otimes |\dot 0\rangle 
\right) &=& S_{0}(u)^{2L}\, (u+i)^{2L} 
\left( |0\rangle \otimes |\dot 0\rangle \right) \,, \non \\
\tilde {\cal T}_{22}(u) \left( |0\rangle \otimes |\dot 0\rangle 
\right) &=& S_{0}(u)^{2L}\, e^{i \gamma_{1} 
L}\, (u+i)^{L} u^{L}
\left( |0\rangle \otimes |\dot 0\rangle \right) \,, \non \\
\tilde {\cal T}_{33}(u) \left( |0\rangle \otimes |\dot 0\rangle 
\right) &=& S_{0}(u)^{2L}\, e^{-i \gamma_{1} 
L}\, (u+i)^{L} u^{L}
\left( |0\rangle \otimes |\dot 0\rangle \right) \,, \non \\
\tilde {\cal T}_{44}(u) \left( |0\rangle \otimes |\dot 0\rangle 
\right) &=&  S_{0}(u)^{2L}\, u^{2L}
\left( |0\rangle \otimes |\dot 0\rangle \right) \,. 
\label{vaceigenvaltwisted}
\ee

The twisted transfer matrix (\ref{twistedtransferPCM}) is given by 
\be
\tilde t(u) = e^{i (\gamma_{2}+\gamma_{3})/2} \tilde {\cal T}_{11}(u) 
+ e^{i (\gamma_{2}-\gamma_{3})/2} \tilde {\cal T}_{22}(u) 
+ e^{i(\gamma_{3}-\gamma_{2})/2} \tilde {\cal T}_{33}(u) 
+ e^{-i(\gamma_{2}+\gamma_{3})/2} \tilde {\cal T}_{44}(u) \,.
\ee

Using a similar ansatz as before (\ref{ansatz2}), namely,
\be
|\tilde \Lambda \rangle = \prod_{j=1}^{m} \tilde {\cal T}_{13}(u_{j})  
\prod_{k=1}^{\dot m} \tilde {\cal T}_{12}(\dot u_{k})   
\left( |0\rangle \otimes |\dot 0\rangle \right) \,,
\ee
we find that the eigenvalues
of the twisted transfer matrix are given by 
\be
\tilde \Lambda(u) &=& S_{0}(u)^{2L}\Bigg\{
(u+i)^{2L} e^{i (\gamma_{2}+\gamma_{3})/2} e^{i \gamma_{1} (m - \dot m)}
\prod_{j=1}^{m}\left(\frac{u-u_{j}-i}{u-u_{j}}\right)  
\prod_{k=1}^{\dot m} \left(\frac{u-\dot u_{k}-i}{u-\dot u_{k}}\right) 
\non \\
&&+
(u+i)^{L}u^{L} e^{i (\gamma_{2}-\gamma_{3})/2} e^{i \gamma_{1} L}\, e^{-i \gamma_{1} (m + \dot 
m)} \prod_{j=1}^{m} \left(\frac{u-u_{j}-i}{u-u_{j}}\right)  
\prod_{k=1}^{\dot m} \left(\frac{u-\dot u_{k}+i}{u-\dot u_{k}}\right)
\non \\
&& +
(u+i)^{L}u^{L} e^{i(\gamma_{3}-\gamma_{2})/2} e^{-i \gamma_{1} L}\, e^{i \gamma_{1} (m + \dot 
m)} 
\prod_{j=1}^{m} \left(\frac{u-u_{j}+i}{u-u_{j}}\right)  
\prod_{k=1}^{\dot m} \left(\frac{u-\dot u_{k}-i}{u-\dot u_{k}}\right) 
\non \\
&&+
u^{2L} e^{-i(\gamma_{2}+\gamma_{3})/2} e^{-i \gamma_{1} (m - \dot m)} 
 \prod_{j=1}^{m} \left(\frac{u-u_{j}+i}{u-u_{j}}\right)  
\prod_{k=1}^{\dot m} \left(\frac{u-\dot u_{k}+i}{u-\dot 
u_{k}}\right)  \Bigg\} \label{eigenvalsPCM}
\,.
\ee 
Remarkably, although the transfer matrix does not seem to factor into two 
pieces, the eigenvalues do:
\be
\tilde \Lambda(u) &=& S_{0}(u)^{2L} \left[ 
c_{1}
(u+i)^{L}\prod_{j=1}^{m} 
\left(\frac{u-u_{j}-i}{u-u_{j}}\right) + 
c_{1}^{-1}
u^{L} \prod_{j=1}^{m} 
\left(\frac{u-u_{j}+i}{u-u_{j}}\right) \right] \non  \\
& & \times \left[ 
c_{2}
(u+i)^{L}\prod_{k=1}^{\dot m} 
\left(\frac{u-\dot u_{k}-i}{u-\dot u_{k}}\right) + 
c_{2}^{-1}
u^{L} \prod_{k=1}^{\dot m} 
\left(\frac{u-\dot u_{k}+i}{u-\dot u_{k}}\right) \right] \,,  
\label{eigenvals2}
\ee
where
\be
c_{1} = e^{i\gamma_{2}/2} e^{i \gamma_{1} L/2} e^{-i \gamma_{1} \dot m}\,, \quad
c_{2} = e^{i\gamma_{3}/2} e^{-i \gamma_{1} L/2} e^{i \gamma_{1} m}\,.
\label{csPCM}
\ee
We can obtain the Bethe equations (as 
in the untwisted case) using the shortcut of requiring that the poles 
cancel, \footnote{The result (\ref{BAEs2}) is similar in structure to
the Bethe ansatz result of quantum braided algebras \cite{referee}.
This may indicate that quantum braided algebras can be equivalently
described by unbraided algebras, but with twisted $R$-matrices.}
\be
\left( \frac{u_{j} +i}{u_{j}}\right)^{L} &=& e^{-i\gamma_{2}} e^{-i \gamma_{1} L}\, e^{2 
i \gamma_{1}  \dot m} \prod_{j' \ne j}^{m}
\frac{u_{j} - u_{j'}+i}{u_{j} - u_{j'}-i}  \,, \non \\
\left( \frac{\dot u_{k} +i}{\dot u_{k}}\right)^{L} &=& e^{-i\gamma_{3}} e^{i \gamma_{1} L}\, 
e^{-2i \gamma_{1}  m} \prod_{k' \ne k}^{\dot m}
\frac{\dot u_{k} - \dot u_{k'}+i}{\dot u_{k} - \dot u_{k'}-i}  \,.
\label{BAEs2}
\ee
The eigenvalues (\ref{eigenvals2}) and 
Bethe equations (\ref{BAEs2}) are the main results of this section.
Gratifyingly, these results are simple deformations of the 
corresponding untwisted results (\ref{eigenvals1}) and (\ref{BAEs1}), 
respectively.

The generalization to the case of an inhomogeneous chain, with 
inhomogeneity $\theta_{l}$ at site $l$, is now straightforward. It 
amounts to making the replacements
\be
(u+i)^{L} \mapsto \prod_{l=1}^{L} (u-\theta_{l}+i) \,, \qquad
u^{L} \mapsto \prod_{l=1}^{L} (u-\theta_{l})\,, \qquad 
S_{0}(u)^{2L}  \mapsto \prod_{l=1}^{L} S_{0}(u-\theta_{l})^{2}
\ee
in the expression (\ref{eigenvals2}) for the eigenvalues. Thus,
the eigenvalues are given by
\be
\tilde \Lambda(u) &=& \prod_{l=1}^{L} S_{0}(u-\theta_{l})^{2}  \\
&\times& \Bigg[
c_{1}
\prod_{l=1}^{L} (u-\theta_{l}+i)\prod_{j=1}^{m} 
\left(\frac{u-u_{j}-i}{u-u_{j}}\right)  + 
c_{1}^{-1}
\prod_{l=1}^{L} (u-\theta_{l}) \prod_{j=1}^{m} 
\left(\frac{u-u_{j}+i}{u-u_{j}}\right) \Bigg] \non  \\
&\times & \Bigg[
c_{2}
\prod_{l=1}^{L} (u-\theta_{l}+i)\prod_{k=1}^{\dot m} 
\left(\frac{u-\dot u_{k}-i}{u-\dot u_{k}}\right)  + 
c_{2}^{-1}
\prod_{l=1}^{L} (u-\theta_{l}) \prod_{k=1}^{\dot m} 
\left(\frac{u-\dot u_{k}+i}{u-\dot u_{k}}\right) \Bigg] \,. \non 
\label{eigenvals3}
\ee
The equations for the auxiliary Bethe roots become
\be
\prod_{l=1}^{L}\left( \frac{u_{j}-\theta_{l} +i}{u_{j}-\theta_{l}}\right) &=& 
e^{-i\gamma_{2}} e^{-i \gamma_{1} L}\, e^{2 
i \gamma_{1}  \dot m} \prod_{j' \ne j}^{m}
\frac{u_{j} - u_{j'}+i}{u_{j} - u_{j'}-i}  \,, \non \\
\prod_{l=1}^{L}\left( \frac{\dot u_{k}-\theta_{l} +i}{\dot u_{k}-\theta_{l}}\right) &=& 
e^{-i\gamma_{3}} e^{i \gamma_{1} L}\, 
e^{-2i \gamma_{1}  m} \prod_{k' \ne k}^{\dot m}
\frac{\dot u_{k} - \dot u_{k'}+i}{\dot u_{k} - \dot u_{k'}-i}  \,.
\label{BAEs3}
\ee
Finally, the Bethe-Yang equations corresponding to the middle node are 
given by 
\be
e^{-i p_{k} {\cal L}} &=& \tilde \Lambda(\theta_{k}) \non \\
&=& 
-e^{i(\gamma_{2}+\gamma_{3})/2}  e^{i \gamma_{1}(m-\dot m)}
\prod_{l\ne k}^{L} S_{0}(\theta_{k}-\theta_{l})^{2} 
(\theta_{k}-\theta_{l}+i)^{2} \non \\
&  & \times
\prod_{j=1}^{m} 
\left(\frac{\theta_{k}-u_{j}-i}{\theta_{k}-u_{j}}\right)
\prod_{k=1}^{\dot m} 
\left(\frac{\theta_{k}-\dot u_{k}-i}{\theta_{k}-\dot u_{k}}\right) 
\,,  \label{BYPCM}
\ee 
where ${\cal L}$ is the length of the ring with the $L$ 
particles of rapidities $\theta_{1}, \ldots, \theta_{L}$.

\section{Twisting two copies of the Hubbard model}\label{sec:Hubbard}

We have seen that, for the $su(2)$ principal chiral model, we were
able to obtain the Bethe equations for the case of a non-factoring
twist by developing an algebraic Bethe ansatz based on the ``full''
monodromy matrix of the two $S$-matrix factors.  In this section we
shall follow the same approach for two copies of the Hubbard model with a
non-factoring twist, which is technically very similar to the twisted
$AdS_{5}/CFT_{4}$ problem.  For a single copy of the Hubbard model,
the algebraic Bethe ansatz was worked out in the paper by Martins and Ramos
\cite{MR}, which hereafter we denote by MR.

Let $S(\lambda,\mu)$ be the $16 \times 16$ $R$-matrix of the Hubbard model, which 
was found by Shastry \cite{Shastry}. In the notation of MR
\be
S(\lambda,\mu) = {\cal P} R_{g}(\lambda,\mu) \,,
\ee
where $R_{g}(\lambda,\mu)$ is given in  MR (18), and ${\cal P}$ 
is the graded permutation matrix,
\be
{\cal P} = \sum_{a,b=1}^{4}(-1)^{p(a) p(b)} e_{a b} \otimes e_{b a} 
\,,
\ee
where the gradings are given by
\be
p(1) = p(4) = 0\,, \qquad p(2) = p(3) = 1 \,,
\label{gradings}
\ee
and 
$e_{a b}$ are the standard elementary matrices with matrix elements
\be
(e_{a b})_{ij} = \delta_{a,i} \delta_{b,j} \,.
\ee

Paralleling our discussion of the $su(2)$ principal chiral model, we
introduce the tensor product of two such $S$-matrices,
\be
{\cal S}_{a\, \dot a\, b\, \dot b }(\lambda,\mu) = S_{ab}(\lambda,\mu)\, 
S_{\dot a  \dot b}(\lambda,\mu) \,.
\ee 
We consider the Drinfeld-Reshetikhin-twisted $S$-matrix
\be 
\tilde {\cal S}(\lambda,\mu) = F\, {\cal S}(\lambda,\mu)\, F \,, 
\ee 
where the twist matrix $F$ is again given by
\be 
F = e^{i \gamma_{1} (h \otimes \id \otimes \id \otimes h - 
\id \otimes h \otimes h \otimes \id )} \,, 
\label{twist}
\ee 
except $h$ is now the diagonal matrix
\be
h=\diag(\frac{1}{2},0,0,-\frac{1}{2}) \,,
\label{h}
\ee
and $\id$ is the $4 \times 4$ unit matrix.  Note that $F$ cannot be factored 
into matrices acting separately on the 13 space and the 24 space.

The main problem is to determine the eigenvalues of the transfer 
matrix
\be
\tilde t(\lambda) = \str_{a \dot a}  M_{a \dot a} \tilde {\cal T}_{a \dot a}(\lambda) \,,
\label{Hubbardtransfertwisted}
\ee
where the monodromy matrix is given by 
\be
\tilde {\cal T}_{a \dot a}(\lambda)
= \tilde {\cal S}_{a\, \dot a\, 1\, \dot 1 
}(\lambda,0) \cdots \tilde {\cal S}_{a\, \dot a\, L\, \dot L 
}(\lambda,0) \,,
\label{Hubbardmonodromtwisted}
\ee
and the matrix $M_{a \dot a}$ is given by
\be
e^{i \gamma_{2} h} \otimes e^{i \gamma_{3} h} \,.
\label{MHubbard}
\ee 
We follow the same approach as for the twisted $su(2)$ principal
chiral model.  The first step is to understand the untwisted case.

\subsection{Untwisted case}

We now consider the untwisted case; i.e., $\gamma_{i} = 0$, and therefore 
both $F$ and $M$ are 1.
In this case, the monodromy matrix is given by (see (\ref{monoprod})),
\be
{\cal T}_{a \dot a}(\lambda)
= T_{a}(\lambda)\ T_{\dot a}(\lambda)  \,.
\ee
We label the matrix elements of $T_{a}(\lambda)$ as in MR (21);
and similarly for the matrix elements of $T_{\dot a}(\lambda)$, except 
we also decorate those with a dot. Hence,
\be
{\cal T}(\lambda) = \left( \begin{array}{cccc}
B & B_{1} & B_{2} & F \\
C_{1} & A_{11} & A_{12} & B_{1}^{*} \\
C_{2} & A_{21} & A_{22} & B_{2}^{*} \\
C & C_{1}^{*} & C_{2}^{*} & D 
\end{array}\right) \otimes
\left( \begin{array}{cccc}
\dot B & \dot B_{1} & \dot B_{2} & \dot F \\
\dot C_{1} &\dot A_{11} &\dot A_{12} &\dot B_{1}^{*} \\
\dot C_{2} &\dot A_{21} &\dot A_{22} &\dot B_{2}^{*} \\
\dot C &\dot C_{1}^{*} &\dot C_{2}^{*} &\dot D 
\end{array}\right) \,.
\ee
We regard ${\cal T}(\lambda)$ as a $16 \times 16$ matrix of 
operators. We denote these operators by their matrix 
elements, i.e., ${\cal T}_{j,k}(\lambda)$, where $j, k \in \{1, 
\ldots, 16\}$. In particular,
\be
{\cal T}_{1,1}(\lambda) &=& B(\lambda)\, \dot B(\lambda) \,, \qquad
{\cal T}_{1,2}(\lambda) = B(\lambda)\, \dot B_{1}(\lambda) \,,  \qquad
{\cal T}_{1,3}(\lambda) = B(\lambda)\, \dot B_{2}(\lambda) \,, \non \\
{\cal T}_{1,5}(\lambda) &=& B_{1}(\lambda)\, \dot B(\lambda) \,, \qquad
{\cal T}_{1,9}(\lambda) = B_{2}(\lambda)\, \dot B(\lambda) \,. 
\ee 

Since the one-particle states are given by MR (44), it is reasonable 
to consider the tensor-product states
\be
|\Phi \rangle =  \left[ {\cal T}_{1,5}( \lambda_{1})   {\cal F}^{1} 
+ {\cal T}_{1,9}( \lambda_{1})   {\cal F}^{2} \right]
\left[ {\cal T}_{1,2}(\dot \lambda_{1})  \dot {\cal F}^{1} 
+ {\cal T}_{1,3}(\dot \lambda_{1}) \dot {\cal F}^{2}  \right] 
\left(|0\rangle  \otimes |\dot 0\rangle \right) \,.
\label{ansatz}
\ee

The exchange relations of the 
diagonal elements ${\cal T}_{j,j}(\lambda)$ with the creation 
operators ${\cal T}_{1,2}(\mu), {\cal T}_{1,3}(\mu),  {\cal T}_{1,5}(\mu), {\cal 
T}_{1,9}(\mu)$ are discussed in Appendix \ref{sec:exchange}. We explicitly record here just the 
``wanted'' (diagonal) terms,
\be
{\cal T}_{j,j}(\lambda)\, {\cal T}_{1,2}(\mu) &=& f_{l(j)}(\lambda, 
\mu)\, {\cal T}_{1,2}(\mu)\, {\cal T}_{j,j}(\lambda) + \ldots \,, \non \\
{\cal T}_{j,j}(\lambda)\, {\cal T}_{1,3}(\mu) &=& g_{l(j)}(\lambda, 
\mu)\, {\cal T}_{1,3}(\mu)\, {\cal T}_{j,j}(\lambda) + \ldots \,, \non \\
{\cal T}_{j,j}(\lambda)\, {\cal T}_{1,5}(\mu) &=& f_{k(j)}(\lambda, 
\mu)\, {\cal T}_{1,5}(\mu)\, {\cal T}_{j,j}(\lambda) + \ldots \,, \non \\
{\cal T}_{j,j}(\lambda)\, {\cal T}_{1,9}(\mu) &=& g_{k(j)}(\lambda, 
\mu)\, {\cal T}_{1,9}(\mu)\, {\cal T}_{j,j}(\lambda) + \ldots \,,
\quad j = 1, \ldots, 16, 
\label{exrlnHubbard}
\ee
where $k(j)$ and $l(j)$ are functions of $j$ such that 
\footnote{Hence, $(k,l)=(1,1)$ for $j=1$; and $(k,l)=(1,2)$ for 
$j=2$, etc.}
\be
j = 4(k-1)+l\,,  \qquad j = 1, \ldots, 16, \qquad k\,, l \in \{1,2,3,4\} \,,
\label{kl}
\ee
and we have defined $f_{i}(\lambda, \mu)$ and $g_{i}(\lambda, \mu)$ by
\be
f_{1}(\lambda, \mu) &=& g_{1}(\lambda, \mu) = \frac{i 
\alpha_{2}(\mu,\lambda)}{\alpha_{9}(\mu,\lambda)} \,, \non \\
f_{2}(\lambda, \mu) &=& g_{3}(\lambda, \mu) = -\frac{i 
\alpha_{1}(\lambda,\mu)}{\alpha_{9}(\lambda,\mu)} \,, \non \\
f_{3}(\lambda, \mu) &=& g_{2}(\lambda, \mu) = - \frac{i 
\alpha_{1}(\lambda,\mu)}{\alpha_{9}(\lambda,\mu)}\bar b(\lambda,\mu) \,, \non \\
f_{4}(\lambda, \mu) &=& g_{4}(\lambda, \mu) = -\frac{i 
\alpha_{8}(\lambda,\mu)}{\alpha_{7}(\lambda,\mu)} \,.
\label{effs}
\ee
Moreover, $\alpha_{i}(\lambda,\mu)$ are defined in MR (A.1)-(A.9),
and $\bar b(\lambda,\mu)$ is defined in MR (27).
 
We observe that the pseudovacuum state $|0\rangle  \otimes |\dot 0\rangle$
is an eigenstate of the diagonal elements of the monodromy matrix,
\be
{\cal T}_{j,j}(\lambda) \left(|0\rangle \otimes |\dot 0\rangle\right)
= \left[ \phi_{k(j)}(\lambda)\, \phi_{l(j)}(\lambda) \right]^{L}
\left(|0\rangle \otimes |\dot 0\rangle\right) \,, \quad j = 1, 
\ldots, 16, 
\label{psuedovaceigenvals}
\ee 
where we have defined $\phi_{i}(\lambda)$ by
\be
\phi_{1}(\lambda) = \omega_{1}(\lambda)\,, \quad
\phi_{2}(\lambda) = \phi_{3}(\lambda) = \omega_{2}(\lambda)\,, \quad
\phi_{4}(\lambda) = \omega_{3}(\lambda)\,,
\label{phi}
\ee 
and $\omega_{i}(\lambda)$ are given by 
\be
\omega_{1}(\lambda) = \alpha_{2}(\lambda,0)\,, \quad
\omega_{2}(\lambda) = -i\alpha_{9}(\lambda,0)\,, \quad
\omega_{3}(\lambda) = \alpha_{7}(\lambda,0)\,.
\ee

The transfer matrix is given by 
\be
t(\lambda) &=& {\cal T}_{1,1}(\lambda) - {\cal T}_{2,2}(\lambda) - {\cal 
T}_{3,3}(\lambda) +{\cal T}_{4,4}(\lambda) \non \\
&-& \left[ {\cal T}_{5,5}(\lambda) - {\cal T}_{6,6}(\lambda) - {\cal 
T}_{7,7}(\lambda) +{\cal T}_{8,8}(\lambda) \right] \non \\
&-& \left[ {\cal T}_{9,9}(\lambda) - {\cal T}_{10,10}(\lambda) - {\cal 
T}_{11,11}(\lambda) +{\cal T}_{12,12}(\lambda) \right] \non \\
&+&  {\cal T}_{13,13}(\lambda) - {\cal T}_{14,14}(\lambda) - {\cal 
T}_{15,15}(\lambda) +{\cal T}_{16,16}(\lambda)  
\,,
\ee
where the signs are due to the supertrace.
Acting with this operator on the states (\ref{ansatz}),
we use (in the standard way) the first term of the exchange 
relations (\ref{exrlnHubbard}), together with the 
pseudovacuum eigenvalues (\ref{psuedovaceigenvals}), to obtain the eigenvalue. Upon factoring 
the result, we obtain
\be
\Lambda(\lambda) &=& \Big[ 
\omega_{1}(\lambda)^{L}
\left( \frac{i\alpha_{2}(\lambda_{1},\lambda)}{\alpha_{9}(\lambda_{1},\lambda)}\right)
+ 
\omega_{3}(\lambda)^{L} 
\left( \frac{-i\alpha_{8}(\lambda,\lambda_{1})}{\alpha_{7}(\lambda,\lambda_{1})}\right)
\non \\
& & - 
\omega_{2}(\lambda)^{L}
\left( \frac{-i\alpha_{1}(\lambda,\lambda_{1})}{\alpha_{9}(\lambda,\lambda_{1})}\right)
\Lambda^{(1)}(\lambda,\lambda_{1}) \Big] \non \\
&\times&
\Big[ 
\omega_{1}(\lambda)^{L}
\left( \frac{i\alpha_{2}(\dot \lambda_{1},\lambda)}{\alpha_{9}(\dot\lambda_{1},\lambda)}\right)
+ 
\omega_{3}(\lambda)^{L}
\left( \frac{-i\alpha_{8}(\lambda,\dot\lambda_{1})}{\alpha_{7}(\lambda,\dot\lambda_{1})}\right)
\non \\
& & - 
\omega_{2}(\lambda)^{L}
\left( \frac{-i\alpha_{1}(\lambda,\dot\lambda_{1})}{\alpha_{9}(\lambda,\dot\lambda_{1})}\right)
\Lambda^{(1)}(\lambda,\dot\lambda_{1})  \Big] 
\,,
\ee
where, as in MR (51),
\be
\Lambda^{(1)}(\lambda,\lambda_{1}) = 1 + \bar 
b(\lambda,\lambda_{1})\,, \qquad
\Lambda^{(1)}(\lambda,\dot\lambda_{1}) = 1 + \bar 
b(\lambda,\dot\lambda_{1})\,.
\ee 
On the basis of MR, we assume that this result can be extrapolated to the general case:
\be
\Lambda(\lambda) &=& \Big[ 
\omega_{1}(\lambda)^{L}
\prod_{j=1}^{n}\left( 
\frac{i\alpha_{2}(\lambda_{j},\lambda)}{\alpha_{9}(\lambda_{j},\lambda)}\right)
+ 
\omega_{3}(\lambda)^{L}
\prod_{j=1}^{n}\left(  
\frac{-i\alpha_{8}(\lambda,\lambda_{j})}{\alpha_{7}(\lambda,\lambda_{j})}\right)
\non \\
& & - 
\omega_{2}(\lambda)^{L}
\prod_{j=1}^{n}\left( 
\frac{-i\alpha_{1}(\lambda,\lambda_{j})}{\alpha_{9}(\lambda,\lambda_{j})}\right)
\Lambda^{(1)}(\lambda,\{\lambda_{k}\})\ \Big] \non \\
&\times&
\Big[ 
\omega_{1}(\lambda)^{L}
\prod_{j=1}^{\dot n}\left( \frac{i\alpha_{2}(\dot 
\lambda_{j},\lambda)}{\alpha_{9}(\dot\lambda_{j},\lambda)}\right)
+ 
\omega_{3}(\lambda)^{L} 
\prod_{j=1}^{\dot n}\left( 
\frac{-i\alpha_{8}(\lambda,\dot\lambda_{j})}{\alpha_{7}(\lambda,\dot\lambda_{j})}\right)
\non \\
& & - 
\omega_{2}(\lambda)^{L}
\prod_{j=1}^{\dot n}\left( 
\frac{-i\alpha_{1}(\lambda,\dot\lambda_{j})}{\alpha_{9}(\lambda,\dot\lambda_{j})}\right)
\Lambda^{(1)}(\lambda,\{\dot \lambda_{k}\}) \Big] 
\,. 
\ee
As expected, this is just the product of two copies of the result for 
a single copy of Hubbard, given in MR (89). Note that
$\Lambda^{(1)}(\lambda,\{\lambda_{k}\})$ are the eigenvalues of the 
auxiliary transfer matrix MR (92). They are given by MR (99), and 
there is a similar expression for $\Lambda^{(1)}(\lambda,\{\dot 
\lambda_{k}\})$. 

The Bethe equations can again be 
obtained using the shortcut that there should be a cancellation of the
poles in the eigenvalues of the transfer matrix; this leads 
to MR (90), (102),
\be
\left(\frac{\omega_{1}(\lambda_{j})}{\omega_{2}(\lambda_{j})}\right)^{L}
&=&\Lambda^{(1)}(\lambda_{j},\{\lambda_{k}\}) = 
\prod_{l=1}^{m}\frac{1}{\bar b(\mu_{l},\lambda_{j})}
\,, \non \\
\left(\frac{\omega_{1}(\dot \lambda_{j})}{\omega_{2}(\dot\lambda_{j})}\right)^{L}
&=&\Lambda^{(1)}(\dot\lambda_{j},\{\dot \lambda_{k}\}) = 
\prod_{l=1}^{\dot m}\frac{1}{\bar b(\dot \mu_{l},\dot \lambda_{j})}\,.
\ee 
To obtain this result, one needs the following properties of the 
functions $\alpha_{i}$
\be
\alpha_{9}(\mu,\lambda) = - \alpha_{9}(\lambda,\mu) \,, \quad
\alpha_{1}(\mu,\lambda) = \alpha_{2}(\lambda,\mu)  \,,
\ee 
which follow from their definitions, given in MR Appendix A. The 
Bethe equations for the auxiliary problem are given by MR (100), and a similar set 
for the dotted roots.

\subsection{Twisted case}

We turn now to the twisted case, for which the monodromy matrix is 
given by (\ref{Hubbardmonodromtwisted}).  The exchange relations 
suffer only deformations of the coefficients. In 
particular, the ``wanted'' terms become (cf. (\ref{exrlnHubbard}))
\be
 \tilde {\cal T}_{j,j}(\lambda)\,  \tilde {\cal T}_{1,2}(\mu) &=& e^{-i \zeta_{k(j)} 
\gamma_{1}/2} f_{l(j)}(\lambda, 
\mu)\,  \tilde {\cal T}_{1,2}(\mu)\,  \tilde {\cal T}_{j,j}(\lambda) + \ldots \,, \non \\
 \tilde {\cal T}_{j,j}(\lambda)\,  \tilde {\cal T}_{1,3}(\mu) &=& e^{-i \zeta_{k(j)} 
\gamma_{1}/2} g_{l(j)}(\lambda, 
\mu)\,  \tilde {\cal T}_{1,3}(\mu)\,  \tilde {\cal T}_{j,j}(\lambda) + \ldots \,, \non \\
 \tilde {\cal T}_{j,j}(\lambda)\,  \tilde {\cal T}_{1,5}(\mu) &=& e^{i \zeta_{l(j)} 
\gamma_{1}/2} f_{k(j)}(\lambda, 
\mu)\,  \tilde {\cal T}_{1,5}(\mu)\,  \tilde {\cal T}_{j,j}(\lambda) + \ldots \,, \non \\
 \tilde {\cal T}_{j,j}(\lambda)\,  \tilde {\cal T}_{1,9}(\mu) &=& e^{i \zeta_{l(j)} 
\gamma_{1}/2} g_{k(j)}(\lambda, 
\mu)\,  \tilde {\cal T}_{1,9}(\mu)\,  \tilde {\cal T}_{j,j}(\lambda) + \ldots \,,
\quad j = 1, \ldots, 16, 
\label{exrlnHubbardtwisted}
\ee
where we have defined
\be
\zeta_{1} = 1\,, \quad \zeta_{2} = \zeta_{3} = 0\,, \quad \zeta_{4} = 
-1\,.
\ee 
The pseudovacuum eigenvalues are now given by (cf. (\ref{psuedovaceigenvals}))
\be
 \tilde {\cal T}_{j,j}(\lambda) \left(|0\rangle \otimes |\dot 0\rangle\right)
= e^{i (\zeta_{k(j)} - \zeta_{l(j)}) \gamma_{1} L/2}
\left[ \phi_{k(j)}(\lambda)\, \phi_{l(j)}(\lambda) \right]^{L}
\left(|0\rangle \otimes |\dot 0\rangle\right) \,, \quad j = 1, 
\ldots, 16.
\ee 
We remind the reader that $k(j)$ and $l(j)$ are defined by (\ref{kl}).

Proceeding as in the untwisted case, one easily finds that the 
eigenvalues of the twisted transfer matrix 
(\ref{Hubbardtransfertwisted}) are given by
\be
\tilde \Lambda(\lambda) &=& \Big[ 
c_{1}\,
\omega_{1}(\lambda)^{L}
\prod_{j=1}^{n}\left( 
\frac{i\alpha_{2}(\lambda_{j},\lambda)}{\alpha_{9}(\lambda_{j},\lambda)}\right)
+ 
c_{1}^{-1}\,
\omega_{3}(\lambda)^{L}
\prod_{j=1}^{n}\left(  
\frac{-i\alpha_{8}(\lambda,\lambda_{j})}{\alpha_{7}(\lambda,\lambda_{j})}\right)
\non \\
& & - 
\omega_{2}(\lambda)^{L}
\prod_{j=1}^{n}\left( 
\frac{-i\alpha_{1}(\lambda,\lambda_{j})}{\alpha_{9}(\lambda,\lambda_{j})}\right)
\Lambda^{(1)}(\lambda,\{\lambda_{k}\})\ \Big] \non \\
&\times&
\Big[ 
c_{2}\,
\omega_{1}(\lambda)^{L}
\prod_{j=1}^{\dot n}\left( \frac{i\alpha_{2}(\dot 
\lambda_{j},\lambda)}{\alpha_{9}(\dot\lambda_{j},\lambda)}\right)
+ 
c_{2}^{-1}\,
\omega_{3}(\lambda)^{L} 
\prod_{j=1}^{\dot n}\left( 
\frac{-i\alpha_{8}(\lambda,\dot\lambda_{j})}{\alpha_{7}(\lambda,\dot\lambda_{j})}\right)
\non \\
& & - 
\omega_{2}(\lambda)^{L}
\prod_{j=1}^{\dot n}\left( 
\frac{-i\alpha_{1}(\lambda,\dot\lambda_{j})}{\alpha_{9}(\lambda,\dot\lambda_{j})}\right)
\Lambda^{(1)}(\lambda,\{\dot \lambda_{k}\}) \Big] 
\label{eigenvalstwisted}
\,,
\ee
where
\be
c_{1} = e^{i \gamma_{2}/2} e^{i \gamma_{1} L/2} e^{-i \gamma_{1} \dot n/2} \,, \quad
c_{2} = e^{i \gamma_{3}/2} e^{-i \gamma_{1} L/2} e^{i \gamma_{1}  n/2} \,.
\ee 

It is very important to observe that the matrix $\hat r$ given in MR 
(26) does {\em not} get deformed by the twist. Hence, the eigenvalues 
of the auxiliary transfer matrix also do not get deformed. Again 
using the shortcut, we find from (\ref{eigenvalstwisted}) that the 
Bethe equations are given by
\be
e^{i \gamma_{2}/2} e^{i \gamma_{1} L/2} e^{-i \gamma_{1} \dot n/2}
\left(\frac{\omega_{1}(\lambda_{j})}{\omega_{2}(\lambda_{j})}\right)^{L}
&=&\Lambda^{(1)}(\lambda_{j},\{\lambda_{k}\}) = 
\prod_{l=1}^{m}\frac{1}{\bar b(\mu_{l},\lambda_{j})}
\,, \non \\
e^{i \gamma_{3}/2} e^{-i \gamma_{1} L/2} e^{i \gamma_{1} n/2}
\left(\frac{\omega_{1}(\dot \lambda_{j})}{\omega_{2}(\dot\lambda_{j})}\right)^{L}
&=&\Lambda^{(1)}(\dot\lambda_{j},\{\dot \lambda_{k}\}) = 
\prod_{l=1}^{\dot m}\frac{1}{\bar b(\dot \mu_{l},\dot \lambda_{j})}\,.
\label{BAEstwisted}
\ee 
Their structure is similar to those for the 
twisted $su(2)$ principal chiral model (\ref{BAEs2}).
The Bethe equations for the auxiliary problem are again given by MR (100).

\section{Twisting $AdS_{5}/CFT_{4}$}\label{sec:AdS/CFT}

We finally come to the AdS/CFT case. Let $S(p_{1},p_{2})$ be the 
graded $su(2|2)$ $S$-matrix in \cite{AFZ, MM}, and let  
${\cal S}(p_{1},p_{2})$ be the $su(2|2)^{2}$ $S$-matrix,
\be
{\cal S}_{a\, \dot a\, b\, \dot b }(p_{1},p_{2}) = S_{ab}(p_{1},p_{2})\, 
S_{\dot a  \dot b}(p_{1},p_{2}) \,.
\label{calS}
\ee 
We consider the Drinfeld-Reshetikhin twist of this $S$-matrix
\be 
\tilde {\cal S}(p_{1},p_{2}) = F\, {\cal S}(p_{1},p_{2})\, F \,, 
\label{drinfeldtwist}
\ee 
where the twist matrix $F$ is given by (see (\ref{nonfactoring}))
\be 
F = e^{i \gamma_{1} (h \otimes \id \otimes \id \otimes h - 
\id \otimes h \otimes h \otimes \id )} \,. 
\label{twistagain}
\ee 
Here $h$ is the diagonal matrix \footnote{The difference between (\ref{h})
and (\ref{hh}) is due to a difference in basis: as already noted in 
(\ref{gradings}), in \cite{MR} the gradings are (0,1,1,0), while in 
\cite{AFZ} the gradings are (0,0,1,1).}
\be
h=\diag(\frac{1}{2},-\frac{1}{2},0,0) \,,
\label{hh}
\ee
and $\id$ is again the $4 \times 4$ unit matrix. In Appendix 
\ref{sec:crossing} we verify that the twisted $S$-matrix has the 
standard crossing symmetry; hence, the scalar factor $S_0$ is exactly the same as the
untwisted one and independent of the deformation parameter.
The (inhomogeneous) transfer matrix is given by 
\be
\tilde t(\lambda) = \str_{a \dot a} M_{a \dot a} \tilde S_{a \dot a 1 \dot 1}(\lambda,p_{1}) 
\ldots  \tilde S_{a \dot a N \dot N}(\lambda,p_{N}) \,,
\label{twistedBCagain}
\ee
where the matrix $M_{a \dot a}$ is given by (see (\ref{TBC}))
\be
M = e^{i (\gamma_{3}-\gamma_{2}) J h}  
\otimes e^{i (\gamma_{3}+\gamma_{2}) J h}  \,,
\label{MAdSCFT}
\ee 
and $J$ is the angular momentum charge.

Let us briefly recall the untwisted case $\gamma_{i}=0$. Based on 
the Hubbard-model results in MR \cite{MR},  Martins and Melo 
obtain in \cite{MM} the eigenvalues of the $su(2|2)$ transfer matrix.
Hereafter we denote the latter reference by MM. Indeed, 
the eigenvalues of the transfer matrix for a single copy of 
Hubbard are given by MR (89), MR (99), which can be rewritten as MM (30).
Martins and Melo argue that MM (30) leads to the eigenvalues of the 
$su(2|2)$ transfer matrix in MM (32).

Turning now to the twisted case, we use the same logic to conclude
that our result (\ref{eigenvalstwisted}) for twisted Hubbard implies
that the eigenvalues of the twisted AdS/CFT transfer matrix (\ref{twistedBCagain}) 
are given by (cf. MM (32))
\bear
\label{lambdaexplic}
&&\tilde\Lambda(\lambda)=\prod_{i=1}^{N} S_{0}(\lambda,p_{i})^{2}
\Bigg[ c_{1}
\prod_{i=1}^{N} \left[\frac{x^{-}(p_{i})-x^{+}(\lambda)}
{x^{+}(p_{i})-x^{-}(\lambda)}\right] \frac{\eta(p_{i})}{\eta(\lambda)}
\prod_{j=1}^{m_1} \eta(\lambda) \frac{x^{-}(\lambda)-x^{+}(\lambda_j)}{x^{+}(\lambda)-x^{+}(\lambda_j)}
\nonumber\\
&-&\prod_{i=1}^{N} \frac{x^{+}(\lambda)-x^{+}(p_{i})}{x^{-}(\lambda)-x^{+}(p_{i})} \frac{1}{\eta{(\lambda)}}
\left \{
\prod_{j=1}^{m_1} \eta(\lambda) \left[\frac{x^{-}(\lambda)-x^{+}(\lambda_j)}{x^{+}(\lambda)-x^{+}(\lambda_j)} \right]
\prod_{l=1}^{m_2} \frac{x^{+}(\lambda)+\frac{1}{x^{+}(\lambda)}-\tilde{\mu_l}+\frac{i}{2 g}}
{x^{+}(\lambda)+\frac{1}{x^{+}(\lambda)}-\tilde{\mu_l}-\frac{i}{2 g}} \right.
\nonumber\\
&+&
\left.\prod_{j=1}^{m_1} \eta(\lambda)\left[\frac{x^{+}(\lambda_j)-\frac{1}{x^{+}(\lambda)}}{x^{+}(\lambda_j)-\frac{1}{x^{-}(\lambda)}}\right]
\prod_{l=1}^{m_2} \frac{x^{-}(\lambda)+\frac{1}{x^{-}(\lambda)}-\tilde{\mu_l}-\frac{i}{2 g}}
{x^{-}(\lambda)+\frac{1}{x^{-}(\lambda)}-\tilde{\mu_l}+\frac{i}{2 g}} \right \}
\nonumber\\
&+&
c_{1}^{-1}
\prod_{i=1}^{N} \left[\frac{1-\frac{1}{x^{-}(\lambda) x^{+}(p_i)}}{1-\frac{1}{x^{-}(\lambda) x^{-}(p_i)}}\right]
\left[\frac{x^{+}(p_i)-x^{+}(\lambda)}{x^{+}(p_i)-x^{-}(\lambda)} \right] \frac{\eta(p_i)}{\eta(\lambda)}
\prod_{j=1}^{m_1} \eta(\lambda) \left [ 
\frac{x^{+}(\lambda_j)-\frac{1}{x^{+}(\lambda)}}{x^{+}(\lambda_j)-\frac{1}{x^{-}(\lambda)}} \right ] \Bigg]
\nonumber\\
&\times&\Bigg[ c_{2}
\prod_{i=1}^{N} \left[\frac{x^{-}(p_{i})-x^{+}(\lambda)}
{x^{+}(p_{i})-x^{-}(\lambda)}\right] \frac{\eta(p_{i})}{\eta(\lambda)}
\prod_{j=1}^{\dot m_1} \eta(\lambda) \frac{x^{-}(\lambda)-x^{+}(\dot\lambda_j)}{x^{+}(\lambda)-x^{+}(\dot\lambda_j)}
\nonumber\\
&-&\prod_{i=1}^{N} \frac{x^{+}(\lambda)-x^{+}(p_{i})}{x^{-}(\lambda)-x^{+}(p_{i})} \frac{1}{\eta{(\lambda)}}
\left \{
\prod_{j=1}^{\dot m_1} \eta(\lambda) \left[\frac{x^{-}(\lambda)-x^{+}(\dot\lambda_j)}{x^{+}(\lambda)-x^{+}(\dot\lambda_j)} \right]
\prod_{l=1}^{\dot m_2} 
\frac{x^{+}(\lambda)+\frac{1}{x^{+}(\lambda)}-\dot{\tilde{\mu}}_l+\frac{i}{2 g}}
{x^{+}(\lambda)+\frac{1}{x^{+}(\lambda)}-\dot{\tilde{\mu}}_l-\frac{i}{2 g}} \right.
\nonumber\\
&+&
\left.\prod_{j=1}^{\dot m_1} \eta(\lambda)\left[\frac{x^{+}(\dot\lambda_j)-\frac{1}{x^{+}(\lambda)}}
{x^{+}(\dot\lambda_j)-\frac{1}{x^{-}(\lambda)}}\right]
\prod_{l=1}^{\dot m_2} \frac{x^{-}(\lambda)+\frac{1}{x^{-}(\lambda)}-\dot{\tilde{\mu}}_l-\frac{i}{2 g}}
{x^{-}(\lambda)+\frac{1}{x^{-}(\lambda)}-\dot{\tilde{\mu}}_l+\frac{i}{2 g}} \right \}
\nonumber\\
&+&
c_{2}^{-1}
\prod_{i=1}^{N} \left[\frac{1-\frac{1}{x^{-}(\lambda) x^{+}(p_i)}}{1-\frac{1}{x^{-}(\lambda) x^{-}(p_i)}}\right]
\left[\frac{x^{+}(p_i)-x^{+}(\lambda)}{x^{+}(p_i)-x^{-}(\lambda)} \right] \frac{\eta(p_i)}{\eta(\lambda)}
\prod_{j=1}^{\dot m_1} \eta(\lambda) \left [ 
\frac{x^{+}(\dot\lambda_j)-\frac{1}{x^{+}(\lambda)}}{x^{+}(\dot\lambda_j)-\frac{1}{x^{-}(\lambda)}} \right ] \Bigg]\,,
\ear
where
\be
c_{1} =  e^{i\gamma_{1} N/2} e^{-i \gamma_{1} \dot m_{1}/2} 
e^{i(\gamma_{3}-\gamma_{2}) J/2}\,, \quad 
c_{2} = e^{-i \gamma_{1} N/2} e^{i \gamma_{1} m_{1}/2} 
e^{i(\gamma_{3}+\gamma_{2}) J/2}\,.
\label{cs}
\ee 

The corresponding Bethe equations are therefore (cf. MM (33)):
\bear
\label{betheexp}
c_{1}
\prod_{i=1}^{N} \left[ \frac{x^{+}(\lambda_j)-x^{-}(p_{i})}{x^{+}(\lambda_j)-x^{+}(p_{i})} \right] \eta(p_{i}) &=&
\prod_{l=1}^{m_2} \frac{x^{+}(\lambda_j)+\frac{1}{x^{+}(\lambda_j)}-\tilde{\mu_l}+\frac{i}{2 g}}
{x^{+}(\lambda_j)+\frac{1}{x^{+}(\lambda_j)}-\tilde{\mu_l}-\frac{i}{2 
g}}\,, \non \\
& &  ~~~~~ j=1,\dots,m_1,
\nonumber \\
c_{2}
\prod_{i=1}^{N} \left[ \frac{x^{+}(\dot\lambda_j)-x^{-}(p_{i})}{x^{+}(\dot\lambda_j)-x^{+}(p_{i})} \right] \eta(p_{i}) &=&
\prod_{l=1}^{\dot m_2} \frac{x^{+}(\dot\lambda_j)+\frac{1}{x^{+}(\dot\lambda_j)}-\dot{\tilde{\mu}}_l+\frac{i}{2 g}}
{x^{+}(\dot\lambda_j)+\frac{1}{x^{+}(\dot\lambda_j)}-\dot{\tilde{\mu}}_l-\frac{i}{2 g}}\,, \non \\
& &  ~~~~~ j=1,\dots,\dot m_1,
\nonumber \\
\prod_{j=1}^{m_1} \frac{\tilde{\mu}_{l}-x^{+}(\lambda_j)-\frac{1}{x^{+}(\lambda_j)}+\frac{i}{2 g}}
{\tilde{\mu}_{l}-x^{+}(\lambda_j)-\frac{1}{x^{+}(\lambda_j)}-\frac{i}{2 g}} & =& 
\prod_{\stackrel{k=1}{k \neq l}}^{m_2} 
\frac{\tilde{\mu}_{l}-\tilde{\mu}_{k}+\frac{i}{g}}
{\tilde{\mu}_{l}-\tilde{\mu}_{k}-\frac{i}{g}}\,,   ~l=1,\dots,m_2, 
\non \\
\prod_{j=1}^{\dot m_1} \frac{\dot{\tilde{\mu}}_l-x^{+}(\dot\lambda_j)-\frac{1}{x^{+}(\dot\lambda_j)}+\frac{i}{2 g}}
{\dot{\tilde{\mu}}_l-x^{+}(\dot\lambda_j)-\frac{1}{x^{+}(\dot\lambda_j)}-\frac{i}{2 g}} & =& 
\prod_{\stackrel{k=1}{k \neq l}}^{m_2} 
\frac{\dot{\tilde{\mu}}_l-\dot{\tilde{\mu}}_k+\frac{i}{g}}
{\dot{\tilde{\mu}}_l-\dot{\tilde{\mu}}_k-\frac{i}{g}}\,,   
~l=1,\dots,\dot m_2.
\ear
In terms of the notation in MM (45), (46), and using the fact 
$\eta(\lambda) = e^{i\lambda/2}$, these Bethe equations become 
\footnote{Note that $m_{j} \mapsto m_{j}^{(1)}\,, \ \dot m_{j} \mapsto 
m_{j}^{(2)}$ for $j = 1, 2$.}
\bear
c_{1} 
e^{i \frac{P}{2}}
\prod_{i=1}^{N} \left[ 
\frac{x^{+}(\lambda_j^{(1)})-x^{-}(p_{i})}{x^{+}(\lambda_j^{(1)})-x^{+}(p_{i})} \right] &=&
\prod_{l=1}^{m_2^{(1)}} 
\frac{x^{+}(\lambda_j^{(1)})+\frac{1}{x^{+}(\lambda_j^{(1)})}-\tilde{\mu_l}^{(1)}+\frac{i}{2 g}}
{x^{+}(\lambda_j^{(1)})+\frac{1}{x^{+}(\lambda_j^{(1)})}-\tilde{\mu_l}^{(1)}-\frac{i}{2 g}}\,, \non \\
& &  
~~~~j=1,\dots,m_1^{(1)},
\label{bethe2a}
\ear
\bear
c_{2}
e^{i \frac{P}{2}} 
\prod_{i=1}^{N} \left[ 
\frac{x^{+}(\lambda_j^{(2)})-x^{-}(p_{i})}{x^{+}(\lambda_j^{(2)})-x^{+}(p_{i})} \right] &=&
\prod_{l=1}^{m_2^{(2)}} 
\frac{x^{+}(\lambda_j^{(2)})+\frac{1}{x^{+}(\lambda_j^{(2)})}-\tilde{\mu_l}^{(2)}+\frac{i}{2 g}}
{x^{+}(\lambda_j^{(2)})+\frac{1}{x^{+}(\lambda_j^{(2)})}-\tilde{\mu_l}^{(2)}-\frac{i}{2 g}}\,, \non \\
& &  
~~~~j=1,\dots,m_1^{(2)},
\label{bethe2b}
\ear
\bear
\prod_{j=1}^{m_1^{(\alpha)}} \frac{\tilde{\mu}_{l}^{(\alpha)}-x^{+}(\lambda_j^{(\alpha)})-\frac{1}{x^{+}(\lambda_j^{(\alpha)})}+\frac{i}{2 g}}
{\tilde{\mu}_{l}^{(\alpha)}-x^{+}(\lambda_j^{(\alpha)})-\frac{1}{x^{+}(\lambda_j^{(\alpha)})}-\frac{i}{2 g}} &=&
\prod_{\stackrel{k=1}{k \neq l}}^{m_2^{(\alpha)}} 
\frac{\tilde{\mu}_{l}^{(\alpha)}-\tilde{\mu}_{k}^{(\alpha)}+\frac{i}{g}}
{\tilde{\mu}_{l}^{(\alpha)}-\tilde{\mu}_{k}^{(\alpha)}-\frac{i}{g}}\,,   
~~~~~l=1,\dots,m_2^{(\alpha)};\ \alpha=1,2\,,
\nonumber \\
\label{bethe3}
\ear
where $P=\sum_{k=1}^{N}p_{k}$ is the total momentum.
Following the change in notation in MM (47)-(49), (51), so that
\be
N = K_{4}\,, \quad m_{1}^{(1)}= K_{1} + K_{3}\,, \quad m_{1}^{(2)}= 
K_{5} + K_{7}\,, \quad m_{2}^{(1)}= K_{2}\,, \quad m_{2}^{(2)}= 
K_{6} \,,
\ee
and the coefficients (\ref{cs}) are given by
\be
c_{1} =  e^{i\gamma_{1} \left(K_{4}-K_{5}-K_{7}\right)/2}  
e^{i(\gamma_{3}-\gamma_{2}) J/2}\,, \quad 
c_{2} = e^{-i \gamma_{1} \left(K_{4}-K_{1}-K_{3}\right)/2}  
e^{i(\gamma_{3}+\gamma_{2}) J/2}\,,
\label{cs2}
\ee 
Eqs. (\ref{bethe2a}), (\ref{bethe2b}) become (cf. MM (52))
\bear
c_{1} 
e^{-i P/2} 
\prod_{i=1}^{K_4}  
\frac{1-\frac{g^2}{x^{-}_{4,i} x_{1,j}}}
{1-\frac{g^2}{x^{+}_{4,i} x_{1,j}}} & =& 
\prod_{l=1}^{K_2} \frac{u_{1,j}-u_{2,l}+\frac{i}{2}}{u_{1,j}-u_{2,l}-\frac{i}{2}},~~j=1,\dots,K_1
\label{bet2K1} \\
c_{1}  
e^{i P/2} 
\prod_{i=1}^{K_4}  
\frac{x^{-}_{4,i}-x_{3,j}}{x^{+}_{4,i}-x_{3,j}} &=&
\prod_{l=1}^{K_2} \frac{u_{3,j}-u_{2,l}+\frac{i}{2}}{u_{3,j}-u_{2,l}-\frac{i}{2}},~~j=1,\dots,K_3
\label{bet2K3} \\
c_{2}  
e^{i P/2} 
\prod_{i=1}^{K_4}  
\frac{x^{-}_{4,i}-x_{5,j}}{x^{+}_{4,i}-x_{5,j}} & =&
\prod_{l=1}^{K_6} \frac{u_{5,j}-u_{6,l}+\frac{i}{2}}{u_{5,j}-u_{6,l}-\frac{i}{2}},~~j=1,\dots,K_5
\label{bet2K5} \\
c_{2} 
e^{-i P/2} 
\prod_{i=1}^{K_4}  
\frac{1-\frac{g^2}{x^{-}_{4,i} x_{7,j}}}
{1-\frac{g^2}{x^{+}_{4,i} x_{7,j}}} & =& 
\prod_{l=1}^{K_6} \frac{u_{7,j}-u_{6,l}+\frac{i}{2}}{u_{7,j}-u_{6,l}-\frac{i}{2}},~~j=1,\dots,K_7
\label{bet2K7}
\ear
The undeformed eqs. (\ref{bethe3}) become the same as MM (53), namely,
\bear
\prod_{j=1}^{K_1} \frac{u_{2,l}-u_{1,j}+\frac{i}{2}}
{u_{2,l}-u_{1,j}-\frac{i}{2}}
\prod_{j=1}^{K_3} \frac{u_{2,l}-u_{3,j}+\frac{i}{2}}
{u_{2,l}-u_{3,j}-\frac{i}{2}} & =& 
\prod_{\stackrel{k=1}{k \neq l}}^{K_2} 
\frac{u_{2,l}-u_{2,k}+i}
{u_{2,l}-u_{2,k}-i},~~~~l=1,\dots,K_2
\nonumber \\
\prod_{j=1}^{K_5} \frac{u_{6,l}-u_{5,j}+\frac{i}{2}}
{u_{6,l}-u_{5,j}-\frac{i}{2}}
\prod_{j=1}^{K_7} \frac{u_{6,l}-u_{7,j}+\frac{i}{2}}
{u_{6,l}-u_{7,j}-\frac{i}{2}} & =& 
\prod_{\stackrel{k=1}{k \neq l}}^{K_6} 
\frac{u_{6,l}-u_{6,k}+i}
{u_{6,l}-u_{6,k}-i},~~~~l=1,\dots,K_6
\label{bet3}
\ear

Finally, we consider the equations for the type-4 Bethe roots (i.e.,
corresponding to the middle node of the Dynkin diagram).  These come
from the Bethe-Yang equations (cf. (\ref{BYPCM})) 
\be
e^{-i p_{k} L} &=& \tilde\Lambda(p_{k})\non \\
&=& c_{1} c_{2} 
\prod_{i=1}^{N} \left[S_{0}(p_{k},p_{i}) \frac{x^{-}(p_{i})-x^{+}(p_{k})}
{x^{+}(p_{i})-x^{-}(p_{k})} \frac{\eta(p_{i})}{\eta(p_{k})} 
\right]^{2}\non \\ 
&  &\times \prod_{j=1}^{m_1} \eta(p_{k}) \frac{x^{-}(p_{k})-x^{+}(\lambda_j)}{x^{+}(p_{k})-x^{+}(\lambda_j)}
\prod_{j=1}^{\dot m_1} \eta(p_{k}) 
\frac{x^{-}(p_{k})-x^{+}(\dot\lambda_j)}{x^{+}(p_{k})-x^{+}(\dot\lambda_j)}\,, 
\label{bethe4}
\ee
where we have used our result (\ref{lambdaexplic}) for 
$\Lambda(\lambda)$. Setting 
\be
L=-J
\label{JL}
\ee
as proposed by MM, substituting the result for the scalar factor from 
MM (36), and changing notations as above, we obtain (cf. MM (50))
\bear
e^{i p_k \left[J +  K_{4} - \frac{1}{2}(K_{3}-K_{1}) - 
\frac{1}{2}(K_{5}-K_{7}) \right]} 
& = & 
c_{1} c_{2}
e^{i P}
\prod_{\stackrel{i=1}{i \neq k}}^{K_4} 
\left[ \frac{x^{+}_{4,k}-x^{-}_{4,i}}{x^{-}_{4,k}-x^{+}_{4,i}} \right]
\left[ \frac{1-\frac{g^2}{x^{+}_{4,k} x^{-}_{4,i}}}
{1-\frac{g^2}{x^{-}_{4,k} x^{+}_{4,i}}} \right ]
[\sigma(p_k,p_i)]^2 
\nonumber \\
&& \times 
\prod_{j=1}^{K_3}  
\frac{x^{-}_{4,k}-x_{3,j}}{x^{+}_{4,k}-x_{3,j}} 
\prod_{j=1}^{K_1}  
\frac{1-\frac{g^2}{x^{-}_{4,k} x_{1,j}}}
{1-\frac{g^2}{x^{+}_{4,k} x_{1,j}}} 
\nonumber \\
&& \times 
\prod_{j=1}^{K_5}  
\frac{x^{-}_{4,k}-x_{5,j}}{x^{+}_{4,k}-x_{5,j}} 
\prod_{j=1}^{K_7}  
\frac{1-\frac{g^2}{x^{-}_{4,k} x_{7,j}}}
{1-\frac{g^2}{x^{+}_{4,k} x_{7,j}}},~k=1,\dots,K_4\,.
\label{bet1}
\ear
For the undeformed case $\gamma_{i}=0$, one recovers the corresponding Bethe equations of Beisert and
Staudacher \cite{BeiSta} by setting  $P=0$  and recalling the result 
for the angular momentum charge
\be
J = {\cal L} - K_{4} + \frac{1}{2}(K_{3}-K_{1}) + 
\frac{1}{2}(K_{5}-K_{7})  \,,
\label{J}
\ee 
where we denote by ${\cal L}$ the parameter used in \cite{BeiSta, BR} to 
identify the length of the chain.

The twisted transfer matrix eigenvalue (\ref{lambdaexplic}), which we
have obtained from diagonalizing the twisted transfer matrix
(\ref{twistedBCagain}) based on the twisted scattering matrix
(\ref{drinfeldtwist}) and the twisted boundary condition (\ref{MAdSCFT}), is
equivalent to the transfer matrix eigenvalue proposed in \cite{GLM};
and for the special case of $\beta$-deformation, it is equivalent to 
the result in \cite{AdLvT}.

\section{Comparison with BR}\label{sec:BR}

The paper \cite{BR} of Beisert and Roiban, to which we refer by BR,
proposes a three-parameter ($\gamma_{1}\,, \gamma_{2}\,, \gamma_{3}$)
deformation of the all-loop asymptotic Bethe equations of Beisert and
Staudacher \cite{BeiSta}.  The $\beta$-deformation \cite{LeiStr} corresponds to a
special case with ${\cal N}=1$ supersymmetry,
\be
\gamma_{1}= \gamma_{2}=\gamma_{3}=2\pi\beta \,.
\label{betacase}
\ee 

The deformed all-loop Bethe equations are given by BR (5.39)
\be
e^{i(\mat{A}\vect{K})_0}
\,U_0
=1,\qquad
e^{i(\mat{A}\vect{K})_j}\,
U_j(x_{j,k})
\mathop{\prod_{j'=1}^7\prod_{k'=1}^{K_{j'}}}_{(j',k')\neq(j,k)}
\frac{u_{j,k}-u_{j',k'}+\ihalf M_{j,j'}}{u_{j,k}-u_{j',k'}-\ihalf M_{j,j'}}
=1\,, \label{BR}
\ee
where 
\be
U_0=
\prod_{k=1}^{K_4}
\frac{x^+_{4,k}}{x^-_{4,k}}\,,
\qquad
U_1(x)=U_3^{-1}(x)=U_5^{-1}(x)=U_7(x)=
\prod_{k=1}^{K_4}
S\indup{aux}(x_{4,k},x)
\label{Us1}
\ee
and
\be
U_4(x)=
U\indup{s}(x)
\lrbrk{\frac{x^-}{x^+}}^{\cal L}
\prod_{k=1}^{K_1}
S^{-1}\indup{aux}(x,x_{1,k})
\prod_{k=1}^{K_3}
S\indup{aux}(x,x_{3,k})
\prod_{k=1}^{K_5}
S\indup{aux}(x,x_{5,k})
\prod_{k=1}^{K_7}
S^{-1}\indup{aux}(x,x_{7,k}).
\label{Us2}
\ee
Moreover,
\be
S\indup{aux}(x_1,x_2)=
\frac{1-g^2/x^+_1x_2}{1-g^2/x^-_1x_2}\,, \qquad
U\indup{s}(x) = \prod_{k=1}^{K_4}\sigma(x,x_{4,k})^{2} \,.
\label{Us3}
\ee

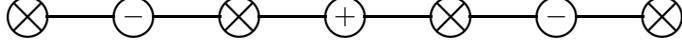
\begin{figure}\centering
\begin{minipage}{260pt}
\setlength{\unitlength}{1pt}%
\small\thicklines%
\begin{picture}(260,20)(-10,-10)
\put(  0,00){\circle{15}}%
\put(  7,00){\line(1,0){26}}%
\put( 40,00){\circle{15}}%
\put( 47,00){\line(1,0){26}}%
\put( 80,00){\circle{15}}%
\put( 87,00){\line(1,0){26}}%
\put(120,00){\circle{15}}%
\put(127,00){\line(1,0){26}}%
\put(160,00){\circle{15}}%
\put(167,00){\line(1,0){26}}%
\put(200,00){\circle{15}}%
\put(207,00){\line(1,0){26}}%
\put(240,00){\circle{15}}%
\put( -5,-5){\line(1, 1){10}}%
\put( -5, 5){\line(1,-1){10}}%
\put( 75,-5){\line(1, 1){10}}%
\put( 75, 5){\line(1,-1){10}}%
\put(155,-5){\line(1, 1){10}}%
\put(155, 5){\line(1,-1){10}}%
\put(235,-5){\line(1, 1){10}}%
\put(235, 5){\line(1,-1){10}}%
\put( 40,00){\makebox(0,0){$-$}}%
\put(120,00){\makebox(0,0){$+$}}%
\put(200,00){\makebox(0,0){$-$}}%
\end{picture}
\end{minipage}

\caption{Dynkin diagram of $su(2,2|4)$.}
\label{fig:DynkinHigher}
\end{figure}

\noindent
For the ``$su(2)$'' grading with $\eta_{1}=\eta_{2}=+1$ which we 
consider here, $M_{j,j'}$
is the Cartan matrix specified by Fig.  \ref{fig:DynkinHigher} (see
Eq.  (5.1) in \cite{BeiSta}), and the twist matrix ${\bf A}$ is given
\footnote{This matrix is given in terms of the 
parameters $(\delta_{1},\delta_{2},\delta_{3})$, which are related to 
$(\gamma_{1}\,, \gamma_{2}\,, \gamma_{3})$ through Eqs. BR (5.2) and 
BR (5.3); i.e.,
\be
\gamma_{1}=-\delta_{1}-2\delta_{2}-\delta_{3}\,, \quad
\gamma_{2}=-\delta_{1}-\delta_{3}\,, \quad
\gamma_{3}=-\delta_{1}+\delta_{3}\,. \label{gammas}
\ee} 
by BR (5.24).  It then follows from BR (4.27) that
\be
(\mat{A}\vect{K})_{0} &=& \frac{1}{2}\big[ 
\gamma_{2}\left(-K_{1}-K_{3}+K_{5}+K_{7}\right) \non \\
& & +\gamma_{3}\left(K_{1}+K_{3}-2K_{4}+K_{5}+K_{7}\right) \big] \,,  \label{AK0} \\
(\mat{A}\vect{K})_{2} &=& (\mat{A}\vect{K})_{6} = 0 \,, \label{AK2} \\
(\mat{A}\vect{K})_{1} +\frac{1}{2}(\mat{A}\vect{K})_{0} &=& -\frac{1}{2}\left[
(\gamma_{3}-\gamma_{2}) J + \gamma_{1}\left(K_{4}-K_{5}-K_{7}\right) \right] 
\,, \label{AK1} \\
(\mat{A}\vect{K})_{3} - \frac{1}{2}(\mat{A}\vect{K})_{0} &=&  
(\mat{A}\vect{K})_{1} +\frac{1}{2}(\mat{A}\vect{K})_{0}\,,  \label{AK3} \\
(\mat{A}\vect{K})_{5} - \frac{1}{2}(\mat{A}\vect{K})_{0} &=& -\frac{1}{2}\left[
(\gamma_{3}+\gamma_{2}) J - \gamma_{1}\left(K_{4}-K_{1}-K_{3}\right) \right] 
\,,  \label{AK5} \\
(\mat{A}\vect{K})_{7} +\frac{1}{2}(\mat{A}\vect{K})_{0} &=&  
(\mat{A}\vect{K})_{5} - \frac{1}{2}(\mat{A}\vect{K})_{0}\,,  \label{AK7} \\
(\mat{A}\vect{K})_{4} + (\mat{A}\vect{K})_{0}  &=& \gamma_{3} J - 
\frac{1}{2} \gamma_{1}\left(K_{5}+K_{7}-K_{1}-K_{3}\right) \,.  \label{AK4} 
\ee 
Note that Eqs. (\ref{BR}) and (\ref{Us1}) imply that the total 
momentum $P$ is given by
\be
P = - (\mat{A}\vect{K})_{0}  \,.
\label{P}
\ee

We now compare the BR Bethe equations with the ones which we 
derived in the previous section.

The fact that Eqs. (\ref{bet3}) are not deformed  
matches with (\ref{AK2}) and (\ref{BR}) with $j=2, 6$.

In order to facilitate the further comparisons, let us rewrite Eqs. 
(\ref{bet2K1}) - (\ref{bet2K7}) in the form (\ref{BR}):
\bear
c_{1}^{-1}
e^{i P/2} 
\prod_{i=1}^{K_4}  
\frac{1-\frac{g^2}{x^{+}_{4,i} x_{1,j}}}{1-\frac{g^2}{x^{-}_{4,i} x_{1,j}}}
\prod_{l=1}^{K_2} 
\frac{u_{1,j}-u_{2,l}+\frac{i}{2}}{u_{1,j}-u_{2,l}-\frac{i}{2}} & = & 
1 ,~~j=1,\dots,K_1 \,,
\label{bet2K1new} \\
c_{1}^{-1}
e^{-i P/2} 
\prod_{i=1}^{K_4}  
\frac{x^{+}_{4,i}-x_{3,j}} 
{x^{-}_{4,i}-x_{3,j}}
\prod_{l=1}^{K_2} 
\frac{u_{3,j}-u_{2,l}+\frac{i}{2}}{u_{3,j}-u_{2,l}-\frac{i}{2}}&=& 
1,~~j=1,\dots,K_3 \,,
\label{bet2K3new} \\
c_{2}^{-1}
e^{-i P/2} 
\prod_{i=1}^{K_4}  
\frac{x^{+}_{4,i}-x_{5,j}}
{x^{-}_{4,i}-x_{5,j}}
\prod_{l=1}^{K_6} 
\frac{u_{5,j}-u_{6,l}+\frac{i}{2}}{u_{5,j}-u_{6,l}-\frac{i}{2}}& =& 
1,~~j=1,\dots,K_5 \,,
\label{bet2K5new} \\
c_{2}^{-1} 
e^{i P/2} 
\prod_{i=1}^{K_4}  
\frac{1-\frac{g^2}{x^{+}_{4,i} x_{7,j}}}
{1-\frac{g^2}{x^{-}_{4,i} x_{7,j}}}
\prod_{l=1}^{K_6} 
\frac{u_{7,j}-u_{6,l}+\frac{i}{2}}{u_{7,j}-u_{6,l}-\frac{i}{2}} & =&  
1,~~j=1,\dots,K_7 \,.
\label{bet2K7new}
\ear
Substituting for $P$ using (\ref{P}), and noting the identities 
(proved using (\ref{cs2}), (\ref{J}) and (\ref{AK0})-(\ref{AK7})),
\be
c_{1}^{-1} e^{i P/2} = e^{i(\mat{A}\vect{K})_{1}}\,, \quad
c_{1}^{-1} e^{-i P/2} = e^{i(\mat{A}\vect{K})_{3}}\,, \quad
c_{2}^{-1} e^{-i P/2} = e^{i(\mat{A}\vect{K})_{5}}\,, \quad
c_{2}^{-1} e^{i P/2} = e^{i(\mat{A}\vect{K})_{7}}\,,
\ee
we see that Eqs. (\ref{bet2K1new})-(\ref{bet2K7new}) match with 
(\ref{BR}) with $j=1,3,5,7$ respectively. Moreover, the identity
\be
c_{1} c_{2} e^{i P} = e^{i(\mat{A}\vect{K})_{4}}
\ee
implies that (\ref{bet1}) matches with (\ref{BR}) with $j=4$.
In summary, provided we take $P$ as in (\ref{P}),
the Bethe equations of Sec. \ref{sec:AdS/CFT} match 
with those in BR.

\section{Discussion}\label{sec:disc}

We have shown that the Beisert-Roiban Bethe equations (\ref{BR}) for
the 3-parameter deformation of $AdS_{5}/CFT_{4}$ can be derived from
the $S$-matrix with the Drinfeld-Reshetikhin twist
(\ref{drinfeldtwist})-(\ref{hh}), together with the $c$-number twist
(\ref{MAdSCFT}) of the boundary conditions.  This result places the
twisted Bethe equations on a firmer conceptual footing.  
Our result also reproduces the proposed twisted transfer matrix 
eigenvalue of \cite{GLM}.
As explained in Appendix \ref{sec:Luscher}, this result also justifies
the deformed $S$-matrix elements used in \cite{ABBN} to compute the
anomalous dimension of the Konishi operator in $\beta$-deformed ${\cal
N}=4$ SYM via the L\"uscher formula.  Indeed, we can recover with our
approach the L\"uscher correction for all known cases in the
literature, both $su(2)$ and $sl(2)$.

We demonstrate in Appendix \ref{sec:optwist} that the transfer matrix
is spectrally equivalent to a transfer matrix which is constructed
using instead {\em untwisted} $S$-matrices and boundary conditions
with {\em operatorial} twists.  It is the latter type of transfer
matrix which is considered in \cite{AdLvT}.  A similar spectral
equivalence was noted for the $\beta$-deformed $su(2)$ sector at one
loop in \cite{BerChe}.  Finally, in Appendix \ref{sec:sl2} we
transform our twisted Bethe ansatz results from the ``$su(2)$''
grading to the ``$sl(2)$'' grading, and show that the results agree
with both \cite{BR} and \cite{AdLvT}.

The scattering matrix is a fundamental object in integrable systems and
can be checked by various means.  As its semiclassical limit
corresponds to time delays, it would be nice to check our proposal
against a classical string theory calculation based directly on the
Lunin-Maldacena \cite{LunMal} background.

\section*{Acknowledgements}
We thank Niklas Beisert, Dmitri Bykov, Davide Fioravanti, Romuald
Janik, Nicolai Reshetikhin, Radu Roiban, Marco Rossi, Christoph Sieg,
Matthias Staudacher and Roberto Tateo for useful discussions and/or
correspondence.
Part of this work was performed at the 2010 summer workshops 
``Integrability in String and Gauge Theories'' (Nordita, Stockholm)
and ``Finite-Size Technology in Low-Dimensional Quantum Systems V'' 
(Benasque).
This work was supported in part by KRF-2007-313- C00150, WCU Grant
No. R32-2008-000-101300 (CA, DB); by the Funda\c{c}\~{a}o para a 
Ci\^{e}ncia e Tecnologia fellowship SFRH/BPD/69813/2010 and travel
financial support from the University PRIN 2007JHLPEZ ``Fisica
Statistica dei Sistemi Fortemente Correlati all' Equilibrio e Fuori
Equilibrio: Risultati Esatti e Metodi di Teoria dei Campi'' (DB);
by OTKA 81461 (ZB); and by the National Science Foundation under Grants
PHY-0554821 and PHY-0854366 (RN).

\begin{appendix}

\section{Derivation of the exchange relations}\label{sec:exchange}

An important ingredient of the algebraic Bethe ansatz approach is the
set of exchange relations which are obeyed by the matrix elements of
the monodromy matrix.  Here we explain how to derive the exchange
relations used in this paper.

\subsection{$su(2)$ principal chiral model}

The exchange relations for the spin-1/2 XXX quantum spin chain are 
well known (see, for example, \cite{QISM}). We shall need the 
relations between the diagonal operators $A, D$ and the creation 
operator $B$
\be
A(u)\, B(v) &=& \frac{u-v-i}{u-v} B(v)\, A(u) + \frac{i}{u-v} B(u)\, 
A(v) \,, \label{er1} \\
D(u)\, B(v) &=& \frac{u-v+i}{u-v} B(v)\, D(u) - \frac{i}{u-v} B(u)\, 
D(v) \,, \label{er2} 
\ee
as well as relations among the diagonal operators,
\be 
\left[ A(u) \,, A(v) \right] = \left[  D(u) \,, D(v) \right] = 0 \,, \label{er4}
\ee
\be 
D(u)\, A(v) = A(v)\, D(u) + \frac{i}{u-v} \left( B(v)\, C(u) - 
B(u)\, C(v) \right)
\,. \label{er3}  
\ee
The same relations hold for the corresponding dotted operators $\dot
A, \dot B, \dot C, \dot D$; and the dotted and undotted operators
commute with each other.

The exchange relations between the diagonal operators ${\cal T}_{jj}$ 
and the creation operators ${\cal T}_{12}, {\cal T}_{13}$ can be 
classified into two types, depending on whether they make use of 
(\ref{er4}) or (\ref{er3}). The former are very simple, and are given in 
(\ref{exchange1}); the latter are more complicated, and are given in 
(\ref{exchange2}).

The first exchange relation in (\ref{exchange1}) can be easily 
derived using (\ref{er1}) and (\ref{er4}):
\be
{\cal T}_{11}(u)\, {\cal T}_{13}(v) &=& A(u)\, \dot A(u) \, B(v)\, \dot 
A(v) \non \\
&=& A(u)\,  B(v) \, \dot A(u)\, \dot A(v) \non \\
&=& \left[ \frac{u-v-i}{u-v} B(v)\, A(u) + \frac{i}{u-v} B(u)\, A(v) 
\right]  \dot A(u)\, \dot A(v) \non \\
&=&  \frac{u-v-i}{u-v} {\cal T}_{13}(v) \, {\cal T}_{11}(u) +
\frac{i}{u-v} {\cal T}_{13}(u)\, {\cal T}_{11}(v) \,,
\ee 
and the remaining relations in (\ref{exchange1}) can be derived in a 
similar way.

The derivation of the first exchange relation in (\ref{exchange2}) 
also begins in a similar way:
\be
{\cal T}_{22}(u)\, {\cal T}_{13}(v) &=&   A(u)\, \dot D(u) \, B(v)\, \dot 
A(v) \non \\
&=& A(u)\,  B(v) \, \dot D(u)\, \dot A(v) \non \\
&=& \left[ \frac{u-v-i}{u-v} B(v)\, A(u) + \frac{i}{u-v} B(u)\, A(v) 
\right]  \dot D(u)\, \dot A(v) \non \\
&=& \frac{u-v-i}{u-v} B(v)\, A(u) \dot D(u)\, \dot A(v) 
+ \frac{i}{u-v} {\cal T}_{24}(u)\, {\cal T}_{11}(v)  \,. 
\label{step1}
\ee
We next observe that the first term on the RHS can be re-expressed 
using (\ref{er3}) as follows
\be
\lefteqn{\frac{u-v-i}{u-v} B(v)\, A(u) \dot D(u)\, \dot A(v)}\non \\
&=& \frac{u-v-i}{u-v} B(v)\, A(u) \left[ 
\dot A(v)\, \dot D(u) + \frac{i}{u-v} \dot B(v)\, \dot C(u) 
 - \frac{i}{u-v} \dot B(u)\, \dot C(v) \right] \non \\ 
&=& \frac{u-v-i}{u-v} {\cal T}_{13}(v) \, {\cal T}_{22}(u) +
\frac{i(u-v-i)}{(u-v)^{2}}{\cal T}_{14}(v) \, {\cal T}_{21}(u)  \non\\
& & -\frac{i(u-v-i)}{(u-v)^{2}} B(v)\, A(u)\, \dot B(u)\, \dot C(v) 
\,. \label{step2}
\ee 
Finally, we use again (\ref{er1}) to re-write the final term in 
(\ref{step2}) as 
\be
\lefteqn{-\frac{i(u-v-i)}{(u-v)^{2}} B(v)\, A(u)\, \dot B(u)\, \dot 
C(v)}\non \\
&=&
-\frac{i(u-v-i)}{(u-v)^{2}}\left[ \frac{u-v}{u-v-i} A(u)\, B(v) 
- \frac{i}{u-v-i}B(u)\, A(v) \right] \dot B(u)\, \dot C(v) \non \\ 
&=& - \frac{i}{u-v} {\cal T}_{12}(u)\, {\cal T}_{23}(v) -
\frac{1}{(u-v)^{2}}{\cal T}_{14}(u) \, {\cal T}_{21}(v) \,.
\label{step3}
\ee 
Combining the results (\ref{step1})-(\ref{step3}), we arrive at 
the first exchange relation in (\ref{exchange2}). The remaining relations in 
(\ref{exchange2}) can be derived in a similar way.

We remark that one can generate many other ( ``bad'') exchange relations
which differ from those given in (\ref{exchange2}).  What singles out
those in (\ref{exchange2}) is that the diagonal term gives the
``wanted'' contribution, while the rest of the terms give ``unwanted''
contributions.  To get this right, it helps to know the desired final
result, which was first found by other means in Sec
\ref{subsec:conventional}.  A further useful check is that these
exchange relations have simple deformations (\ref{exchange2twisted}),
which does not seem to be the case for ``bad'' exchange relations.

\subsection{Two copies of the Hubbard model}

The exchange relations between the diagonal operators $B, A_{jj}, D$ 
and the creation operators $B_{j}$ are given in MR (34)-(36). For 
example, 
\be
B(\lambda)\, B_{j}(\mu) &=& \frac{i 
\alpha_{2}(\mu,\lambda)}{\alpha_{9}(\mu,\lambda)} B_{j}(\mu)\, 
B(\lambda) - \frac{i 
\alpha_{5}(\mu,\lambda)}{\alpha_{9}(\mu,\lambda)} B_{j}(\lambda)\, 
B(\mu) \,, \quad j = 1,2, \label{exrltnHubbard1} \\
A_{jj}(\lambda)\, B_{j}(\mu) &=& -\frac{i 
\alpha_{1}(\lambda,\mu)}{\alpha_{9}(\lambda,\mu)} B_{j}(\mu)\, 
A_{jj}(\lambda) + \frac{i 
\alpha_{5}(\lambda,\mu)}{\alpha_{9}(\lambda,\mu)} B_{j}(\lambda)\, 
A_{jj}(\mu) \,, \quad j = 1,2, \label{exrltnHubbard2} 
\ee 
where there is {\em no} sum over repeated indices.
We shall also need exchange relations among the diagonal 
operators (see (12.D.1) in \cite{EFGKK}),
\be
B(\lambda)\, B(\mu) &=& B(\mu)\, B(\lambda) \,, \label{diagHubbard1} \\
A_{jj}(\lambda)\, B(\mu) &=& B(\mu)\, A_{jj}(\lambda) - 
\frac{i\alpha_{5}(\lambda,\mu)}{\alpha_{9}(\lambda,\mu)}
\left( B_{j}(\mu)\, C_{j}(\lambda) - B_{j}(\lambda)\, C_{j}(\mu) 
\right)  \,, \quad j = 1, 2, \label{diagHubbard2} \\
A_{jj}(\lambda)\, D(\mu) &=& D(\mu)\, A_{jj}(\lambda) - 
\frac{i\alpha_{5}(\lambda,\mu)}{\alpha_{9}(\lambda,\mu)}
\left( C^{*}_{j}(\mu)\, B^{*}_{j}(\lambda) - C^{*}_{j}(\lambda)\, 
B^{*}_{j}(\mu) \right)  \,, \quad j = 1, 2, \label{diagHubbard3} \\
D(\lambda)\, B(\mu) &=& B(\mu)\, D(\lambda) - 
\frac{\alpha_{4}(\lambda,\mu)}{\alpha_{7}(\lambda,\mu)}
\left(F(\lambda)\, C(\mu) - F(\mu)\, C(\lambda) \right) 
\label{diagHubbard4} \\
&-& \frac{i\alpha_{10}(\lambda,\mu)}{\alpha_{7}(\lambda,\mu)}
\left(
B^{*}_{1}(\lambda)\, C_{2}(\mu) - B^{*}_{2}(\lambda)\, C_{1}(\mu) +
B_{1}(\mu)\, C^{*}_{2}(\lambda) - B_{2}(\mu)\, C^{*}_{1}(\lambda) 
\right) \,. \non 
\ee
The same relations hold for the corresponding dotted operators.

The exchange relations between the diagonal operators ${\cal T}_{j,j}$
and the creation operators ${\cal T}_{1,2}, {\cal T}_{1,3}, {\cal
T}_{1,5}, {\cal T}_{1,9}$ can be classified into four types, depending
on which of the four relations 
(\ref{diagHubbard1})-(\ref{diagHubbard4}) they make use of.  The
relations of the first type which make use of (\ref{diagHubbard1}) are
the simplest; while the relations of the fourth type which make use of
(\ref{diagHubbard4}) are the most complicated.  All of these relations
can be derived using the same procedure which we used for the
principal chiral model.

Here is an example of the first type:
\be
{\cal T}_{1,1}(\lambda)\, {\cal T}_{1,2}(\mu) &=& B(\lambda)\, \dot B(\lambda)\, 
B(\mu)\, \dot B_{1}(\mu) \non \\
&=& B(\lambda)\, B(\mu)\,
\dot B(\lambda)\, \dot B_{1}(\mu) \non \\
&=& B(\lambda)\, B(\mu) \left[
\frac{i 
\alpha_{2}(\mu,\lambda)}{\alpha_{9}(\mu,\lambda)} \dot B_{1}(\mu)\, 
\dot B(\lambda) - \frac{i 
\alpha_{5}(\mu,\lambda)}{\alpha_{9}(\mu,\lambda)} \dot B_{1}(\lambda)\, 
\dot B(\mu) \right] \non \\
&=& \frac{i 
\alpha_{2}(\mu,\lambda)}{\alpha_{9}(\mu,\lambda)}{\cal 
T}_{1,2}(\mu)\, {\cal T}_{1,1}(\lambda)
- \frac{i 
\alpha_{5}(\mu,\lambda)}{\alpha_{9}(\mu,\lambda)}{\cal 
T}_{1,2}(\lambda)\, {\cal T}_{1,1}(\mu) \,,
\ee
where we have used (\ref{exrltnHubbard1}) to pass to the third line.

For an example of the second type, let us consider
\be
{\cal T}_{2,2}(\lambda)\, {\cal T}_{1,5}(\mu) &=& B(\lambda)\, \dot 
A_{11}(\lambda)\, 
B_{1}(\mu)\, \dot B(\mu) \non \\
&=& B(\lambda)\, B_{1}(\mu)\,
\dot A_{11}(\lambda)\, \dot B(\mu) \non \\
&=& \left[
\frac{i 
\alpha_{2}(\mu,\lambda)}{\alpha_{9}(\mu,\lambda)} B_{1}(\mu)\, 
B(\lambda) - \frac{i 
\alpha_{5}(\mu,\lambda)}{\alpha_{9}(\mu,\lambda)} B_{1}(\lambda)\, 
B(\mu) \right] \dot A_{11}(\lambda)\, \dot B(\mu) \non \\
&=& \frac{i 
\alpha_{2}(\mu,\lambda)}{\alpha_{9}(\mu,\lambda)} B_{1}(\mu)\, 
B(\lambda)\, \dot A_{11}(\lambda)\, \dot B(\mu) 
- \frac{i \alpha_{5}(\mu,\lambda)}{\alpha_{9}(\mu,\lambda)}
{\cal T}_{2,6}(\lambda)\, {\cal T}_{1,1}(\mu) \,. \label{Hubbardstep1}
\ee
The first term on the RHS can be re-expressed using (\ref{diagHubbard2}) as 
follows
\be
\lefteqn{\frac{i 
\alpha_{2}(\mu,\lambda)}{\alpha_{9}(\mu,\lambda)} B_{1}(\mu)\, 
B(\lambda)\, \dot A_{11}(\lambda)\, \dot B(\mu)}\non \\
&=& \frac{i 
\alpha_{2}(\mu,\lambda)}{\alpha_{9}(\mu,\lambda)} B_{1}(\mu)\, 
B(\lambda) \left[
\dot B(\mu)\, \dot A_{11}(\lambda) - 
\frac{i\alpha_{5}(\lambda,\mu)}{\alpha_{9}(\lambda,\mu)}
\left( \dot B_{1}(\mu)\, \dot C_{1}(\lambda) - \dot B_{1}(\lambda)\, 
\dot C_{1}(\mu) \right) \right] \non \\
&=&\frac{i 
\alpha_{2}(\mu,\lambda)}{\alpha_{9}(\mu,\lambda)} {\cal 
T}_{1,5}(\mu)\, {\cal T}_{2,2}(\lambda) + 
\frac{\alpha_{2}(\mu,\lambda)\alpha_{5}(\lambda,\mu)}{\alpha_{9}(\mu,\lambda)\alpha_{9}(\lambda,\mu)}
{\cal T}_{1,6}(\mu)\, {\cal T}_{2,1}(\lambda) \non \\
& & - \frac{\alpha_{2}(\mu,\lambda)\alpha_{5}(\lambda,\mu)}{\alpha_{9}(\mu,\lambda)\alpha_{9}(\lambda,\mu)}
B_{1}(\mu)\, B(\lambda)\, \dot B_{1}(\lambda)\, \dot C_{1}(\mu) \,. 
\label{Hubbardstep2}
\ee
Finally, we use again  (\ref{exrltnHubbard1}) to re-write the final term  in 
(\ref{Hubbardstep2}) as
\be
\lefteqn{- \frac{\alpha_{2}(\mu,\lambda)\alpha_{5}(\lambda,\mu)}{\alpha_{9}(\mu,\lambda)\alpha_{9}(\lambda,\mu)}
B_{1}(\mu)\, B(\lambda)\, \dot B_{1}(\lambda)\, \dot C_{1}(\mu) } 
\non \\
&=&  \frac{i\alpha_{5}(\lambda,\mu)}{\alpha_{9}(\lambda,\mu)}
\left[ B(\lambda)\, B_{1}(\mu) + 
i\frac{\alpha_{5}(\mu,\lambda)}{\alpha_{9}(\mu,\lambda)}
B_{1}(\lambda)\, B(\mu) \right] 
\dot B_{1}(\lambda)\, \dot C_{1}(\mu) \non \\
&=& \frac{i\alpha_{5}(\lambda,\mu)}{\alpha_{9}(\lambda,\mu)}
{\cal T}_{1,2}(\lambda)\, {\cal T}_{2,5}(\mu) - 
\frac{\alpha_{5}(\mu,\lambda)\alpha_{5}(\lambda,\mu)}{\alpha_{9}(\mu,\lambda)\alpha_{9}(\lambda,\mu)}
{\cal T}_{1,6}(\lambda)\, {\cal T}_{2,1}(\mu)
\label{Hubbardstep3}
\ee
Combining the results (\ref{Hubbardstep1})-(\ref{Hubbardstep3}), we arrive at 
the exchange relation 
\be
{\cal T}_{2,2}(\lambda)\, {\cal T}_{1,5}(\mu) &=&
\frac{i 
\alpha_{2}(\mu,\lambda)}{\alpha_{9}(\mu,\lambda)} {\cal 
T}_{1,5}(\mu)\, {\cal T}_{2,2}(\lambda) + 
\frac{\alpha_{2}(\mu,\lambda)\alpha_{5}(\lambda,\mu)}{\alpha_{9}(\mu,\lambda)\alpha_{9}(\lambda,\mu)}
{\cal T}_{1,6}(\mu)\, {\cal T}_{2,1}(\lambda) \non \\
&+& \frac{i \alpha_{5}(\mu,\lambda)}{\alpha_{9}(\mu,\lambda)}
{\cal T}_{2,6}(\lambda)\, {\cal T}_{1,1}(\mu) + 
\frac{i\alpha_{5}(\lambda,\mu)}{\alpha_{9}(\lambda,\mu)}
{\cal T}_{1,2}(\lambda)\, {\cal T}_{2,5}(\mu) \non \\
&-& \frac{\alpha_{5}(\mu,\lambda)\alpha_{5}(\lambda,\mu)}{\alpha_{9}(\mu,\lambda)\alpha_{9}(\lambda,\mu)}
{\cal T}_{1,6}(\lambda)\, {\cal T}_{2,1}(\mu) \,.
\ee

An example of an exchange relation of the fourth type, which can be 
derived in a similar manner using (\ref{exrltnHubbard2}) and (\ref{diagHubbard4}), is
\be
{\cal T}_{8,8}(\lambda)\, {\cal T}_{1,5}(\mu) &=&
-\frac{i \alpha_{1}(\lambda,\mu)}{\alpha_{9}(\lambda,\mu)} 
{\cal T}_{1,5}(\mu)\, {\cal T}_{8,8}(\lambda) + 
\frac{i\alpha_{5}(\lambda,\mu)}{\alpha_{9}(\lambda,\mu)}
{\cal T}_{4,8}(\lambda)\, {\cal T}_{5,5}(\mu)\non \\
-& &\hspace{-0.4in}\frac{i\alpha_{1}(\lambda,\mu)\alpha_{4}(\lambda,\mu)}{\alpha_{9}(\lambda,\mu)\alpha_{7}(\lambda,\mu)}
{\cal T}_{1,8}(\mu)\, {\cal T}_{8,5}(\lambda) 
-\frac{\alpha_{1}(\lambda,\mu)\alpha_{10}(\lambda,\mu)}{\alpha_{9}(\lambda,\mu)\alpha_{7}(\lambda,\mu)}
\left( {\cal T}_{1,6}(\mu)\, {\cal T}_{8,7}(\lambda) - {\cal 
T}_{1,7}(\mu)\, {\cal T}_{8,6}(\lambda) \right) \non \\
-& &\hspace{-0.4in}\frac{\alpha_{4}(\lambda,\mu)}{\alpha_{7}(\lambda,\mu)}{\cal 
T}_{5,8}(\lambda)\, {\cal T}_{4,5}(\mu) 
-\frac{i\alpha_{10}(\lambda,\mu)}{\alpha_{7}(\lambda,\mu)}
\left( {\cal T}_{6,8}(\lambda)\, {\cal T}_{3,5}(\mu) - {\cal T}_{7,8}(\lambda)\, {\cal T}_{2,5}(\mu)  
\right)  \\
+& &\hspace{-0.4in}\frac{i\alpha_{5}(\lambda,\mu)\alpha_{4}(\lambda,\mu)}{\alpha_{9}(\lambda,\mu)\alpha_{7}(\lambda,\mu)}
{\cal T}_{1,8}(\lambda)\, {\cal T}_{8,5}(\mu)
+ \frac{\alpha_{5}(\lambda,\mu)\alpha_{10}(\lambda,\mu)}{\alpha_{9}(\lambda,\mu)\alpha_{7}(\lambda,\mu)}
 \left( {\cal T}_{2,8}(\lambda)\, {\cal T}_{7,5}(\mu) - {\cal 
 T}_{3,8}(\lambda)\, {\cal T}_{6,5}(\mu) \right)  \,. \non
\ee

It is not feasible to present all 64 exchange relations in their
entirety. Nevertheless, the ``wanted'' (diagonal) terms of all these
exchange relations are given in (\ref{exrlnHubbard}).

\section{Crossing symmetry}\label{sec:crossing}

We verify here that the twisted $S$-matrix (\ref{drinfeldtwist}) has 
the standard crossing 
symmetry property. For a single copy of the untwisted $su(2|2)$ $S$-matrix
in the elliptic parametrization \cite{AF}, 
the crossing relation \cite{Ja} is given by 
\be
C_{1}^{-1} S_{12}^{t_{1}}(z_{1},z_{2}) C_{1} 
S_{12}(z_{1}+\omega_{2},z_{2}) = I_{12} \,, \qquad
C_{1}^{-1} S_{12}^{t_{1}}(z_{1},z_{2}) C_{1} 
S_{12}(z_{1},z_{2}-\omega_{2}) = I_{12} \,,
\ee
where $C$ is the $4 \times 4$ matrix given by 
\be
C=\left( \begin{array}{cc}
\sigma_{2} & 0 \\
0 & i \sigma_{2} 
\end{array} \right) \,,
\ee
and $\sigma_{2}$ is the second Pauli matrix.

We write the full untwisted $su(2|2)^{2}$ $S$-matrix (\ref{calS}) as
\be
{\cal S}_{1234}(z_{1},z_{2}) = S_{13}(z_{1},z_{2})\,  
S_{24}(z_{1},z_{2}) \,.
\ee
One can easily check that it obeys the crossing relation
\be
C_{1}^{-1} C_{2}^{-1} {\cal S}_{1234}^{t_{1} t_{2}}(z_{1},z_{2}) 
C_{1}C_{2} 
{\cal S}_{1234}(z_{1}+\omega_{2},z_{2}) = I_{1234} \,,
\ee
and a similar second relation.
We write the twisted $S$-matrix (\ref{drinfeldtwist}) as 
\be 
\tilde {\cal S}_{1234}(z_{1},z_{2}) = F_{1234}\, {\cal S}_{1234}(z_{1},z_{2})\, F_{1234} \,, 
\ee 
It should obey the same crossing relation, i.e.,
\be
C_{1}^{-1} C_{2}^{-1} \tilde {\cal S}_{1234}^{t_{1} t_{2}}(z_{1},z_{2}) 
C_{1}C_{2} 
\tilde {\cal S}_{1234}(z_{1}+\omega_{2},z_{2}) = I_{1234} \,.
\label{twistedcrossing}
\ee
We find that the twist matrix $F$ (\ref{twistagain}) obeys the relation
\be
C_{1}^{-1} C_{2}^{-1} F_{1234} C_{1}C_{2} = F_{1234}^{-1} \,.
\ee
Using this identity, one can check that the crossing relation 
(\ref{twistedcrossing}) is indeed satisfied.

\section{L\"uscher correction}\label{sec:Luscher}

We show here that the proposed twisted scattering matrix and twisted
boundary condition reproduce the wrapping correction not only for
the Konishi operator \cite{ABBN} but also for generic multiparticle states
both in the $su(2)$ and in the $sl(2)$ sectors analyzed in \cite{GLM, AdLvT}.  

Let us start with the $su(2)$ sector. It consists of identical particles
carrying the labels $1\dot{1}=X$. In the L\"uscher correction, we need
the scattering matrix of the $X$ particle on the mirror boundstates.
Since only the $R_{1}$ charge gives a nonvanishing contribution on
$X$, the twist factor of the scattering matrix will be
\be 
F=e^{i\gamma_{12}\mathbb{I}\otimes R_{2}}=q^{\mathbb{I}\otimes 
R_{2}}\,,
\ee 
where in the last equality we focus on the $\beta$-deformation only:
$\gamma_{12}=\frac{1}{2}\gamma=\pi\beta$, and $q=e^{i \pi \beta}$. Evaluating $R_{2}$ on the
mirror boundstates gives the following twist of the $S$-matrix elements: 

\medskip{}

\begin{tabular}{|c|c|c|c|c|}
\hline 
 & $\vert B_{k}\rangle_{I}$ & $\vert B_{k}\rangle_{II}$ & $\vert F_{k}\rangle_{I}$ & $\vert F_{k}\rangle_{II}$\tabularnewline
\hline
$1$ & $1$ & $1$ & $q$ & $q^{-1}$\tabularnewline
\hline
\end{tabular}~~~~\begin{tabular}{|c|c|c|c|c|}
\hline 
 & $\vert\dot{B}_{k}\rangle_{I}$ & $\vert\dot{B}_{k}\rangle_{II}$ & $\vert\dot{F}_{k}\rangle_{I}$ & $\vert\dot{F}_{k}\rangle_{II}$\tabularnewline
\hline
$\dot{1}$ & $1$ & $1$ & $q^{-1}$ & $q$\tabularnewline
\hline
\end{tabular}

\medskip{}
\noindent
Taking into account that the scattering is diagonal in this sector,
the wrapping correction for $N$ particles of type $1\dot{1}$ will
contain the $N^{th}$ power of the above expressions. 

Additionally, we also have to remember that the mirror boundstate
should satisfy the twisted boundary condition, which for 
$\gamma_{23}=\gamma_{13}=\frac{1}{2}\gamma$
reads as: $q^{4J\,(h\otimes1)}$. In detail, we have\medskip{}

\noindent \begin{center}
\begin{tabular}{|c|c|c|c|c|}
\hline 
 & $\vert B_{k}\rangle_{I}$ & $\vert B_{k}\rangle_{II}$ & $\vert F_{k}\rangle_{I}$ & $\vert F_{k}\rangle_{II}$\tabularnewline
\hline
$BC$ & $1$ & $1$ & $1$ & $1$\tabularnewline
\hline
\end{tabular}~~~~\begin{tabular}{|c|c|c|c|c|}
\hline 
 & $\vert\dot{B}_{k}\rangle_{I}$ & $\vert\dot{B}_{k}\rangle_{II}$ & $\vert\dot{F}_{k}\rangle_{I}$ & $\vert\dot{F}_{k}\rangle_{II}$\tabularnewline
\hline
$BC$ & $1$ & $1$ & $q^{2J}$ & $q^{-2J}$\tabularnewline
\hline
\end{tabular}\medskip{}

\par\end{center}

\noindent 
Combining the two results, we can equivalently describe our
deformation with a different twisted boundary condition ($BC^{\prime}$)
given by \medskip{}

\noindent \begin{center}
\begin{tabular}{|c|c|c|c|c|}
\hline 
 & $\vert B_{k}\rangle_{I}$ & $\vert B_{k}\rangle_{II}$ & $\vert F_{k}\rangle_{I}$ & $\vert F_{k}\rangle_{II}$\tabularnewline
\hline
$BC^{\prime}$ & $1$ & $1$ & $q^{N}$ & $q^{-N}$\tabularnewline
\hline
\end{tabular}~~~~\begin{tabular}{|c|c|c|c|c|}
\hline 
 & $\vert\dot{B}_{k}\rangle_{I}$ & $\vert\dot{B}_{k}\rangle_{II}$ & $\vert\dot{F}_{k}\rangle_{I}$ & $\vert\dot{F}_{k}\rangle_{II}$\tabularnewline
\hline
$BC^{\prime}$ & $1$ & $1$ & $q^{2J-N}$ & $q^{-2J+N}$\tabularnewline
\hline
\end{tabular}\medskip{}

\par\end{center}

\noindent 
which completely agrees with the $su(2)$ part of Table
1 in \cite{AdLvT}. As the scatterings in the $sl(2)$ sectors are not
twisted, our twisted boundary conditions are equivalent to the $sl(2)$
part of Table 1 in \cite{AdLvT}. 


\section{Operatorial twists of the boundary conditions}\label{sec:optwist}

We demonstrate here that our transfer matrix, which is constructed
with twisted $S$-matrices and boundary conditions with $c$-number
twists, is spectrally equivalent to a transfer matrix which is
constructed with {\em untwisted} $S$-matrices and boundary conditions
with {\em operatorial} twists.  It is the latter type of transfer
matrix which is considered in \cite{AdLvT}.  Moreover, we show
directly that the same twisted Bethe equations can also be derived
starting from the latter transfer matrix.

\subsection{$su(2)$ principal chiral model}\label{optwistPCM}

For the case of the $su(2)$ principal chiral model, the transfer matrix 
is given by (\ref{twistedtransferPCM}). Let us now streamline the 
notation, and denote $a \dot a$ by $A$, and $j \dot j$ by $j$ for $j 
= 1, \ldots, L$. The transfer matrix (\ref{twistedtransferPCM}) then 
takes the form
\be
\tilde t(u) = \tr_{A} M_{A}\, \tilde {\cal T}_{A}(u) \,, \qquad 
\tilde {\cal T}_{A}(u) = \prod_{j=1}^{L} \tilde {\cal S}_{A j}(u) \,,
\label{twistedtransferPCM2}
\ee
where 
\be
\tilde {\cal S}_{A j}(u) = F_{A j}\, {\cal S}_{A j}(u)\, F_{A j} \,.
\ee 
The $F$-matrix satisfies \cite{drinfeld2}
\be
F_{12}\, F_{13}\, F_{23} = F_{23}\, F_{13}\, F_{12} \,,
\ee
as well as 
\be
{\cal S}_{12}(u)\, F_{13}\, F_{23} = F_{23}\, F_{13}\, {\cal 
S}_{12}(u) \,.
\ee
This equation means that the twist appears as a seam (defect) in the
spin chain, whose location can be changed without altering the
spectrum \cite{Bajnok:2006bf}.  To see this,
we observe that under spectral equivalence ($\equiv$) the 
$S$-matrix and the $F$-matrix acting in different quantum spaces 
commute,
\be
{\cal S}_{12}(u)\, F_{13} = F_{23} (F_{13}\, {\cal S}_{12}(u)) \, 
F_{23}^{-1} \equiv F_{13}\, {\cal S}_{12}(u)  \,.
\ee
The same is true for $F$-matrices,
\be
F_{12}\, F_{13} = F_{23} (F_{13}\, F_{12}) \, 
F_{23}^{-1} \equiv F_{13}\, F_{12}  \,.
\ee
The transfer matrix (\ref{twistedtransferPCM2}) can therefore be 
written as
\be
\tilde t(u) &=& \tr_{A} M_{A} \prod_{j=1}^{L} F_{A j}\, {\cal S}_{A 
j}(u)\, F_{A j} \non \\
&\equiv& \tr_{A} M_{A} \prod_{j=1}^{L} F_{A j}^{2} \prod_{j=1}^{L} {\cal S}_{A j}(u)  \non \\
&=& \tr_{A} \tilde M_{A}\, {\cal T}_{A}(u) \,, 
\label{twistedtransferPCMnew}
\ee
where
\be
\tilde M_{A} = M_{A} \prod_{j=1}^{L}  F_{A j}^{2} \,, \qquad
{\cal T}_{A}(u) = \prod_{j=1}^{L} {\cal S}_{A j}(u) \,.
\label{tMA}
\ee
This shows that the transfer matrix (\ref{twistedtransferPCM2}) (which
is constructed with the twisted $S$-matrices $\tilde {\cal S}$ and the
matrix $M_{A}$ which acts only on the auxiliary space) is spectrally
equivalent to the transfer matrix (\ref{twistedtransferPCMnew}) (which is
constructed with the {\em untwisted} $S$-matrices ${\cal S}$ and the
matrix $\tilde M_{A}$ which acts also on all the quantum spaces). 
\footnote{A similar observation has been made by Foerster, Links and 
Roditi \cite{drinfeld3}.}

Let us now explicitly evaluate $\tilde M_{A}$. The $F$-matrix (\ref{twist1})
can be rewritten as
\be
F_{Aj}=e^{i \gamma_{1}(H_{A} \dot H_{j} - \dot H_{A} H_{j})} \,,
\ee
where we have defined $H = h \otimes \id$ and $\dot H = \id \otimes h$. 
Hence,
\be
\prod_{j=1}^{L}  F_{A j}^{2} = e^{i2\gamma_{1}\left[ H_{A} 
\sum_{j=1}^{L}\dot H_{j} - \dot H_{A}\sum_{j=1}^{L}H_{j}\right]} \,.
\ee
Moreover, $M_{A}$ (\ref{MPCM}) can be rewritten as 
\be
M_{A} = e^{i\gamma_{2} H_{A}+ i\gamma_{3}\dot H_{A}} \,.
\ee 
Hence, $\tilde M_{A}$ in (\ref{tMA}) is given by
\be
\tilde M_{A} &=& e^{i \left(\gamma_{2}+2\gamma_{1} \sum_{j=1}^{L}\dot H_{j}\right) H_{A}
+ i \left(\gamma_{3}-2\gamma_{1} \sum_{j=1}^{L} H_{j}\right) \dot H_{A}} \non \\
&=& e^{i \left(\gamma_{2}+2\gamma_{1} \dot S^{z}\right) h} \otimes
e^{i \left(\gamma_{3}-2\gamma_{1} S^{z}\right) h}   \,, \label{MopPCM}
\ee
where the spin operators are given by
\be
S^{z} = \sum_{j=1}^{L} H_{j}\,, \qquad \dot S^{z} = 
\sum_{j= 1}^{L} \dot H_{j}\,.
\ee
Evidently, $\tilde M_{A}$ contains spin operators which act
on the quantum space.

Finally, let us derive the Bethe equations corresponding to the 
transfer matrix (\ref{twistedtransferPCMnew}). Using (\ref{MopPCM}),
we see that this transfer matrix is given by
\be
\tilde t(u) &=& e^{\frac{i}{2}\left[ \gamma_{2}+\gamma_{3}+2\gamma_{1}\left(\dot 
S^{z}- S^{z}\right)\right]} {\cal T}_{11}(u) +
e^{\frac{i}{2}\left[ \gamma_{2}-\gamma_{3}+2\gamma_{1}\left(\dot 
S^{z}+ S^{z}\right)\right]} {\cal T}_{22}(u) \non \\
&+&
e^{\frac{i}{2}\left[ \gamma_{3}-\gamma_{2}-2\gamma_{1}\left(\dot 
S^{z}+ S^{z}\right)\right]} {\cal T}_{33}(u) +
e^{-\frac{i}{2}\left[ \gamma_{2}+\gamma_{3}+2\gamma_{1}\left(\dot 
S^{z}- S^{z}\right)\right]} {\cal T}_{44}(u) \,.
\label{twistedtransferPCM3}
\ee
Recall that the spin operators satisfy the commutation relations
\be
\left[ S^{z}\,, B(u) \right] = - B(u)\,, \qquad 
\left[ S^{z}\,, A(u) \right] = \left[ S^{z}\,, D(u) \right] = 0 \,,
\ee 
and similarly for the operators with dots. Hence, the commutation 
relations of the spin operators with the creation operators are given by 
\be
\left[  S^{z}\,,  {\cal T}_{13}(u) \right] &=& - {\cal T}_{13}(u)  \,,  \qquad 
\left[  S^{z}\,, {\cal T}_{12}(u) \right] = 0  \non \\
\left[ \dot S^{z}\,, {\cal T}_{13}(u) \right] &=& 0 \,, \qquad\qquad\quad
\left[ \dot S^{z}\,, {\cal T}_{12}(u) \right] = - {\cal T}_{12}(u) 
\,.
\ee
Moreover, acting on the pseudovacuum,
\be
S^{z} (|0\rangle \otimes |\dot 0\rangle) = \frac{L} {2} (|0\rangle 
\otimes |\dot 0\rangle)
\,,  \qquad
\dot S^{z} (|0\rangle \otimes |\dot 0\rangle) = \frac{L} {2} 
(|0\rangle \otimes |\dot 0\rangle)
\,.   
\ee 
Therefore, acting on a general state (\ref{ansatz2}), 
\be 
e^{i\gamma  S^{z}} | \Lambda \rangle = e^{i\gamma \left( 
\frac{L}{2} - m \right)}| \Lambda \rangle \,, \qquad 
e^{i\gamma \dot S^{z}} | \Lambda \rangle = e^{i\gamma \left( 
\frac{L}{2} - \dot m \right)}| \Lambda \rangle \,.
\ee
Acting with the transfer matrix (\ref{twistedtransferPCM3})
on a general state (\ref{ansatz2}), we see (using also the untwisted exchange 
relations (\ref{exchange1}), (\ref{exchange2}) and the pseudovacuum eigenvalues 
(\ref{vaceigenval})) that the
corresponding eigenvalues are given by the same expression
(\ref{eigenvalsPCM}) which we obtained before. Hence, we arrive at the 
same twisted Bethe equations as before.

\subsection{Two copies of the Hubbard model}\label{optwistHubbard}

For the case of two copies of the Hubbard model, the same argument as 
above implies that the transfer matrix  (\ref{Hubbardtransfertwisted})
is spectrally equivalent to 
\be
\tilde t(\lambda) = \str_{a \dot a}  \tilde M_{a \dot a} {\cal T}_{a \dot a}(\lambda) \,,
\label{Hubbardtransfertwisted2}
\ee
where the monodromy matrix ${\cal T}_{a \dot a}(\lambda)$ is not
twisted, but the diagonal matrix $\tilde M$ contains spin-like operators
which act on the quantum space,
\be
\tilde M = e^{i \left(\gamma_{2}+\gamma_{1} \dot \eta^{z}\right) h} \otimes
e^{i \left(\gamma_{3}-\gamma_{1} \eta^{z}\right) h}   \,,
\label{MopHubbard}
\ee
where the $\eta^{z}$ operator is defined in MR (136). The matrix $h$ 
is now given by (\ref{h}).

The operator $\eta^{z}$ has the following property given in MR (139)
\be
\left[ \eta^{z}\,, \vec B(\lambda) \right] = - \vec B(\lambda)\,.
\ee 
Hence, the commutation 
relations of $\eta^{z}$ and $\dot\eta^{z}$ with the creation operators are given by
\be
\left[  \eta^{z}\,,  {\cal T}_{1,2}(\lambda) \right] &=& 
\left[  \eta^{z}\,,  {\cal T}_{1,3}(\lambda) \right] = 0   \,,  \quad 
\left[  \eta^{z}\,,  {\cal T}_{1,5}(\lambda) \right] = - 
{\cal T}_{1,5}(\lambda)  \,,  \quad 
\left[  \eta^{z}\,,  {\cal T}_{1,9}(\lambda) \right] = - 
{\cal T}_{1,9}(\lambda)  \,, \non \\
\left[  \dot\eta^{z}\,,  {\cal T}_{1,2}(\lambda) \right] &=& - 
{\cal T}_{1,2}(\lambda) \,,  \quad 
\left[  \dot\eta^{z}\,,  {\cal T}_{1,3}(\lambda) \right] = - 
{\cal T}_{1,3}(\lambda)   \,,  \quad 
\left[  \dot\eta^{z}\,,  {\cal T}_{1,5}(\lambda) \right] =
\left[  \dot\eta^{z}\,,  {\cal T}_{1,9}(\lambda) \right] = 0  \,. \non \\
\ee
Moreover, acting on the pseudovacuum,
\be
\eta^{z} (|0\rangle \otimes |\dot 0\rangle) = L (|0\rangle 
\otimes |\dot 0\rangle)
\,,  \qquad
\dot \eta^{z} (|0\rangle \otimes |\dot 0\rangle) = L
(|0\rangle \otimes |\dot 0\rangle)
\,.   
\ee 
Therefore, acting on a general state, 
\be 
e^{i\gamma  \eta^{z}} | \Lambda \rangle = e^{i\gamma \left( 
L - n \right)}| \Lambda \rangle \,, \qquad 
e^{i\gamma \dot \eta^{z}} | \Lambda \rangle = e^{i\gamma \left( 
L - \dot n \right)}| \Lambda \rangle \,.
\ee
Acting with the transfer matrix (\ref{Hubbardtransfertwisted2})
on a general state, we find (using also the untwisted exchange 
relations (\ref{exrlnHubbard}) and the pseudovacuum eigenvalues 
(\ref{psuedovaceigenvals}))
that the
corresponding eigenvalues are given by the same expression (\ref{eigenvalstwisted}) 
which we obtained before. Hence, we arrive at the 
same twisted Bethe equations as before.

\subsection{AdS/CFT}\label{optwistAdSCFT}

For the AdS/CFT case, it now follows that the transfer matrix
(\ref{twistedBCagain}) is spectrally equivalent to 
\be
\tilde t(\lambda) = \str_{a \dot a}  \tilde M_{a \dot a}  S_{a \dot a 1 \dot 1}(\lambda,p_{1}) 
\ldots  S_{a \dot a N \dot N}(\lambda,p_{N}) \,,
\label{twistedBCagain2}
\ee
where the matrix $\tilde M_{a \dot a}$ is given by 
\be
\tilde M = e^{i \left[(\gamma_{3}-\gamma_{2}) J +\gamma_{1} \dot 
\eta^{z}\right] h}  
\otimes e^{i \left[ (\gamma_{3}+\gamma_{2}) J -\gamma_{1} 
\eta^{z}\right] h}  \,,
\label{MAdSCFTop}
\ee 
and $h$ is given by (\ref{hh}). This leads to the same eigenvalues 
(\ref{lambdaexplic}), and therefore the BR Bethe equations.

\section{$sl(2)$ grading}\label{sec:sl2}

Here we transform our twisted Bethe ansatz results from the ``$su(2)$'' grading 
to the ``$sl(2)$'' grading, and show that the results
agree with both \cite{BR} and \cite{AdLvT}.

We recall that the eigenvalues of the twisted AdS/CFT transfer matrix (\ref{twistedBCagain})
in the $su(2)$ grading are given by (\ref{lambdaexplic}), which 
we now abbreviate as follows
\be
\tilde\Lambda_{su2}(\lambda)=\prod_{i=1}^{N} S_{0}(\lambda,p_{i})^{2} 
\left[\Lambda_{1}(\lambda)-\Lambda_{2}(\lambda)-\Lambda_{3}(\lambda)+\Lambda_{4}(\lambda)\right]
\left[\dot{\Lambda}_{1}(\lambda)-\dot{\Lambda}_{2}(\lambda)
-\dot{\Lambda}_{3}(\lambda)+\dot{\Lambda}_{4}(\lambda)\right] 
\ee
In order to obtain the corresponding expression in the $sl(2)$
grading, we perform a dualization on the fermionic roots
$x^{+}(\lambda_{j})$ \cite{Beisert, Beisert:2005di} by noting
that the first Bethe equation in (\ref{betheexp}) is an algebraic equation 
$q(x^{+}(\lambda_{j}))=0$, where $q(x)$ is given by 
\be
\label{q1}
q(x) &=& c_{1}\prod_{i=1}^{N}\eta(p_{i})(x-x^{-}(p_{i}))
\prod_{l=1}^{m_{2}}x\left(x+\frac{1}{x}-\tilde{\mu}_{l}-\frac{i}{2g}\right) \non \\
&-& \prod_{i=1}^{N}(x-x^{+}(p_{i}))
\prod_{l=1}^{m_{2}}x\left(x+\frac{1}{x}-\tilde{\mu}_{l}+\frac{i}{2g}\right)\,.
\ee 
Note that we have included a factor $x^{m_{2}}$ to ensure that $q(x)$ is a polynomial in $x$ of 
degree $N+2m_{2}$.
This polynomial has $m_{1}$ roots $x^{+}(\lambda_{j})$ and $\tilde{m}_{1}$ 
additional roots $x^{+}(\tilde{\lambda}_{j})$, where $\tilde{m}_{1}=N+2m_{2}-m_{1}$.
It can therefore be written also in the following factorized form 
\be
\label{q2}
q(x)=c
\prod_{j=1}^{m_{1}}\left(x-x^+(\lambda_j)\right)
\prod_{j=1}^{\tilde{m}_{1}}\left(x-x^+(\tilde{\lambda}_j)\right) \,,
\ee 
where $c$ is some non-vanishing constant.
The equality of (\ref{q1}) and (\ref{q2}) implies that the function 
$Q(x)$ defined by
\be
Q(x)&=&\prod_{j=1}^{m_{1}}\frac{1}{\left(x-x^+(\lambda_j)\right)}
\prod_{j=1}^{\tilde{m}_{1}}\frac{1}{\left(x-x^+(\tilde{\lambda}_j)\right)}
\left[c_{1}\prod_{i=1}^{N}\eta(p_{i})(x-x^{-}(p_{i}))
\prod_{l=1}^{m_{2}}x\left(x+\frac{1}{x}-\tilde{\mu}_{l}-\frac{i}{2g}\right)\right.\non\\
&-&\left.\prod_{i=1}^{N}(x-x^{+}(p_{i}))
\prod_{l=1}^{m_{2}}x\left(x+\frac{1}{x}-\tilde{\mu}_{l}+\frac{i}{2g}\right)\right]
\ee
is independent of $x$. The identities following from 
$Q\left(x^{+}(\lambda)\right)=Q\left(x^{-}(\lambda)\right)$ and 
$Q\left(1/x^{+}(\lambda)\right)=Q\left(1/x^{-}(\lambda)\right)$ 
imply that 
\be
\label{dual1}
\Lambda_{1}(\lambda)-\Lambda_{2}(\lambda)&=&\eta(\lambda)^{m_{1}-N}
\prod_{i=1}^{N}\frac{x^{+}(\lambda)-x^{-}(p_{i})}{x^{-}(\lambda)-x^{+}(p_{i})}
\prod_{j=1}^{m_{1}}\frac{x^{-}(\lambda)-x^+(\lambda_j)}{x^{+}(\lambda)-x^+(\lambda_j)}\non \\
&\times& \left[c_{1}\prod_{i=1}^{N}\eta(p_{i})
-\prod_{i=1}^{N}\frac{x^{+}(\lambda)-x^{+}(p_{i})}{x^{+}(\lambda)-x^{-}(p_{i})}
\prod_{l=1}^{m_{2}}\frac{x^{+}(\lambda)+\frac{1}{x^{+}(\lambda)}-\tilde{\mu}_{l}+\frac{i}{2g}}
{x^{+}(\lambda)+\frac{1}{x^{+}(\lambda)}-\tilde{\mu}_{l}-\frac{i}{2g}}\right]\non\\
&=&
-\eta(\lambda)^{2m_{2}}\left[\frac{x^{-}(\lambda)}{x^{+}(\lambda)}\right]^{m_{2}}
\prod_{j=1}^{\tilde{m}_{1}}\frac{1}{\eta(\lambda)}\frac{x^{+}(\lambda)-x^+(\tilde{\lambda}_j)}{x^{-}(\lambda)-x^+(\tilde{\lambda}_j)}\non\\
&\times&\left[1-c_{1}\prod_{i=1}^{N}\eta(p_{i})\frac{x^{-}(\lambda)-x^{-}(p_{i})}{x^{-}(\lambda)-x^{+}(p_{i})}
\prod_{l=1}^{m_{2}}\frac{x^{-}(\lambda)+\frac{1}{x^{-}(\lambda)}-\tilde{\mu}_{l}-\frac{i}{2g}}
{x^{-}(\lambda)+\frac{1}{x^{-}(\lambda)}-\tilde{\mu}_{l}+\frac{i}{2g}}\right] \non \\
&\equiv& \tilde{\Lambda}_{1}(\lambda)-\tilde{\Lambda}_{2}(\lambda) \,,
\ee
and
\be
\label{dual2}
-\Lambda_{3}(\lambda)+\Lambda_{4}(\lambda) &=& \eta(\lambda)^{m_{1}-N}
\prod_{i=1}^{N}\frac{x^{+}(\lambda)-x^{+}(p_{i})}{x^{-}(\lambda)-x^{+}(p_{i})}
\prod_{j=1}^{m_{1}}\frac{\frac{1}{x^{+}(\lambda)}-x^+(\lambda_j)}{\frac{1}{x^{-}(\lambda)}-x^+(\lambda_j)}\non\\
&\times&\left[-\prod_{l=1}^{m_{2}}\frac{x^{-}(\lambda)+\frac{1}{x^{-}(\lambda)}-\tilde{\mu}_{l}-\frac{i}{2g}}
{x^{-}(\lambda)+\frac{1}{x^{-}(\lambda)}-\tilde{\mu}_{l}+\frac{i}{2g}}
+c_{1}^{-1}\prod_{i=1}^{N}\frac{1-\frac{1}{x^{-}(\lambda)x^{+}(p_{i})}}{1-\frac{1}{x^{-}(\lambda)x^{-}(p_{i})}}\eta(p_{i})\right]\non\\
&=&-\eta(\lambda)^{2m_{2}}\left[\frac{x^{-}(\lambda)}{x^{+}(\lambda)}\right]^{m_{2}}
\prod_{j=1}^{\tilde{m}_{1}}\frac{1}{\eta(\lambda)}
\frac{\frac{1}{x^{-}(\lambda)}-x^+(\tilde{\lambda}_j)}{\frac{1}{x^{+}(\lambda)}-x^+(\tilde{\lambda}_j)}
\prod_{i=1}^{N}\frac{x^{-}(\lambda)-x^{-}(p_{i})}{x^{-}(\lambda)-x^{+}(p_{i})}\non\\
&\times&\left[-c_{1}^{-1}\prod_{i=1}^{N}\eta(p_{i})
\prod_{l=1}^{m_{2}}\frac{x^{+}(\lambda)+\frac{1}{x^{+}(\lambda)}-\tilde{\mu}_{l}+\frac{i}{2g}}
{x^{+}(\lambda)+\frac{1}{x^{+}(\lambda)}-\tilde{\mu}_{l}-\frac{i}{2g}}
+\prod_{i=1}^{N}\frac{1-\frac{1}{x^{+}(\lambda)x^{-}(p_{i})}}{1-\frac{1}{x^{+}(\lambda)x^{+}(p_{i})}}\right] \non\\
&\equiv& -\tilde{\Lambda}_{3}(\lambda)+\tilde{\Lambda}_{4}(\lambda)\,,
\ee 
respectively. Performing an analogous dualization for the roots 
$x^{+}(\dot{\lambda}_{j})$ in $\dot{\Lambda}_{1}(\lambda)-\dot{\Lambda}_{2}(\lambda)$ 
and $-\dot{\Lambda}_{3}(\lambda)+\dot{\Lambda}_{4}(\lambda)$, and recalling that 
$\eta(\lambda)=\sqrt{\frac{x^{+}(\lambda)}{x^{-}(\lambda)}}$, 
we arrive at the desired result for the dual eigenvalues of the twisted AdS/CFT transfer matrix
\be
\label{duallambdaexplic}
&&\tilde{\Lambda}_{sl2}(\lambda)=\prod_{i=1}^{N} S_{0}(\lambda,p_{i})^{2}
\Bigg[
\prod_{j=1}^{\tilde{m}_1} \frac{1}{\eta(\lambda)} \frac{x^{+}(\lambda)-x^{+}(\tilde{\lambda}_j)}{x^{-}(\lambda)-x^{+}(\tilde{\lambda}_j)}-\prod_{i=1}^{N} \frac{x^{-}(\lambda)-x^{-}(p_{i})}{x^{-}(\lambda)-x^{+}(p_{i})} \eta(p_{i})\non\\
&\times&\left \{c_{1}
\prod_{j=1}^{\tilde{m}_1} \frac{1}{\eta(\lambda) }\left[\frac{x^{+}(\lambda)-x^{+}(\tilde{\lambda}_j)}{x^{-}(\lambda)-x^{+}(\tilde{\lambda}_j)} \right]
\prod_{l=1}^{m_2} \frac{x^{-}(\lambda)+\frac{1}{x^{-}(\lambda)}-\tilde{\mu_l}-\frac{i}{2 g}}
{x^{-}(\lambda)+\frac{1}{x^{-}(\lambda)}-\tilde{\mu_l}+\frac{i}{2 g}} \right.
\nonumber\\
&+&
c_{1}^{-1}\left.\prod_{j=1}^{\tilde{m}_1} \frac{1}{\eta(\lambda)}\left[\frac{x^{+}(\tilde{\lambda}_j)-\frac{1}{x^{-}(\lambda)}}{x^{+}(\tilde{\lambda}_j)-\frac{1}{x^{+}(\lambda)}}\right]
\prod_{l=1}^{m_2} \frac{x^{+}(\lambda)+\frac{1}{x^{+}(\lambda)}-\tilde{\mu_l}+\frac{i}{2 g}}
{x^{+}(\lambda)+\frac{1}{x^{+}(\lambda)}-\tilde{\mu_l}-\frac{i}{2 g}} \right \}
\nonumber\\
&+&
\prod_{i=1}^{N} \left[\frac{1-\frac{1}{x^{+}(\lambda) x^{-}(p_i)}}{1-\frac{1}{x^{+}(\lambda) x^{+}(p_i)}}\right]
\left[\frac{x^{-}(p_i)-x^{-}(\lambda)}{x^{+}(p_i)-x^{-}(\lambda)} \right] \prod_{j=1}^{\tilde{m}_1} \frac{1}{\eta(\lambda) }\left [ 
\frac{x^{+}(\tilde{\lambda}_j)-\frac{1}{x^{-}(\lambda)}}{x^{+}(\tilde{\lambda}_j)-\frac{1}{x^{+}(\lambda)}} \right ] \Bigg]
\nonumber\\
&\times&\Bigg[
\prod_{j=1}^{\dot{\tilde{m}}_1} \frac{1}{\eta(\lambda)} \frac{x^{+}(\lambda)-x^{+}(\dot{\tilde{\lambda}}_j)}{x^{-}(\lambda)-x^{+}(\dot{\tilde{\lambda}}_j)}-\prod_{i=1}^{N} \frac{x^{-}(\lambda)-x^{-}(p_{i})}{x^{-}(\lambda)-x^{+}(p_{i})} \eta(p_{i})\non\\
&\times&\left \{c_{2}
\prod_{j=1}^{\dot{\tilde{m}}_1} \frac{1}{\eta(\lambda) }\left[\frac{x^{+}(\lambda)-x^{+}(\dot{\tilde{\lambda}}_j)}{x^{-}(\lambda)-x^{+}(\dot{\tilde{\lambda}}_j)} \right]
\prod_{l=1}^{\dot{m}_2} \frac{x^{-}(\lambda)+\frac{1}{x^{-}(\lambda)}-\dot{\tilde{\mu}}_l-\frac{i}{2 g}}
{x^{-}(\lambda)+\frac{1}{x^{-}(\lambda)}-\dot{\tilde{\mu}}_l+\frac{i}{2 g}} \right.
\nonumber\\
&+&
c_{2}^{-1}
\left.\prod_{j=1}^{\dot{\tilde{m}}_1} \frac{1}{\eta(\lambda)}\left[\frac{x^{+}(\dot{\tilde{\lambda}}_j)-\frac{1}{x^{-}(\lambda)}}{x^{+}(\dot{\tilde{\lambda}}_j)-\frac{1}{x^{+}(\lambda)}}\right]
\prod_{l=1}^{\dot{m}_2} \frac{x^{+}(\lambda)+\frac{1}{x^{+}(\lambda)}-\dot{\tilde{\mu}}_l+\frac{i}{2 g}}
{x^{+}(\lambda)+\frac{1}{x^{+}(\lambda)}-\dot{\tilde{\mu}}_l-\frac{i}{2 g}} \right \}
\nonumber\\
&+&
\prod_{i=1}^{N} \left[\frac{1-\frac{1}{x^{+}(\lambda) x^{-}(p_i)}}{1-\frac{1}{x^{+}(\lambda) x^{+}(p_i)}}\right]
\left[\frac{x^{-}(p_i)-x^{-}(\lambda)}{x^{+}(p_i)-x^{-}(\lambda)} \right] \prod_{j=1}^{\dot{\tilde{m}}_1} \frac{1}{\eta(\lambda) }\left [ 
\frac{x^{+}(\dot{\tilde{\lambda}}_j)-\frac{1}{x^{-}(\lambda)}}{x^{+}(\dot{\tilde{\lambda}}_j)-\frac{1}{x^{+}(\lambda)}} \right ] \Bigg]\,,
\ee
where the twist factors can be expressed in terms of $\tilde{m}_{1}$ and $\dot{\tilde{m}}_{1}$ as follows
\be
\label{ctildes}
c_{1}=e^{i(\gamma_{3}-\gamma_{2})J/2}e^{i \gamma_{1} 
(\dot{\tilde{m}}_{1}-2\dot{m_{2}})/2}\,,
\qquad c_{2}=e^{i(\gamma_{3}+\gamma_{2})J/2}e^{-i\gamma_{1} (\tilde{m}_{1}-2m_{2})/2}\,.
\ee

If we identify $c_{1}$ ($c_{2}$) with $e^{i\alpha_{l}}$ ($e^{i\alpha_{r}}$), 
the eigenvalue of the ``left (right) wing'' transfer matrix matches 
the expression (8.1) of \cite{AdLvT} with $Q=1$, 
once we map $x^{\pm}\rightarrow x^{\mp}$, $g\rightarrow -g/2$, $x^{+}(\tilde{\lambda}_{j})\rightarrow y^{1}_{j}$ ($x^{+}(\dot{\tilde{\lambda}}_{j})\rightarrow y^{2}_{j}$), 
$\tilde{\mu}_{l}\rightarrow w^{1}_{l}$ ($\dot{\tilde{\mu}}_{l}\rightarrow w^{2}_{l}$), and $N=K^{\rm{I}},\ 
\tilde{m}_{1}=K^{\rm{II}}_{1},\ 
m_{2}=K^{\rm{III}}_{1}$ ($\dot{\tilde{m}}_{1}=K^{\rm{II}}_{2},\ \dot{m}_{2}=K^{\rm{III}}_{2}$):
\begin{eqnarray}
    \label{eqn;FullEignvalue}
&&T_{1,1}^{l}(v\,|\,\vec{u})=\prod_{i=1}^{K^{\rm{II}}_{1}}{\textstyle{\frac{y^{1}_i-x^-}
{y^{1}_i-x^+}\sqrt{\frac{x^+}{x^-}}
\, +}}\non \\
&&{\textstyle{+}}\prod_{i=1}^{K^{\rm{II}}_{1}}{\textstyle{\frac{y^{1}_i-x^-}{y^{1}_i-x
^+}\sqrt{\frac{x^+}{x^-}}\left[
\frac{x^++\frac{1}{x^+}-y^{1}_i-\frac{1}{y^{1}_i}}{x^++\frac{1}{x^+}-y^{1}_i-\frac{1
}{y^{1}_i}-\frac{2i}{g}}\right]}}\prod_{i=1}^{K^{\rm{I}}}
{\textstyle{\left[\frac{(x^--x^-_i)(1-x^-
x^+_i)}{(x^+-x^-_i)(1-x^+
x^+_i)}\frac{x^+}{x^-}  \right]}}\nonumber\\
&&\quad -\prod_{i=1}^{K^{\rm{II}}_{1}}
{\textstyle{\frac{y^{1}_i-x^-}{y^{1}_i-x^+}\sqrt{\frac{x^+}{x^-}}}}\prod_{i=1}^
{K^{\rm{I}}}{\textstyle{\frac{x^+-x^+_i}{x^+-x^-_i}\sqrt{\frac{x^-_i}{x^
+_i}} }}\times\nonumber\\
&&\quad\times
\left\{e^{i\alpha_{l}}\prod_{i=1}^{K^{\rm{III}}_{1}}{\textstyle{\frac{w^{1}_i-x^+-\frac{1}{x^+}
-\frac{i}{g}}{w^{1}_i-x^+-\frac{1}{x^+}+\frac{i}{g}}+ }}
e^{-i\alpha_{l}}\prod_{i=1}^{K^{\rm{II}}_{1}}{\textstyle{\frac{y^{1}_i+\frac{1}{y^{1}_i}-x^+-\frac
{1}{x^+}}{y^{1}_i+\frac{1}{y^{1}_i}-x^+-\frac{1}{x^+}+\frac{2i}{g}}}}\prod_{i=1}^{K^{\rm{III}}_{1}}{\textstyle{\frac{w^{!}_i-x^+-\frac{1}{x^+
}+\frac{3i}{g}}{w^{1}_i-x^+-\frac{1}{x^+}+\frac{i}{g}}}}\right\},
\end{eqnarray}
where we use the identities:
\be
&{\textstyle{\frac{y_i-x^-}{y_i-x
^+}\left[
\frac{x^++\frac{1}{x^+}-y_i-\frac{1}{y_i}}{x^++\frac{1}{x^+}-y_i-\frac{1
}{y_i}-\frac{2i}{g}}\right]}=\frac{y_{i}-\frac{1}{x^{+}}}{y_{i}-\frac{1}{x^{-}}}\,,}\quad{\textstyle{\frac{(x^--x^-_i)(1-x^-
x^+_i)}{(x^+-x^-_i)(1-x^+
x^+_i)}\frac{x^+}{x^-}}}={\textstyle{\frac{(x^+-x^+_i)\left(1-\frac{1}{x^-
x^+_i}\right)}{(x^+-x^-_i)\left(1-\frac{1}{x^-
x^-_i}\right)}}\,,}&
 \non \\
&{\textstyle{\frac{w_i-x^+-\frac{1}{x^+
}+\frac{3i}{g}}{w_i-x^+-\frac{1}{x^+}+\frac{i}{g}}}}={\textstyle{\frac{w_i-x^--\frac{1}{x^-
}+\frac{i}{g}}{w_i-x^--\frac{1}{x^-}-\frac{i}{g}}}\,.}&
\ee

The Bethe equations corresponding to the eigenvalues (\ref{duallambdaexplic}) are 
given by:
\bear
\label{dualbetheexp}
c_{1}\prod_{i=1}^{N} \left[ \frac{x^{+}(\tilde{\lambda}_j)-x^{-}(p_{i})}{x^{+}(\tilde{\lambda}_j)-x^{+}(p_{i})} \right] \eta(p_{i}) &=&
\prod_{l=1}^{m_2} \frac{x^{+}(\tilde{\lambda}_j)+\frac{1}{x^{+}(\tilde{\lambda}_j)}-\tilde{\mu_l}+\frac{i}{2 g}}
{x^{+}(\tilde{\lambda}_j)+\frac{1}{x^{+}(\tilde{\lambda}_j)}-\tilde{\mu_l}-\frac{i}{2 
g}}\,, \non \\
& &  ~~~~~ j=1,\dots, \tilde m_1,
\nonumber \\
c_{2}\prod_{i=1}^{N} \left[ \frac{x^{+}(\dot{\tilde{\lambda}}_j)-x^{-}(p_{i})}{x^{+}(\dot{\tilde{\lambda}}_j)-x^{+}(p_{i})} \right] \eta(p_{i}) &=&
\prod_{l=1}^{\dot m_2} \frac{x^{+}(\dot{\tilde{\lambda}}_j)+\frac{1}{x^{+}(\dot{\tilde{\lambda}}_j)}-\dot{\tilde{\mu}}_l+\frac{i}{2 g}}
{x^{+}(\dot{\tilde{\lambda}}_j)+\frac{1}{x^{+}(\dot{\tilde{\lambda}}_j)}-\dot{\tilde{\mu}}_l-\frac{i}{2 g}}\,, \non \\
& &  ~~~~~ j=1,\dots,\dot {\tilde m}_1,
\nonumber \\
c_{1}^{2}\prod_{j=1}^{\tilde{m}_1} \frac{\tilde{\mu}_{l}-x^{+}(\tilde{\lambda}_j)-\frac{1}{x^{+}(\tilde{\lambda}_j)}+\frac{i}{2 g}}
{\tilde{\mu}_{l}-x^{+}(\tilde{\lambda}_j)-\frac{1}{x^{+}(\tilde{\lambda}_j)}-\frac{i}{2 g}} & =& 
\prod_{\stackrel{k=1}{k \neq l}}^{m_2} 
\frac{\tilde{\mu}_{l}-\tilde{\mu}_{k}+\frac{i}{g}}
{\tilde{\mu}_{l}-\tilde{\mu}_{k}-\frac{i}{g}}\,,   ~l=1,\dots,m_2, 
\non \\
c_{2}^{2}\prod_{j=1}^{\dot{\tilde{m}}_1} \frac{\dot{\tilde{\mu}}_l-x^{+}(\dot{\tilde{\lambda}}_j)-\frac{1}{x^{+}(\dot{\tilde{\lambda}}_j)}+\frac{i}{2 g}}
{\dot{\tilde{\mu}}_l-x^{+}(\dot{\tilde{\lambda}}_j)-\frac{1}{x^{+}(\dot{\tilde{\lambda}}_j)}-\frac{i}{2 g}} & =& 
\prod_{\stackrel{k=1}{k \neq l}}^{m_2} 
\frac{\dot{\tilde{\mu}}_l-\dot{\tilde{\mu}}_k+\frac{i}{g}}
{\dot{\tilde{\mu}}_l-\dot{\tilde{\mu}}_k-\frac{i}{g}}\,,   
~l=1,\dots,\dot m_2.
\ear
which match the equations (8.2) in \cite{AdLvT}, via the 
identifications used above.
Following the change in notation in MM (47)-(49), (51), so that 
$\tilde m_{1} \mapsto \tilde m_{1}^{(1)}\,, m_{2} \mapsto  
m_{2}^{(1)}\,, \dot {\tilde m}_{1} \mapsto \tilde m_{1}^{(2)}\,, \dot m_{2} \mapsto  
m_{2}^{(2)}$ and 
\be
N = K_{4}\,, \quad \tilde{m}_{1}^{(1)}= \tilde{K}_{1} + \tilde{K}_{3}\,, \quad \tilde{m}_{1}^{(2)}= 
\tilde{K}_{5} + \tilde{K}_{7}\,, \quad m_{2}^{(1)}= K_{2}\,, \quad m_{2}^{(2)}= 
K_{6} \,,
\ee
then the coefficients (\ref{ctildes}) are given by
\be
c_{1}=e^{i(\gamma_{3}-\gamma_{2})J/2}e^{i \gamma_{1} 
(\tilde{K}_{5}+\tilde{K}_{7}-2K_{6})/2}\,, \qquad 
c_{2}=e^{i(\gamma_{3}+\gamma_{2})J/2}e^{-i\gamma_{1} 
(\tilde{K}_{1}+\tilde{K}_{3}-2K_{2})/2}\,,
\ee
and the first two equations in (\ref{dualbetheexp}) become
\bear
c_{1}e^{-i P/2} 
\prod_{i=1}^{K_4}  
\frac{1-\frac{g^2}{x^{-}_{4,i} x_{\tilde 1,j}}}
{1-\frac{g^2}{x^{+}_{4,i} x_{\tilde 1,j}}}\prod_{l=1}^{K_2} 
\frac{u_{\tilde 1,j}-u_{2,l}-\frac{i}{2}}{u_{\tilde 1,j}-u_{2,l}+\frac{i}{2}} & =& 
1,~~j=1,\dots,\tilde{K}_1
\label{dualbet2K1} \\
c_{1}
e^{i P/2} 
\prod_{i=1}^{K_4}  
\frac{x^{-}_{4,i}-x_{\tilde 3,j}}{x^{+}_{4,i}-x_{\tilde 3,j}}
\prod_{l=1}^{K_2} \frac{u_{\tilde 3,j}-u_{2,l}-\frac{i}{2}}{u_{\tilde 3,j}-u_{2,l}+\frac{i}{2}} &=&
1,~~j=1,\dots,\tilde{K}_3
\label{dualbet2K3} \\
c_{2} 
e^{i P/2} 
\prod_{i=1}^{K_4}  
\frac{x^{-}_{4,i}-x_{\tilde 5,j}}{x^{+}_{4,i}-x_{\tilde 5,j}}
\prod_{l=1}^{K_6} \frac{u_{\tilde 5,j}-u_{6,l}-\frac{i}{2}}{u_{\tilde 5,j}-u_{6,l}+\frac{i}{2}} & =&1
,~~j=1,\dots,\tilde{K}_5
\label{dualbet2K5} \\
c_{2} 
e^{-i P/2} 
\prod_{i=1}^{K_4}  
\frac{1-\frac{g^2}{x^{-}_{4,i} x_{\tilde 7,j}}}
{1-\frac{g^2}{x^{+}_{4,i} x_{\tilde 7,j}}}
\prod_{l=1}^{K_6} \frac{u_{\tilde 7,j}-u_{6,l}-\frac{i}{2}}{u_{\tilde 7,j}-u_{6,l}+\frac{i}{2}} & =& 1
,~~j=1,\dots,\tilde{K}_7
\label{dualbet2K7}
\ear
while the last two can be written as
\be
c_{1}^{-2}\prod_{j=1}^{\tilde{K}_1} \frac{u_{2,l}-u_{\tilde 1,j}-\frac{i}{2}}
{u_{2,l}-u_{\tilde 1,j}+\frac{i}{2}}
\prod_{j=1}^{\tilde{K}_3} \frac{u_{2,l}-u_{\tilde 3,j}-\frac{i}{2}}
{u_{2,l}-u_{\tilde 3,j}+\frac{i}{2}}\prod_{\stackrel{k=1}{k \neq l}}^{K_2} 
\frac{u_{2,l}-u_{2,k}+i}
{u_{2,l}-u_{2,k}-i} & =& 1,~~l=1,\dots,K_2 \\
c_{2}^{-2}\prod_{j=1}^{\tilde{K}_5} \frac{u_{6,l}-u_{\tilde 5,j}-\frac{i}{2}}
{u_{6,l}-u_{\tilde 5,j}+\frac{i}{2}}
\prod_{j=1}^{\tilde{K}_7} \frac{u_{6,l}-u_{\tilde 7,j}-\frac{i}{2}}
{u_{6,l}-u_{\tilde 7,j}+\frac{i}{2}}\prod_{\stackrel{k=1}{k \neq l}}^{K_6} 
\frac{u_{6,l}-u_{6,k}+i}
{u_{6,l}-u_{6,k}-i} & =& 1,~~l=1,\dots,K_6
\label{dualbet3}
\ee
Moreover, the equations for the type-4 Bethe roots turn out to be undeformed:
\be
\label{dualbethe4}
e^{-i p_{k} L} = \tilde \Lambda_{sl2}(p_{k})=\prod_{i=1}^{N} S_{0}^{2}(p_{k},p_{i})\prod_{j=1}^{\tilde{m}_1} \frac{1}{\eta(p_{k})} \frac{x^{+}(p_{k})-x^{+}(\tilde{\lambda}_j)}{x^{-}(p_{k})-x^{+}(\tilde{\lambda}_j)}
\prod_{j=1}^{\dot{\tilde{m}}_1} \frac{1}{\eta(p_{k})} 
\frac{x^{+}(p_{k})-x^{+}(\dot{\tilde{\lambda}}_j)}{x^{-}(p_{k})-x^{+}(\dot{\tilde{\lambda}}_j)}\,.
\ee
One can recover the equations (\ref{dualbetheexp}) and
(\ref{dualbethe4}) also by dualizing directly (using the relations
(\ref{dual1}), (\ref{dual2})) the corresponding Bethe equations in the $su(2)$ 
grading, namely (\ref{betheexp}) and (\ref{bethe4}).
Setting again $L=-J$, substituting the result for the scalar factor in 
MM (36) and changing notations, we obtain
\bear
e^{i p_k 
\left[J+\frac{1}{2}(\tilde{K}_{3}-\tilde{K}_{1})+\frac{1}{2}(\tilde{K}_{5}-\tilde{K}_{7})\right]}
& = & 
\prod_{\stackrel{i=1}{i \neq k}}^{K_4} 
\left[ \frac{x^{-}_{4,k}-x^{+}_{4,i}}{x^{+}_{4,k}-x^{-}_{4,i}} \right]
\left[ \frac{1-\frac{g^2}{x^{+}_{4,k} x^{-}_{4,i}}}
{1-\frac{g^2}{x^{-}_{4,k} x^{+}_{4,i}}} \right ]
[\sigma(p_k,p_i)]^2 
\nonumber \\
&& \times 
\prod_{j=1}^{\tilde{K}_3}  
\frac{x^{+}_{4,k}-x_{\tilde{3},j}}{x^{-}_{4,k}-x_{\tilde{3},j}} 
\prod_{j=1}^{\tilde{K}_1}  
\frac{1-\frac{g^2}{x^{+}_{4,k} x_{\tilde{1},j}}}
{1-\frac{g^2}{x^{-}_{4,k} x_{\tilde{1},j}}} 
\nonumber \\
&& \times 
\prod_{j=1}^{\tilde{K}_5}  
\frac{x^{+}_{4,k}-x_{\tilde{5},j}}{x^{-}_{4,k}-x_{\tilde{5},j}} 
\prod_{j=1}^{\tilde{K}_7}  
\frac{1-\frac{g^2}{x^{+}_{4,k} x_{\tilde{7},j}}}
{1-\frac{g^2}{x^{-}_{4,k} x_{\tilde{7},j}}} ,~~~~k=1,\dots,K_4\,.
\label{dualbet1}
\ear
We can now recover the corresponding untwisted Bethe equations of Beisert and
Staudacher \cite{BeiSta} in the grading $\eta_{1}=\eta_{2}=-1$ by setting
$P=0$ and recalling (see (5.6) in \cite{BeiSta}) the definition of the
angular momentum charge for that grading
\be
J = {\cal L} - \frac{1}{2}(\tilde{K}_{3}-\tilde{K}_{1}) - \frac{1}{2}(\tilde{K}_{5}-\tilde{K}_{7})  \,.
\label{dualJ}
\ee 

\subsection{Comparison with BR}\label{sec:dualBR}

Since BR \cite{BR} does not explicitly consider the all-loop twisted Bethe 
equations in the $sl_{2}$ grading, a little more effort is required 
to make the comparison. The BR Bethe equations in this grading are 
still given by (\ref{BR}), where now
\be
U_{\tilde{1}}(x)=U_{\tilde{3}}^{-1}(x)=U_{\tilde{5}}^{-1}(x)=U_{\tilde{7}}(x)=
\prod_{k=1}^{K_4}
S\indup{aux}^{-1}(x_{4,k},x)
\ee
and 
\be
U_4(x)=
U\indup{s}(x)
\lrbrk{\frac{x^-}{x^+}}^{\cal L}
\prod_{k=1}^{K_{4}}
S^{2}\indup{aux}(x,x_{4,k})
\prod_{k=1}^{\tilde{K}_1}
S\indup{aux}(x,x_{\tilde{1},k})
\prod_{k=1}^{\tilde{K}_3}
S^{-1}\indup{aux}(x,x_{\tilde{3},k})&&\non\\
\times\prod_{k=1}^{\tilde{K}_5}
S^{-1}\indup{aux}(x,x_{\tilde{5},k})
\prod_{k=1}^{\tilde{K}_7}
S\indup{aux}(x,x_{\tilde{7},k}).&&
\ee
(The quantities $U_{0}\,, S\indup{aux}\,, U\indup{s}$ are the same as 
before, see (\ref{Us1}),  (\ref{Us3}).)

\begin{figure}\centering
\begin{minipage}{260pt}
\setlength{\unitlength}{1pt}%
\small\thicklines%
\begin{picture}(260,20)(-10,-10)
\put(  0,00){\circle{15}}%
\put(  7,00){\line(1,0){26}}%
\put( 40,00){\circle{15}}%
\put( 47,00){\line(1,0){26}}%
\put( 80,00){\circle{15}}%
\put( 87,00){\line(1,0){26}}%
\put(120,00){\circle{15}}%
\put(127,00){\line(1,0){26}}%
\put(160,00){\circle{15}}%
\put(167,00){\line(1,0){26}}%
\put(200,00){\circle{15}}%
\put(207,00){\line(1,0){26}}%
\put(240,00){\circle{15}}%
\put( -5,-5){\line(1, 1){10}}%
\put( -5, 5){\line(1,-1){10}}%
\put( 75,-5){\line(1, 1){10}}%
\put( 75, 5){\line(1,-1){10}}%
\put(155,-5){\line(1, 1){10}}%
\put(155, 5){\line(1,-1){10}}%
\put(235,-5){\line(1, 1){10}}%
\put(235, 5){\line(1,-1){10}}%
\put( 40,00){\makebox(0,0){$+$}}%
\put(120,00){\makebox(0,0){$-$}}%
\put(200,00){\makebox(0,0){$+$}}%
\end{picture}
\end{minipage}

\caption{Dynkin diagram of $su(2,2|4)$ for $\eta_{1}=\eta_{2}=-1$.}
\label{fig:DynkinHighersl2}
\end{figure}
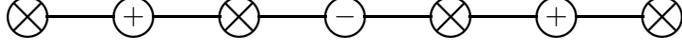

For the $sl_{2}$ grading with $\eta_{1}=\eta_{2}=-1$ which we now 
consider, $M_{j,j'}$
is the Cartan matrix specified by Fig.  \ref{fig:DynkinHighersl2} 
(see Eq.  (5.1) in \cite{BeiSta}).
The twist matrix ${\bf A}$ is given by BR (5.19) \footnote{We note that Eqs. (\ref{BR}) 
for $\eta_{1}=\eta_{2}=-1$ can be alternatively obtained by writing Eqs. (6.1) in \cite{Gromov:2007ky} 
with $\eta=-1$, setting their twists as
\be 
e^{i(\phi_{i}-\phi_{i+1})}=e^{-i(\mat{A}\vect{K})_{i}}\,,\non
\ee 
and, as suggested there, exchanging the twists $\phi_{1}\leftrightarrow\phi_{2}$,  
$\phi_{3}\leftrightarrow\phi_{4}$, $\phi_{5}\leftrightarrow\phi_{6}$, $\phi_{7}\leftrightarrow\phi_{8}$.
}
\be
\label{eq:TwistDynkin}
\mat{A}=
\delta_1
\lrbrk{\vect{q}_{p}^{}\vect{q}_{q_2}^\trans
-\vect{q}_{q_2}^{}\vect{q}_{p}^\trans}
+\delta_2
\lrbrk{\vect{q}_{q_2}^{}\vect{q}_{q_1}^\trans
-\vect{q}_{q_1}^{}\vect{q}_{q_2}^\trans}
+\delta_3
\lrbrk{\vect{q}_{q_1}^{}\vect{q}_{p}^\trans
-\vect{q}_{p}^{}\vect{q}_{q_1}^\trans} \,,
\ee 
where we take the three charge vectors to be
\bear\label{eq:ChargeDynkin}
\vect{q}_{q_1}\eq(\phantom{+}0\mathpunct{|}+1,-2,+1,\phantom{+}0,\phantom{+}0,\phantom{+}0,\phantom{+}0),\non\\
\vect{q}_{p_{\hphantom{1}}}  \eq(+1\mathpunct{|}\phantom{+}0,+1,-1,\phantom{+}0,-1,+1,\phantom{+}0),\non\\
\vect{q}_{q_2}\eq(\phantom{+}0\mathpunct{|}\phantom{+}0,\phantom{+}0,\phantom{+}0,\phantom{+}0,+1,-2,+1),
\ear
such that the Dynkin labels $[q_{1},p,q_{2}]$ in the grading $\eta_{1}=\eta_{2}=-1$
(see, for instance, Eq. (5.3) in \cite{BeiSta}) can be extracted as 
\be
\vect{q}_{q_1}\cdot \vect{K}=q_1,\qquad
\vect{q}_{p}\cdot \vect{K}=p,\qquad
\vect{q}_{q_2}\cdot \vect{K}=q_2\,, 
\ee
where now $\vect{K}= (L\mathpunct{|}\tilde K_{1}, K_{2}, \tilde K_{3}, 
K_{4}, \tilde K_{5}, K_{6}, \tilde K_{7})$. Explicitly, the twisting matrix {\bf A} reads
\[
\label{eq:TwistDual}
\mat{A}=
\left(
\mbox{\scriptsize$\displaystyle
\begin{array}{c|ccccccc}
0&-\delta_3&+2\delta_{3}&-\delta_3&0&+\delta_1&-2\delta_{1}&+\delta_1\\\hline
+\delta_3&0&+\delta_{3}&-\delta_3&0&-\delta_2-\delta_3&+2\delta_{2}+\delta_{3}&-\delta_2\\
-2\delta_{3}&-\delta_{3}&0&+\delta_{3}&0&+\delta_{1}+2\delta_{2}+2\delta_{3}&-2\delta_{1}-4\delta_{2}-2\delta_{3}&+\delta_{1}+2\delta_{2}\\
+\delta_3&+\delta_3&-\delta_{3}&0&0&-\delta_1-\delta_2-\delta_3&+2\delta_{1}+2\delta_{2}+\delta_{3}&-\delta_1-\delta_2\\ 
0&0&0&0&0&0&0&0\\ 
-\delta_1&+\delta_2+\delta_3&-\delta_{1}-2\delta_{2}-2\delta_{3}&+\delta_1+\delta_2+\delta_3&0&0&+\delta_{1}&-\delta_1\\
+2\delta_{1}&-2\delta_{2}-\delta_{3}&+2\delta_{1}+4\delta_{2}+2\delta_{3}&-2\delta_{1}-2\delta_{2}-\delta_{3}&0&-\delta_{1}&0&+\delta_{1}\\
-\delta_1&+\delta_2&-\delta_{1}-2\delta_{2}&+\delta_1+\delta_2&0&+\delta_1&-\delta_{1}&0
\end{array}$}
\right).
\]
It then follows from (\ref{gammas}) and BR (4.27) that
\be
(\mat{A}\vect{K})_{0} &=& \frac{1}{2}\big[ 
\gamma_{2}\left(\tilde{K}_{1}+\tilde{K}_{3}-\tilde{K}_{5}-\tilde{K}_{7}-2K_{2}+2K_{6}\right) \non \\
& & 
+\gamma_{3}\left(-\tilde{K}_{1}-\tilde{K}_{3}-\tilde{K}_{5}-\tilde{K}_{7}+2K_{2}+2K_{6}\right) \big] \,,  \label{dualAK0} \\
(\mat{A}\vect{K})_{1} -\frac{1}{2}(\mat{A}\vect{K})_{0} &=& \frac{1}{2}\left[
(\gamma_{3}-\gamma_{2}) J + \gamma_{1}\left(\tilde K_{5}+ \tilde K_{7}-2K_{6}\right) \right] 
\,, \label{dualAK1} \\
(\mat{A}\vect{K})_{2} &=&-2 \left[(\mat{A}\vect{K})_{1} -\frac{1}{2}(\mat{A}\vect{K})_{0}\right]\,, \label{dualAK2} \\
(\mat{A}\vect{K})_{3} + \frac{1}{2}(\mat{A}\vect{K})_{0} &=&  
(\mat{A}\vect{K})_{1} -\frac{1}{2}(\mat{A}\vect{K})_{0}\,,  \label{dualAK3} \\
(\mat{A}\vect{K})_{4}  &=& 0 \,.  \label{dualAK4} \\
(\mat{A}\vect{K})_{5} + \frac{1}{2}(\mat{A}\vect{K})_{0} &=& \frac{1}{2}\left[
(\gamma_{3}+\gamma_{2}) J - \gamma_{1}\left(\tilde K_{1}+ \tilde K_{3}-2K_{2}\right) \right] 
\,,  \label{dualAK5} \\
(\mat{A}\vect{K})_{6} &=&-2 \left[(\mat{A}\vect{K})_{5} + \frac{1}{2}(\mat{A}\vect{K})_{0} \right]\,, \label{dualAK6} \\
(\mat{A}\vect{K})_{7} -\frac{1}{2}(\mat{A}\vect{K})_{0} &=&  
(\mat{A}\vect{K})_{5} + \frac{1}{2}(\mat{A}\vect{K})_{0}\,.  \label{dualAK7} 
\ee 
As already noted in (\ref{P}), the total momentum is given by $P = - 
(\mat{A}\vect{K})_{0}$.

We now compare the BR Bethe equations with the ones which we 
derived above by dualization.
The fact that Eqs. (\ref{dualbet1}) are not deformed  
matches with (\ref{dualAK4}) and (\ref{BR}) with $j=4$.
Substituting for $P$ using (\ref{P}), and noting the following identities 
(proved using (\ref{ctildes}), (\ref{dualJ}) and 
(\ref{dualAK0})-(\ref{dualAK7})),
\be
&c_{1} e^{-i P/2} = e^{i(\mat{A}\vect{K})_{1}}\,, \quad
c_{1} e^{i P/2} = e^{i(\mat{A}\vect{K})_{3}}\,, \quad
c_{2} e^{i P/2} = e^{i(\mat{A}\vect{K})_{5}}\,, \quad
c_{2} e^{-i P/2} = e^{i(\mat{A}\vect{K})_{7}}\,,& \non\\
&c_{1}^{-2}=e^{i(\mat{A}\vect{K})_{2}}\,, \quad
c_{2}^{-2}=e^{i(\mat{A}\vect{K})_{6}}\,,&
\ee
we see that Eqs. (\ref{dualbet2K1})-(\ref{dualbet3}) match with 
(\ref{BR}) with $j=1,3,5,7,2,6$ respectively.

\end{appendix}

\end{document}